\documentclass[
        reprint,
 amsmath,amssymb,
 aps,
]{revtex4-1}

\usepackage[english]{babel}
\usepackage{graphicx}
\usepackage{color}
\usepackage{amssymb,amsfonts,amsmath}
\usepackage{xcolor}
\usepackage{gensymb}
\usepackage{natbib}

\begin{document}

\title{
\large{Geometrically Induced Selectivity and Unidirectional Electroosmosis
in Uncharged Nanopores} \\
}


\author{\normalsize{
        Giovanni Di Muccio$^1$,
        Blasco Morozzo della Rocca$^2$,
        Mauro Chinappi$^1$
        }}

 \email{mauro.chinappi@uniroma2.it}

 \affiliation{$^1$ Dipartimento di Ingegneria Industriale, Universit\`a di Roma Tor Vergata, Via del Politecnico 1,
                 00133, Rome, Italy. \\
              $^2$ Dipartimento di Biologia, Universit\`a di Roma Tor Vergata, Via della Ricerca Scientifica 1,
                00133, Rome, Italy.}

\date{\today}


\begin{abstract}
Selectivity towards positive and negative ions in nanopores 
is often associated with electroosmotic flow, 
the control of which is pivotal in several micro-nanofluidic technologies.
Selectivity is traditionally understood to be 
a consequence of surface charges that alter the
ion distribution in the pore lumen.
Here we present a purely geometrical mechanism to induce ionic selectivity 
and electroosmotic flow in uncharged nanopores 
and we tested it via molecular dynamics simulations.
Our approach exploits the accumulation of charges,
driven by an external electric field,
in a coaxial cavity that decorates the membrane close
to the pore entrance.
The selectivity was shown to depend on the applied voltage
and results to be completely inverted when reverting 
the voltage. The simultaneous inversion of ionic selectivity 
and electric field direction causes a unidirectional 
electroosmotic flow.
We developed a quantitatively accurate theoretical model
for designing pore geometry to achieve the desired electroosmotic velocity.
Finally, we show that unidirectional electroosmosis 
also occurs in much more complex scenarios, such as a biological pore whose structure presents 
a coaxial cavity surrounding the pore constriction as well as a complex surface charge pattern.
The capability to induce ion selectivity without altering 
the pore lumen shape or the surface charge may open to a
more flexible design of selective membranes. 
\end{abstract}


\maketitle

\vspace{0.3 cm}

\vspace{0.3 cm}


Transport of ions, water, small molecules and polymers 
through transmembrane protein channels
plays a fundamental role in sustaining cellular life
and it is drawing increasing attention
thanks to the recent progress of nanofluidic 
technology~\cite{bocquet2020nanofluidics}.
High cation or anion selectivity~\cite{hong2017scalable},
diode-like current rectification~\cite{siwy2006ion,karnik2007rectification},
different gating 
mechanisms~\cite{beckstein2001hydrophobic,powell2011electric,wilson2018water,camisasca2020gas},
surprisingly large flow 
rates~\cite{agre2004aquaporin,gravelle2014large,secchi2016massive,holt2006fast}
and other unexpected and {\it exotic} fluid phenomena at the nanoscale
were unveiled in the last two decades~\cite{kavokine2020fluids}.
This fostered the  development of technological applications
based on either biological or synthetic nanopores,
such as single molecule nanopore 
sensing~\cite{betermier2020single,gu1999stochastic}
blue energy harvesting~\cite{feng2016single,siria2013giant}
and high-throughput biomimetic filters~\cite{tu2020rapid}.

The coupling of the extreme fluid confinement, 
geometrical shape and interfacial physico-chemical properties
leads to non-trivial electrohydrodynamic phenomena in 
nanofluidic systems.
For example, cation or anion selectivity in nanopores 
is traditionally understood
to be a consequence of charges present on the pore wall.
Indeed, the electrolyte solution in contact with a charged surface
forms an oppositely charged diffused layer, known as the Debye layer,
at the solid-liquid interface~\cite{schoch2008transport}.
Due to the high surface-to-volume ratio,
the Debye layer often occupies a non-negligible part 
of the lumen 
of charged nanopores. 
When a voltage is applied across the pore,
the total electric current will be mostly formed
by the predominant mobile charges (cations or anions) 
present in the Debye layer,
resulting in a selective ionic transport.
Moreover, the Coulombic force acting on the net charge of the Debye layer
results in a force on the solvent that generates a fluid motion, 
usually indicated as electroosmotic flow (EOF).
EOF plays a relevant role in nanopore sensing technology
since it can compete or cooperate with 
electrophoresis and dielectroforetic forces acting on the 
analyte~\cite{boukhet2016probing,chinappi2020analytical}
and it can be exploited to capture molecules 
independently of their charge~\cite{huang2017electro,asandei2016electroosmotic}.

Many studies aimed at tuning ionic selectivity and EOF
involve the chemical modification of the pore to introduce surface charges
\cite{ramirez2003modeling,small2015nanoporous,zeng2015ion}
but other mechanisms have been exploited.
An example is provided by externally gated nanopores, 
where the pore surface charge 
is controlled via additional 
electrodes~\cite{nishizawa1995metal,kalman2009control,guan2014voltage,cheng2018low,fuest2015three,ren2017nanopore}
applied to the membrane substrate.
External gating allows
 to achieve a good control of the pore selectivity,
although the complex fabrication {\sl de facto} 
limits its application 
for pores of nanometer or sub-nanometer diameter.
Another strategy that can be employed 
to tune pore selectivity 
exploits Induced-Charge Electrokinetic (ICEK) phenomena.
Differently from externally gated 
selectivity control, 
in ICEK the same external electric field that drives the ions 
through the pore also polarizes the
solid membrane inducing a surface charge that,
in turn, alters the Debye layer in the nanochannel
and, hence, the selectivity and 
the EOF~\cite{bazant2010induced,yao2020induced}.
A core ingredient to generate a net EOF by ICEK is the presence
of some asymmetries in the system
that gives rise 
to inhomogeneities of ionic density distributions  
along the pore in response to the applied voltage.
In the nanopore realm, this asymmetry is often introduced in 
the pore geometry (e.g., conical pores~\cite{yao2020induced}) 
or imposing
salt gradients through the membrane~\cite{hsu2018theory}.

\begin{figure*}[!ht]
        \centering
        \includegraphics[width=0.95\textwidth]{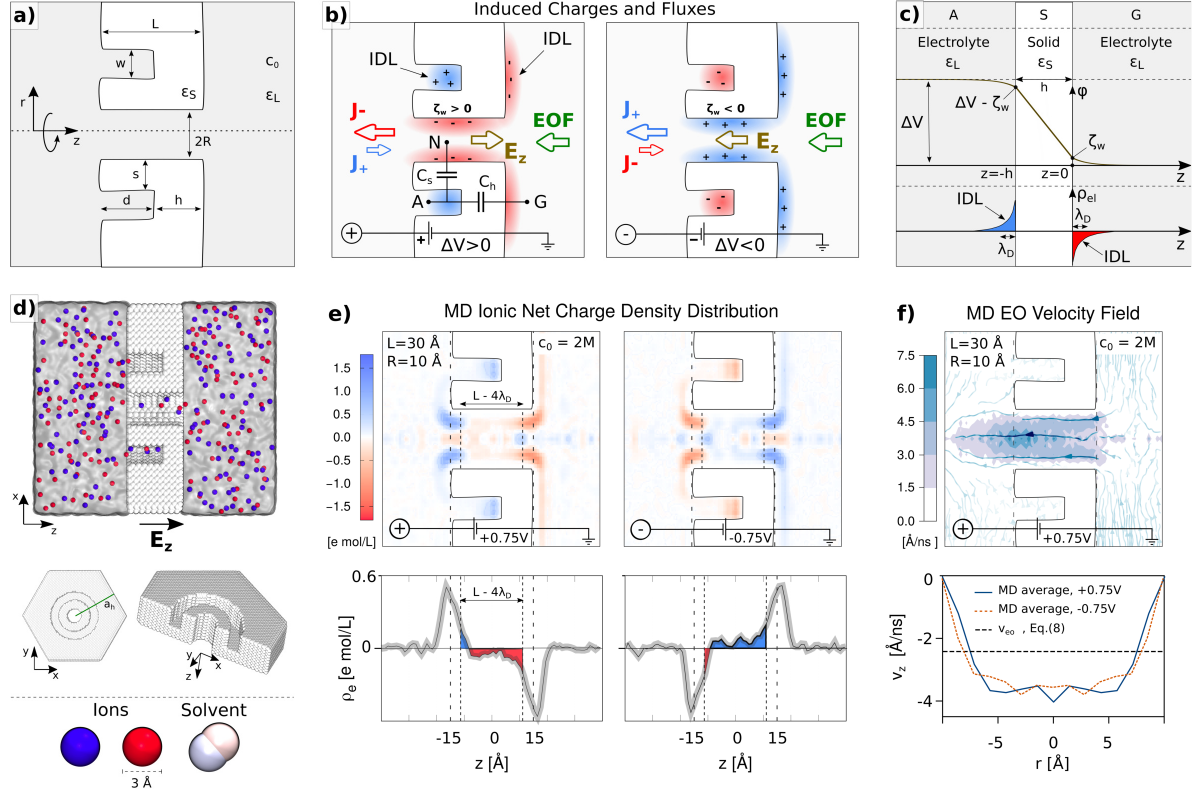}
        \caption{
		\label{fig:nanopore}
		{\bf Geometrically Induced Selectivity Switch.}
		{\bf a)}~Geometry of the system.
                A nanopore of radius $R$ is drilled through a membrane
                of thickness $L$. The channel is surrounded by
                a coaxial cavity of width $w$ and depth $d = L-h$,
		at a distance $s$ from the nanopore wall.
		{\bf b)}~Working principle. 
		An external applied voltage $\Delta V$
		gives rise to induced Debye layers (IDLs) 
		at the solid-liquid interfaces,
		the polarity of which depends on the voltage sign.
		Meanwhile, the electric field $E_z$ 
		drives the ions through the nanopore.
		The presence of a charged IDL 
		inside the nanopore results in 
		a selective ionic transport ($J_+ \ne J_-$),
		causing an electroosmotic flow (EOF).
		Since both the electric field $E_z$ and 
		the selectivity depend on the applied voltage polarity,
		the EOF (green arrow) is
		always oriented in the same direction.
		{\bf c)}~Planar electrolytic capacitance.
		An infinite neutral membrane separates 
		two reservoirs filled by the same 
		electrolyte solution.
		When a voltage $\Delta V$ 
		is imposed across the membrane,
		surface electric potentials~$\pm \zeta_w$ 
		arise at the solid-liquid interfaces 
		and charges are accumulated in the  
		IDLs (blue and red areas) whose
		characteristic size is the Debye length $\lambda_D$.
		{\bf d)}~Molecular Dynamics set-up 
		and tilted views of the membrane.
		White spheres represent the solid membrane atoms,
		blue and red ones the positive and negative ions,
		and the transparent gray background 
		the solvent, composed of dipolar diatomic molecules,
		shown at the bottom.
		{\bf e)} Charge distribution 
		from MD at $\Delta V = \pm 0.75V$, 
		with $c_0=2$~M salt concentration.
		The bottom plots represent the average net charge density 
		in cylindrical sections of radius 
		$R=10~\mathrm{\AA}$ along the pore axis.
		Confidence intervals, calculated using a block average
		with each block corresponding to 
		$10~\mathrm{ns}$, are reported in shaded gray.
		{\bf f)} Electroosmotic velocity field
		from MD at $\Delta V = +0.75V$.
		Bottom panel represents the MD 
		average velocity profile ($v_z$ component)
		inside the pore ($\vert z \vert < L/2-2\lambda_D$)
		at $\Delta V = \pm 0.75~\mathrm{V}$.
		The dashed line represents the model prediction,
		Eq.~\eqref{eq:veo}.
		MD distributions and fluxes are averaged 
		over $800$~ns MD trajectory (16\;000 frames),
		see methods.
                }
\end{figure*}

Here, 
we propose a mechanism to induce a voltage-dependent ionic selectivity 
and EOF in uncharged cylindrical nanopores
by taking advantage of geometrical asymmetries of the membrane
without any external voltage-gating control, salt gradient, 
or chemical modification of the pore surface.
Our system, Fig.~\ref{fig:nanopore}a-b, 
exploits the accumulation of charge
between the pore lumen and a coaxially surrounding cavity.
The induced selectivity 
is completely inverted by reverting the applied electric field. 
The concurrent inversion of ionic selectivity and applied voltage
generates a unidirectional EOF, 
independently of the applied voltage polarity.
Since the same electrical field that induces the pore selectivity
is also responsible for the ion motion, the mechanism we propose can be
included into the broad class of ICEK phenomena.
We developed a theory, based on 
a continuum electrohydrodynamical description,
to assess the dependence of selectivity and EOF
from applied voltage $\Delta V$ and pore geometry.

As a proof of principle, 
we set up molecular dynamics (MD) simulations of 
a model system composed of an uncharged
 solid-state nanopore surrounded by a coaxial cavity, 
Fig.~\ref{fig:nanopore}a,d.
Our MD results show that the EOF 
depends quadratically on $\Delta V$,
in agreement with the theory.
We also explored more complex scenarios 
where a surface charge is present at the pore wall to understand
in which conditions the geometrically induced EOF 
is predominant with respect to EOF due to fixed surface charge.
We finally show that selectivity switch
and unidirectional EOF 
may also occur for the CsgG bacterial amyloid secretion 
channel~\cite{cao2014structure,goyal2014structural},
a protein pore employed in a commercial 
nanopore sequencing device.\cite{van2020dual}
CsgG has a coaxial cavity
like our simplified model and, in addition,
presents a complex surface charge 
pattern, as usual for biopores.

\begin{figure}[]
        \centering
        \includegraphics[width=0.4\textwidth]{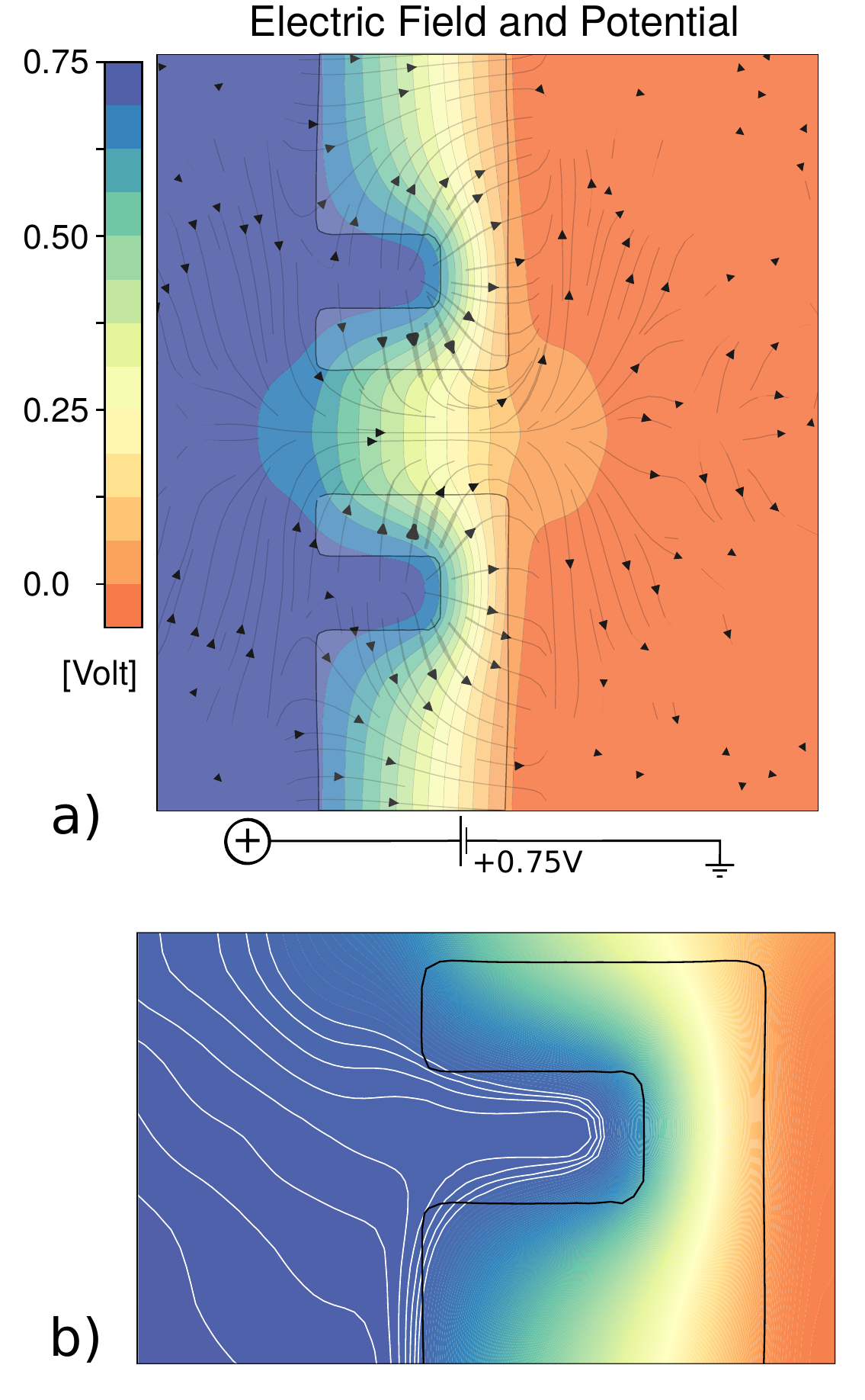}
        \caption{
		\label{fig:efpot}
		{\bf a) Electric potential map.}
		The black arrowed lines represent the electric field 
		$\mathbf{E}(r,z)=-\nabla V$.
		We filtered out the lines where 
		$|\mathrm{E(r,z)}|<13\%$ of the maximum intensity.
		The potential map is averaged 
		over $800$~ns MD trajectory (16\;000 frames),
		see methods and refer to
		the MD simulation of the 2M system shown 
		in Fig.~\ref{fig:nanopore}d-f,
		with $R=10$\AA, $L=30$\AA, $h=10$\AA, $s=9$\AA, $w=12$\AA~
		 at $\Delta V= +0.75\,$V transmembrane applied bias.
		{\bf b)} Zoom on the cavity.
		The isolines roughly follow 
		the solid walls, indicating 
		the presence of the induced Debye layer
		inside the cavity. Selected isolines in the left reservoir 
		were highlighted in white for clarity.
		}
\end{figure}

\begin{figure*}[]
        \centering
        \includegraphics[width=0.95\textwidth]{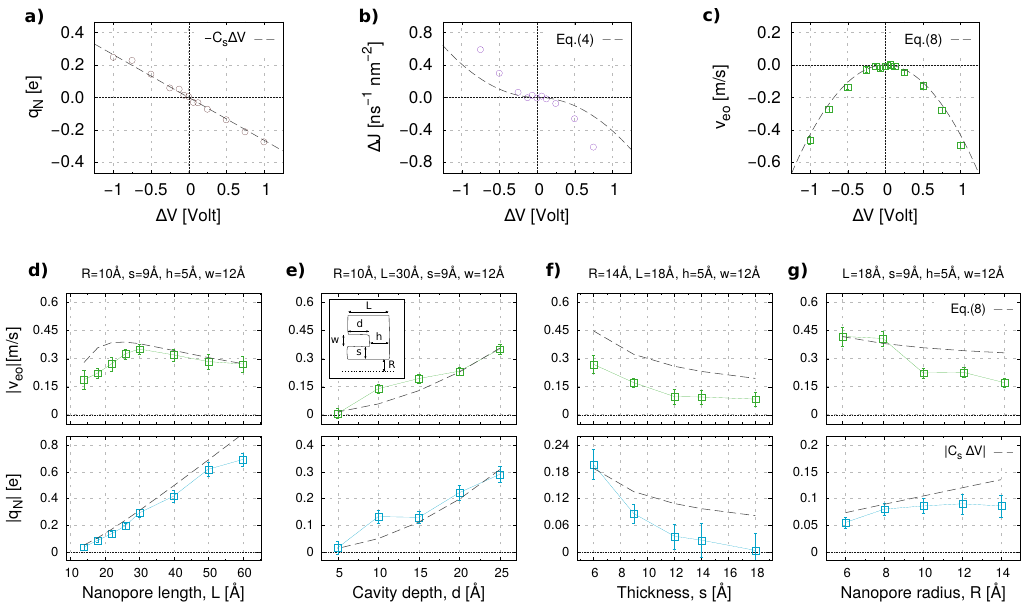}
        \caption{
		\label{fig:EOF-parametric}
		{\bf Electrohydrodynamic fluxes and charges in the nanopore.}
		{\bf a-c)}~ 
		\label{fig:MDresults}
		Charge in the pore ($q_N$), 
		selectivity ($\Delta J$) and 
		average EO velocity ($v_{eo}$)
		from MD simulation of the 2M system shown 
		in Fig.~\ref{fig:nanopore}d-f,
		with $R=10$\AA, $L=30$\AA, $h=10$\AA, $s=9$\AA, $w=12$\AA.
		Dashed lines refer to the analytical model described in the text.
		{\bf d-g)}~Electroosmotic velocity and total charge in the pore 
		as a function of 
		{\bf d)}~pore length $L$, {\bf e)}~depth of the cavity $d$,
		{\bf f)}~thickness $s$ and of {\bf g)}~pore radius $R$. 
		Analytical model results are shown as dashed lines  
		and MD data as colored squares. 
		Each error bar represents the standard error obtained from 
		an $800$~ns MD trajectory (16\;000~frames). 
		Inset in {\bf e)}~recalls the geometric parameters of our model.
                }
\end{figure*}

\section{Results and discussion}

{\bf Geometrically Induced Selectivity Switch: 
working principle and MD simulations.}
\label{sec:model}
Let us consider the system 
represented in Fig.~\ref{fig:nanopore}a,
composed of a solid insulating membrane (white) of thickness $L$
with a cylindrical nanopore of radius $R$,
surrounded by a coaxial cavity of width $w$ and depth $d=L-h$, 
at a distance $s$ from the nanopore wall.
The membrane (relative permittivity $\varepsilon_S$)
is immersed in 1:1 electrolyte solution
(gray background) with relative permittivity $\varepsilon_L$
and oppositely charged ions 
with the same ion mobility
$\mu_{\pm} = \mu$.
The pore is completely uncharged, 
so equilibrium 
(no applied voltage)
ionic concentrations $c_+$ and $c_-$
are homogeneous everywhere and equal to the bulk value
$c_0$.
When a voltage $\Delta V$ 
is applied across the nanopore,
two main effects occur, as sketched in Fig.~\ref{fig:nanopore}b;
i) ions flow through the pore lumen 
($J_+$ and $J_-$ arrows)
and ii) 
Induced Debye Layers (IDL) form at the solid walls 
(blue and red charged clouds), 
depending on the voltage polarity.
The presence of the cavity affects
the IDL shape resulting in an accumulation 
of charges across the cavity and the nanopore lumen,
whose signs depend on the voltage polarity, 
see Fig.~\ref{fig:nanopore}b.
The broken electroneutrality inside the pore
results in ionic selectivity 
(anionic and cationic currents are different) 
and EOF.

In order to catch the dependence of the pore selectivity 
on the applied voltage $\Delta V$ we reasoned as follows.
As a first approximation, 
electrophoretic ionic fluxes are proportional
to the concentration and mobility of each 
species~\cite{schoch2008transport},
${\bf{J}}_{\pm} = \pm \mu \, c_{\pm} \, {\bf{E}}$,
with ${\bf E}$ the driving electric field.
We use the difference between the 
cations and anions fluxes
as a measure of the ionic selectivity
\begin{equation}
	\Delta J = 
	\left \langle |J_+| - |J_-| \right \rangle_N
	\approx 
	\mu 
	\frac{
	\left \langle \rho_{el}\right \rangle_N
	}
	{\nu e} 
	|E_z|
	\; ,
	\label{eq:deltaJ}
\end{equation}
with $\rho_{el} = \nu e (c_+ - c_-)$ 
the net charge density,
$\nu$ the valence of the ions, $e$ the elementary charge
and where
$\left \langle \,..\,\right\rangle_N$
denotes the volumetric average inside the nanopore.
So, selectivity depends on the sign of the charge of the 
IDL inside the nanopore lumen.

To quantify the IDL in the nanopore,  
we focus on the positive voltage case
of Fig.~\ref{fig:nanopore}b, left side.
A potential difference is present between 
the lateral cavity (point A at potential $\Delta V$), 
and the right reservoir of the membrane (point G, grounded),
and between the cavity and the pore lumen (point N).

The planar membrane solution description
is instrumental to understanding the IDL dependence on
voltage, Fig.~\ref{fig:nanopore}c.
In the right reservoir (G), 
due to the potential difference 
($\zeta_w$)
between the bulk and the wall,
negative ions accumulate close to the membrane surface, 
red area.
Similarly, positive ions accumulate   
on the left side (A), blue area.
Inside the membrane the electric potential $\phi(z)$
decays linearly.
$\zeta_w$ is proportional to the applied voltage $\Delta V$,
see Supplementary Note S1 and Supplementary Fig.~S1 for details.
Since the accumulated charge in the IDL
is also linear in $\Delta V$,
the process can be described as a capacitance between A and G.
Extending this reasoning to our nanopore system,
the charge accumulation 
between 
the lateral cavity (point A) 
and the nanopore lumen (point N)
can be modeled as a capacitance.
Actually, the potential difference between 
the lateral cavity and the nanopore lumen 
is a function of the $z$ coordinates since 
the potential inside the pore lumen 
varies along the nanopore axis.
Nevertheless, in a quasi-1D approximation,
see Supplementary Note S1,
the total charge $q_N$ inside the nanopore 
is still proportional to the applied voltage, 
{\sl i.e.}, $q_N=-C_s \Delta V$, 
with 
\begin{equation}
	C_s =
	\pi \varepsilon_0 \varepsilon_S
	\frac
		{(L-h)^2}
		{L \ln\left(1+\frac{s}{R}\right)}
	\;
	\frac
		{L-4 \lambda_d}
		{L}
	\; ,
	\label{eq:cs}
\end{equation}
an equivalent capacitance between 
the cavity and the pore that depends only on 
geometrical parameters.
Therefore, the average net charge density
inside the nanopore is 
\begin{equation}
	\left\langle \rho_{el} \right \rangle_N=
		-\frac{C_s \Delta V}{\pi R^2 L}
	\; ,
	\label{eq:average-charge}
\end{equation}
and, consequently, the ionic selectivity, Eq.~\eqref{eq:deltaJ} reads
\begin{equation}
 \Delta J =
	-\frac{\mu}{\nu e} \, 
	\frac{C_s \vert \Delta V \vert }{\pi R^2 L^2} \Delta V
	\; .
	\label{eq:deltaJ2}
\end{equation}
Eq.~\eqref{eq:deltaJ2} shows that selectivity reverts when 
inverting the applied voltage $\Delta V$ and 
its magnitude depends on $\Delta V$ quadratically.

We tested the validity of the above analytical model at the nanoscale by using
all-atoms molecular dynamics (MD) simulations.
To get rid of any asymmetries of the electrolyte
that may potentially give rise to competing 
selectivity of the nanopore
(e.g., differences between ion mobilities, 
different hydration shells around cations and anions,
preferential interaction of one ion with the solid),
we built a custom symmetric model for the electrolyte solution. 
In particular, we considered 
two monovalent ionic species with the
same mass 
dissolved in a liquid composed of diatomic dipolar molecules. 
The membrane is composed of neutral atoms.
All the atoms have the same van der Waals
radius, and the volume of the solvent
molecule is similar to 
water, see methods for details and 
Supplementary Fig.~S4-S9
for a characterization of the fluid 
in terms of phase diagram, 
relative electrical permittivity, wetting, 
ion mobility, and viscosity.

We first studied a system 
with pore length $L=30~\mathrm{\AA}$, pore radius $R=10~\mathrm{\AA}$,
cavity width $w=12~\mathrm{\AA}$ and depth $d=10~\mathrm{\AA}$
at distance $s=9~\mathrm{\AA}$, 
for a 2M solution Fig.~\ref{fig:nanopore}d.
Ionic net charge densities are reported in Fig.~\ref{fig:nanopore}e
for positive $\Delta V= +0.75V$ and negative
applied voltage $\Delta V = -0.75V$ showing the formation
of IDLs.
It is apparent that when a positive voltage is applied,
positive charges are accumulated inside the cavity
and a corresponding negative IDL arises along the pore.
The opposite happens for negative bias.
The characteristic length scale of the IDL appears to be, 
as expected, of the order of the 
Debye length of the electrolyte solution, 
$\lambda_D\simeq 2$~\AA, in this case.
Moreover, liquid velocity profiles
show an EOF
directed from right to left 
for both positive
and negative voltages,
Fig.~\ref{fig:nanopore}f.
The MD simulations revealed additional
features of the charge distributions,
such as the two opposite charge density peaks
appearing at the nanopore entrance 
and discontinuous patterns along the pore axis.
Nevertheless, the overall IDL formation 
mechanism proposed in Fig.~\ref{fig:nanopore}b is confirmed:
when changing the applied voltage, the selectivity of the
pore switches from cations to anions.
	The electric potential estimated from MD
	simulations, Fig.~\ref{fig:efpot}, 
	further confirms the trend 
	of the voltage drops schematically described 
	in our model. The electric potential decreases quite linearly 
	along the pore, while large part of 
	the cavity is approximatively isopotential
	with respect to the left reservoir ($\Delta V=+0.75\,$V).
	More in detail, the 
	isolines follow the wall surface 
	inside the cavity,
	indicating that the IDL
	contours the wall profile,  
 	Fig.~\ref{fig:efpot}b.
\\

{\bf Parabolic electroosmosis.}
As anticipated in the previous section, 
a major consequence of the selectivity switch is
that the EOF is always negative in our framework 
(Fig.~\ref{fig:nanopore}b), {\sl i.e.}, 
directed from the right to the left side of the membrane,
for both positive and negative voltages.
An analytical insight 
into the dependence of EOF on $\Delta V$
can be derived using 
a continuum electrohydrodynamics approach
based on the Poisson-Nernst-Planck and 
Navier-Stokes (PNP-NS) equations~\cite{schoch2008transport}.
	PNP-NS system is derived under 
several assumptions that are not always 
respected at the nanoscale, such as the continuum
assumption. Moreover, in order to get 
a practical analytical solution, we needed to 
rely on several additional hypotheses, such as
dilute solution limit and homogeneous mobility.
A discussion of these hypotheses and their implications is reported
in Supplementary Note S2.
For $\lambda_D \ll R$ (no Debye layer overlap), PNP-NS 
predicts that the electroosmotic volumetric flow rate ($Q_{eo}$)
through a cylindrical channel of radius $R$ and length $L$ 
can be written as
\begin{equation}
	Q_{eo}  
	= \pi R^2 \vert v_{eo} \vert  
        \quad , \quad
	v_{eo} = - 
		\frac{\varepsilon_0 \varepsilon_L \zeta_w}{\eta} 
		\frac{\Delta V}{L} 
	\; ,
	\label{eq:Qeo}
\end{equation}
with 
$\varepsilon_L$ and $\eta$ 
relative permittivity and viscosity 
of the electrolyte solution;
$\zeta_w$ is the average surface electrokinetic
potential~\cite{herr2000electroosmotic}
and $v_{eo}$ is the Helmholtz-Smoluchowski 
electroosmotic velocity,
{\sl i.e.}, the velocity of the plug flow
obtained when 
$\lambda_D \ll R$.~\cite{bruustheoretical}
Note that, in this work, $v_{eo}$ is positive if directed
from left to right, see, Fig.~\ref{fig:nanopore}a.
In this framework, 
the net charge density $\rho_{el}$ 
and, hence, the total charge $q_N$ inside the nanopore 
are a function of $\zeta_w$
\begin{equation}
	q_N = 2\pi L
	\int_0^R  dr \,r  \rho_{el}(r)
	\approx -2\pi L \varepsilon_0 \varepsilon_L \,  
	\frac{R}{\lambda_D} \, \zeta_w 
	\; ,
	\label{eq:qN}
\end{equation}
where in the rightmost term we considered that 
for $R \gg \lambda_D$ the charge in the pore 
can be approximated as  the product of pore
surface $2\pi R L$ 
times the surface charge of a planar Debye layer 
$\varepsilon_0 \varepsilon_L \zeta_w / \lambda_D$~\cite{schoch2008transport}.
Thus, $\zeta_w$ is
proportional to $q_N$ and,
for Eq.~\eqref{eq:average-charge}, to $\Delta V$.
Combining Eqs.\eqref{eq:qN} and~\eqref{eq:average-charge}
we get
\begin{equation}
	\zeta_w = 
	\frac{\lambda_D}{R}
		\frac
		{C_s \Delta V}
		{2\pi \varepsilon_0 \varepsilon_L L} 
	\; ,
	\label{eq:zetaw}
\end{equation}
that, when introduced into Eq.~\eqref{eq:Qeo},
leads to the parabolic expression for the EOF velocity
\begin{equation*}
	v_{eo} = -
	\frac{\lambda_D C_s}{2\pi \eta R L^2}
	\; 
	\Delta V^2 \;
	=
	\qquad \qquad \qquad \qquad \; \; 
\end{equation*}
\begin{equation}
	= -
	\frac
		{\varepsilon_0 \varepsilon_S} 
		{2 \eta}
	\; 
	\frac
		{\lambda_D}
		{R}
	\;
	\frac
		{(L-h)^2   (L-4 \lambda_d)}
		{\ln\left(1+\frac{s}{R}\right)
		L^4}
	\Delta V^2 
	\;
	. 
	\label{eq:veo}
\end{equation}
Equations (\ref{eq:Qeo}-\ref{eq:veo})
are strictly valid only for $\lambda_D \ll R$,
and therefore, in principle,
accurate quantitative predictions cannot be expected.
Nevertheless, 
for the pore in 
Fig.~\ref{fig:nanopore}d-f 
($L=30$~\AA\, 
and $R=10$~\AA)
the model predictions are in very good 
agreement with MD data.
The capacitance $C_s$, Eq.~\eqref{eq:cs},
well predicts the dependence of 
net pore charge $q_N$ on $\Delta V$,
dashed line in 
Fig.~\ref{fig:MDresults}a.
The MD selectivity $\Delta J$, computed from
the ionic currents shown in Supplementary Fig.~S10,
is reported in Fig.~\ref{fig:MDresults}b,
confirming the selectivity switch
predicted by Eq.~\eqref{eq:deltaJ2} of our model.
The higher MD values may be explained by 
the convective contribution to ion transport
that is not included in Eq.~\eqref{eq:deltaJ}.
Indeed, since the EOF is
directed as the dominant ionic flow,
it always results in an increase of selectivity.
Finally, Eq.~\eqref{eq:veo} 
gives an excellent quantitative estimation of the 
average electroosmotic velocity,
$v_{eo}=Q_{eo} /\pi R^2$,
with $Q_{eo}$ computed from MD simulations,
Fig.~\ref{fig:MDresults}c.
\\

\begin{figure*}
        \centering
        \includegraphics[width=0.98\textwidth]{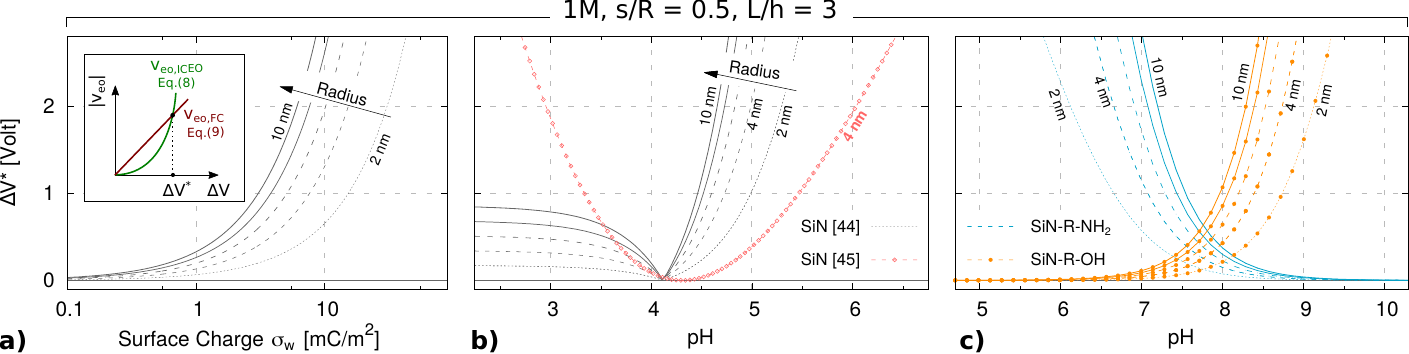}
        \caption{
		\label{fig:soglie}
		{\bf Threshold voltage $\Delta V^*$ in the presence of a fixed surface charge.}
		Threshold voltage $\Delta V^*$ is defined in Eq.~\eqref{eq:dvstar}
		as the voltage where the magnitude of fixed charge EO velocity, Eq.~\eqref{eq:veofc}, and 
		induced charge EO velocity, Eq.~\eqref{eq:veo}, are equal, as sketched in the inset 
		of panel {\bf a}. 
		{\bf a)} $\Delta V^*$ as a function of fixed surface charge $\sigma_w$, 
		for pores of increasing radius from $R = 2$~nm to $R=10$~nm.
		{\bf b)} pH dependence of $\Delta V^*$ for Silicon Nitride pores, for different radii.
		Experimental fit for $\sigma_w=\sigma_w(pH)$ dependency on pH was taken from Lin 
		{\sl et al.}~\cite{lin2021surface} (black curves), 
		or Bandara {\sl et al.}~\cite{bandara2019chemically} (red curve), see Methods. 
		{\bf c)} pH dependence of $\Delta V^*$ for surface-modified Silicon Nitride pores 
		with amine (cyan) or hydroxyl (orange) moieties, 
		$\sigma_w=\sigma_w(pH)$ taken from Bandara {\sl  et al.}~\cite{bandara2019chemically},
		see Methods. 
		Reported examples are with fixed ratios $L/h=3$ and $s/R=0.5$ at 1M KCl. 
		}
\end{figure*}

{\bf Effect of geometric parameters.}
To verify the robustness of the observed phenomenon 
and the accuracy of the proposed quantitative model,
we performed a second set of MD simulations
focusing on the role of geometrical parameters.
Each set of simulations is performed at $\Delta V = +0.75$~V,
by varying one single geometrical parameter 
while keeping fixed all the others.
Results are reported in Fig.~\ref{fig:MDresults}d-g,
with a sketch of the geometry reported 
in the inset of Fig.~\ref{fig:MDresults}e.
The electroosmotic velocity  
$\vert v_{eo} \vert$
is reported on the top panels,
while the total accumulated charge 
inside the nanopore $|q_N|$
is shown in the bottom ones.
We observe induced charge accumulation inside the pore 
and a concomitant EOF in all cases.
The general trends predicted by our model 
are in good agreement with the simulations. 
The quasi-1D capacitance model, Eq.~\eqref{eq:cs},
predicts the MD data within two error bars for almost all cases.
The analytical $v_{eo}$, Eq.~\eqref{eq:veo},
better matches the MD data for longer pores ($L>30$~\AA),
while it slightly overestimates the flow rates for the shorter ones,
see Fig.~\ref{fig:MDresults}d.
Anyhow, the model correctly indicates that
the dependence on $L$ is non-monotonic;
this is due to the competing effect between 
the driving electric field $E_z=\Delta V /L$,
which decreases with $L$,
and the induced capacitance
$C_s$, Eq.~\eqref{eq:cs}, that increases with $L$.
The induced charge effect and EOF 
increase with the cavity depth $d=L-h$,
Fig.~\ref{fig:MDresults}e,
consistently with the increase 
of the voltage drop
between the pore lumen and the deeper portion 
of the cavity,
see the quasi-1D pore capacitance model in Supplementary Note~S1,
and the electric potential maps in Supplementary Fig.~S11.
The geometrically induced selectivity vanishes for $d \to 0$, 
as trivially expected since the system becomes symmetric.
The MD data of 
Fig.~\ref{fig:MDresults}e 
refer to
a pore with $L=30~\mathrm{\AA}$ and, as for 
Fig.~\ref{fig:MDresults}d,
are in quantitative agreement with the model.
We also ran simulations for $L=18~\mathrm{\AA}$,
at different thickness $s$ and radius $R$.
In both cases, the model overestimates 
$q_N$ and $v_{eo}$ although capturing the trends
of the MD data, 
{\sl e.g.}, for increasing $s$ the lateral capacitance
decreases and so do $q_N$ and $v_{eo}$.
The apparent quantitative agreement 
for $R<10~\mathrm{\AA}$ 
could be more probably ascribed to fortuitous compensation
of different sources of atomistic effects 
more than to a correct description of such extremely confined conditions.

The geometrically induced selectivity and the unidirectional EOF
are not limited to nanometer and subnanometer scale. 
Eq.~\eqref{eq:veo}
 allows quantifying EOF for pores of any size
and can hence be employed for nanopore system design.
As an example, in Supplementary Fig.~S12, we report
$v_{eo}$ for a water electrolyte solution through a silicon nitride pore 
of radius $R = 20~\mathrm{nm}$.
Such relatively large pores are widely used in experimental 
studies~\cite{zeng2019rectification,houghtaling2019estimation}
and the required surface patterning can be achieved with well-established  
techniques~\cite{chou2020lifetime}.
Eq.~\eqref{eq:veo} indicates that 
as the system size increases, $\vert v_{eo} \vert$ decreases.
This decrease can be partially compensated using 
materials with larger dielectric constants 
or increasing the Debye length, 
as both $\lambda_D$ and $\varepsilon_S$ appear of Eq.~\eqref{eq:veo} numerator, 
but with some caveats discussed in Supplementary Note S2.
Briefly, for $\lambda_D$, 
Eq.~\eqref{eq:veo} can reasonably estimate
the flux only until $\lambda_D/R \ll 1$ (no Debye layer overlap).
Similarly, the low concentrations needed to
achieve relatively large $\lambda_D$ will result in 
a small number of ions in the nanopore, an occurrence 
which may lead to the failure 
of the PNP-NS model to yield quantitative predictions.
For a pore of radius $R= 20~\mathrm{nm}$,
Eq.~\eqref{eq:veo} indicates that  
a $\vert v_{eo} \vert \simeq 0.1~\mathrm{m/s}$ can be obtained, see Supplementary Fig. S12.
This EOF can be in principle experimentally measured.
A possible technique is the one proposed by 
Secchi et al.~\cite{secchi2016massive}, where 
the velocity field far from the pore is measured 
following the trajectory of tracers.
This approach allows to measure the 
flow only at a distance of a few 
$\mathrm{\mu m}$ but not close to the pore. 
Nevertheless, 
a $\vert v_{eo} \vert \simeq 0.1~\mathrm{m/s}$ 
at the exit of a pore of $R=20~\mathrm{nm}$
would result in a
velocity of magnitude $v \sim 0.4 \cdot 10^{-4}~\mathrm{m/s}$
at a distance of $1~\mathrm{\mu m}$ from the pore 
(fluid velocity scales as $1/r^2$, with $r$ the distance from the pore).
This value appears to be within reach of the 
proposed experimental technique~\cite{secchi2016massive}
and can be generated under an applied voltage of 
$1 \le  \Delta V \le 2$~Volt, depending on the 
salt concentration (0.2 or 0.02M) and the geometry,
see Supplementary Fig.~S12.

Another approach to experimentally validate
our results is to infer
the EOF from its effect on the capture 
of nanoparticles by a nanopore.
Indeed, the capture rate 
is ruled by 
the competition/cooperation of different effects,
the most relevant being electrophoresis, electroosmosis
and dielectrophoresis~\cite{huang2017electro,asandei2016electroosmotic,chinappi2020analytical}.
Analytical expressions for the capture rate have 
been recently proposed~\cite{chinappi2020analytical}
and, in principle, they allow directly to  
relate EOF and capture rate, if pore and particle geometry, charge
and dielectric properties are known.
Due to the difficulties in modeling pore entrance effects,
quantitatively accurate estimations of EOF are not expected;
nevertheless, a clear indication of the EOF direction 
and of the dependence of $v_{eo}$ on $\Delta V$ should be achievable.
\\

\begin{figure*}[]
	\centering
	\includegraphics[width=0.95\textwidth]{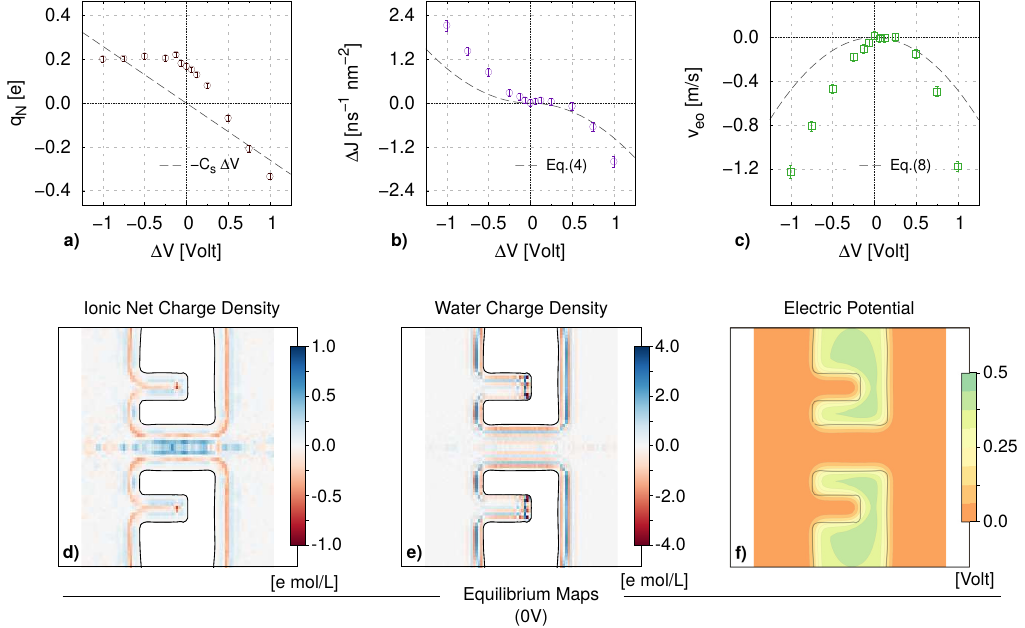}
        \caption{
		\label{fig:asymmetric}
		{\bf Effect of asymmetric electrolyte.}
		{\bf a-c)} Charge in the pore $q_N$, 
		selectivity $\Delta J$,
		and average EO velocity $v_{eo}$ from MD simulation 
		of a nanopore  
		with $R=10$\AA, $L=30$\AA, $h=10$\AA, $s=9$\AA, $w=12$\AA~
		(same as Fig.~\ref{fig:nanopore}d-f)
		in a 2M KCl water solution (symbols).
		Gray dashed lines represent the theoretical 
		predictions for a symmetric case, {\sl i.e.}, $q_N = -C_s \Delta V$
		for the nanopore charge and
		Eqs.~(\ref{eq:average-charge}-\ref{eq:veo}) for $\Delta J$,
                and $v_{eo}$;
		The other parameters used 
		are 
		$\mu = 1.0\times10^3\,\mathrm{\AA^2/(V\,ns)}$,
		$\lambda_D = 2.1\,$\AA,
		$\varepsilon_S=1$;
		$\eta=0.3\,$mPa\,s (TIP3P viscosity $\simeq 1/3$ experimental water~\cite{yeh2004system}).
		{\bf d-f)} 
		Ionic and water charge density and electric potential 
		at equilibrium ($\Delta V = 0$), 
		showing the intrinsic polarization
		and layering at the solid-liquid interface, 
		despite the zero charge of the solid membrane.
		The potential difference 
		between the bulk liquid and the membrane interior
		is related to the presence of interfacial charge dipoles.
		MD distributions and fluxes are averaged
		over 800 ns MD trajectory (16 000 frames).
		Errors are calculated using a block average 
		protocol with a block length of 10 ns.
	}
\end{figure*}
%
{\bf Application to weakly-charged solid-state nanopores.}
The theoretical model we developed is valid for 
neutral pores, {\sl i.e.}, no intrinsic surface charge 
is present at the pore walls.
For Silicon Nitride, a widely used material 
for solid-state nanopores, the zero-charge condition 
is achieved at $\mathrm{pH} \simeq 4.1$.~\cite{bandara2019chemically,hoogerheide2009probing,lin2021surface}
Moreover, coatings can be used to alter the zero-charge 
pH making it possible 
to get weakly charged pores 
(a few $\mathrm{mC/m^2}$) for 
wide ranges of pH.~\cite{bandara2019chemically} 
Instead, for $\mathrm{Hf O_2}$, another material used 
for nanopores~\cite{larkin2014high}, the zero-charge pH is 
$\simeq 7.5$.~\cite{kosmulski1997attempt}
A partial list of materials and conditions where
the nanopore surface is neutral and, hence, 
geometrically induced selectivity and EOF can
be effectively employed is reported in Supplementary Table S1.

The capability to control surface charge 
in solid-state pores naturally raise a question 
on the relative impact of 
EOF due to fixed surface charge  
and to the geometrically induced 
mechanism presented in this work.  
As a first approximation, EO velocity 
due to fixed surface charge density $\sigma_w$ can be expressed
as
\begin{equation}
v_{eo,FC} 
=
- 
\frac{\sigma_w \lambda_D}{\eta} 
\frac{\Delta V}{L} 
\; ,
\label{eq:veofc}
\end{equation}
that, in essence, is Eq.~\eqref{eq:Qeo} 
with $\sigma_w=\epsilon_0 \epsilon_L \zeta_w/\lambda_D$,
see Supplementary Note S2.
Since $v_{eo,FC}$ scales with $\Delta V$, while geometrically 
induced electroosmotic velocity, Eq.~\eqref{eq:veo}, scales
as $\Delta V^2$, at large enough $\Delta V$ the latter 
becomes dominant, see inset in Fig.~\ref{fig:soglie}a.
The magnitude of the threshold voltage $\Delta V^*$
where the intensity of two contributions is equal
can be obtained by combining Eqs.~\eqref{eq:veofc} and \eqref{eq:veo},
resulting in
\begin{equation}
\Delta V^* 
=
\vert \sigma_w \vert
\frac
{2 R L^3 \ln \left (1 + s/R \right )}
{\epsilon_0 \epsilon_S 
	\left (L-h \right)^2 
	\left(L - 4\lambda_D \right)}  
\, .
\label{eq:dvstar}
\end{equation}
$\Delta V^*$ depends not only on geometrical parameters
but also on surface charge $\sigma_w$ and Debye length 
$\lambda_D$ that,
in turn, depend on pore material, pH, and ionic strength.
As a first example, Fig.~\ref{fig:soglie}a reports $\Delta V^*$ as
a function of $\sigma_w$ in pores of radii between $2-10$~nm.
It is evident that, for $\sigma_w < 5~\mathrm{mC/m^2}$, 
$\Delta V^* \le 2$~V even for quite large nanopores ($R=10$~nm),
while $\Delta V^* \le 0.5$~V for the narrower one ($R=2$~nm).
Instead,
Fig.~\ref{fig:soglie}b shows $\Delta V^*$ as a function
of pH for bare SiN nanopores.
We employed 
two analytical models
describing $\sigma_w$ as a function of 
pH~\cite{lin2021surface,bandara2019chemically}, 
based on fitted experimental data, see Methods.
For both of them, $\Delta V^*$ is below $1$~V in
a relatively wide range of pH.
Indeed, in bare SiN nanopores
both silanol groups and amines
are usually 
exposed on the surface~\cite{lin2021surface},
and $\sigma_w$ changes sign around pH 4.1-4.3 (point of zero charge).
By using surface modification,
it is possible to keep a low $\sigma_w$,
and thus low $\Delta V^*$,
for a wider range of pH,
Fig.~\ref{fig:soglie}c.~\cite{bandara2019chemically}
In particular,  
for the reported SiN-R-OH modified nanopore, 
with R alkane linker,
the pore is essentially neutral for $\mathrm{pH<7}$.
Conversely,
the amine modified SiN-R-$\mathrm{NH_2}$ nanopores
is, in essence, neutral for $\mathrm{pH > 8.5}$. 
In these pH ranges, $\Delta V^* < 150~\mathrm{mV}$
for $10$~nm radius pores and is even smaller for smaller radii.

The above arguments implicitly assume a superposition of effects, 
{\sl i.e.}, the total EOF can be decomposed as the sum of
fixed charge and induced charge contributions. 
This hypothesis is quite strong so the estimation provided 
by Eq.~\eqref{eq:dvstar} should be understood
as way to determine approximate voltage ranges where the 
intrinsic selectivity or the induced charge mechanism dominate the EOF.
The above theoretical arguments are supported by 
MD simulations of a model pore (similar to the one shown in Fig.~\ref{fig:nanopore}),
modified with a surface charge of
$\sigma_w = 2.5$ or $5~\mathrm{mC/m^2}$,
see Supplementary Fig.~S13.
For these two systems, 
MD simulations confirm that above 
the theoretical $\Delta V^*$
the geometrically induced EOF dominates 
on the EOF due to fixed charges.
The $v_{eo}$ dependence on the voltage 
is still parabolic although shifted,
in line with the superposition of effects hypothesis underlying 
Eq.~\eqref{eq:dvstar}.

\vspace{0.4 cm}

%
{\bf Effect of asymmetric electrolyte.}
We then performed MD simulations of a nanopore system
releasing one of the model hypotheses: the molecular symmetry of the 
electrolyte. 
Instead of using
our custom perfectly symmetric electrolyte 
employed for the MD simulation data in 
Figs.~\ref{fig:nanopore}-\ref{fig:EOF-parametric},
in this section we used a 2M KCl water solution.
Now the mobilities of the two ions are different,
as well as the structure of the first shell of water molecules around them.
The overall behavior of the system 
is similar to the symmetric electrolyte case.
In particular, a selectivity switch and a unidirectional EOF are observed, 
see Fig.~\ref{fig:asymmetric}a-c.
Some asymmetries are evident, as expected.
At equilibrium, $\Delta V = 0$, 
the system exhibits an intrinsic net positive charge accumulation
inside the nanopore lumen
($q_N \simeq 0.2\,$e, Fig.~\ref{fig:asymmetric}a,d),
despite the zero surface charge of the solid.
Indeed, the asymmetric electrolyte develops 
an equilibrium charge layering
at the solid-liquid interface,
Fig.~\ref{fig:asymmetric}d.
This is also evident from the peculiar orientation 
of the water molecules at the wall, forming surface dipoles,
Fig.~\ref{fig:asymmetric}e.
The presence of interfacial dipoles generates
an intrinsic polarization of the membrane
and, hence, a non-zero surface potential,
Fig.~\ref{fig:asymmetric}f.
The formation of a non-zero surface potential in uncharged nanopores
due to electrolyte asymmetries was proposed 
by Dukhin et al.~\cite{dukhin2005electrokinetics}
and later investigated 
by other authors~\cite{kim2009high,mucha2005unified}.
For instance, in Kim et al.~\cite{kim2009high} it was shown that
the different hydration forces among cations and anions 
lead to a slightly different equilibrium position 
of positive and negative charges ({\sl i.e.}, a charge layering) 
at the solid/liquid interface of uncharged hydrophobic nanopores.
The charge layering results in a non-zero surface potential and EOF.
A similar layering was also found 
in Mucha et al.~\cite{mucha2005unified}
at liquid/air interfaces.

Hence, for an asymmetric electrolyte, two effects 
rule the pore charge accumulation: the pore lumen's 
equilibrium surface potential that leads to an intrinsic selectivity 
(cation, in the present case) and the induced charge 
mechanism due to the presence of the lateral cavity.
We observe different behaviors 
under opposite $\Delta V$, see Fig.~\ref{fig:asymmetric}a-c.
For $\Delta V<0$, the charge inside the nanopore, $q_N$, 
remains relatively constant
and the selectivity and EOF 
are both roughly proportional to $\Delta V$.
For $\Delta V>0$, instead,
$q_N$ decreases linearly with $\Delta V$,
and, coherently to the induced charge mechanism,
the selectivity and EOF are quadratic.
In such a complex scenario, the theoretical expressions derived 
for the perfect symmetric case 
(dashed gray lines in Fig.~\ref{fig:asymmetric}a-c) 
fall short in predicting quantitatively the 
selectivity and EOF intensity. Nevertheless, they
still provide 
the order of magnitude of the effect.
\\

\begin{figure*}[!ht]
        \centering
        \includegraphics[width=0.95\textwidth]{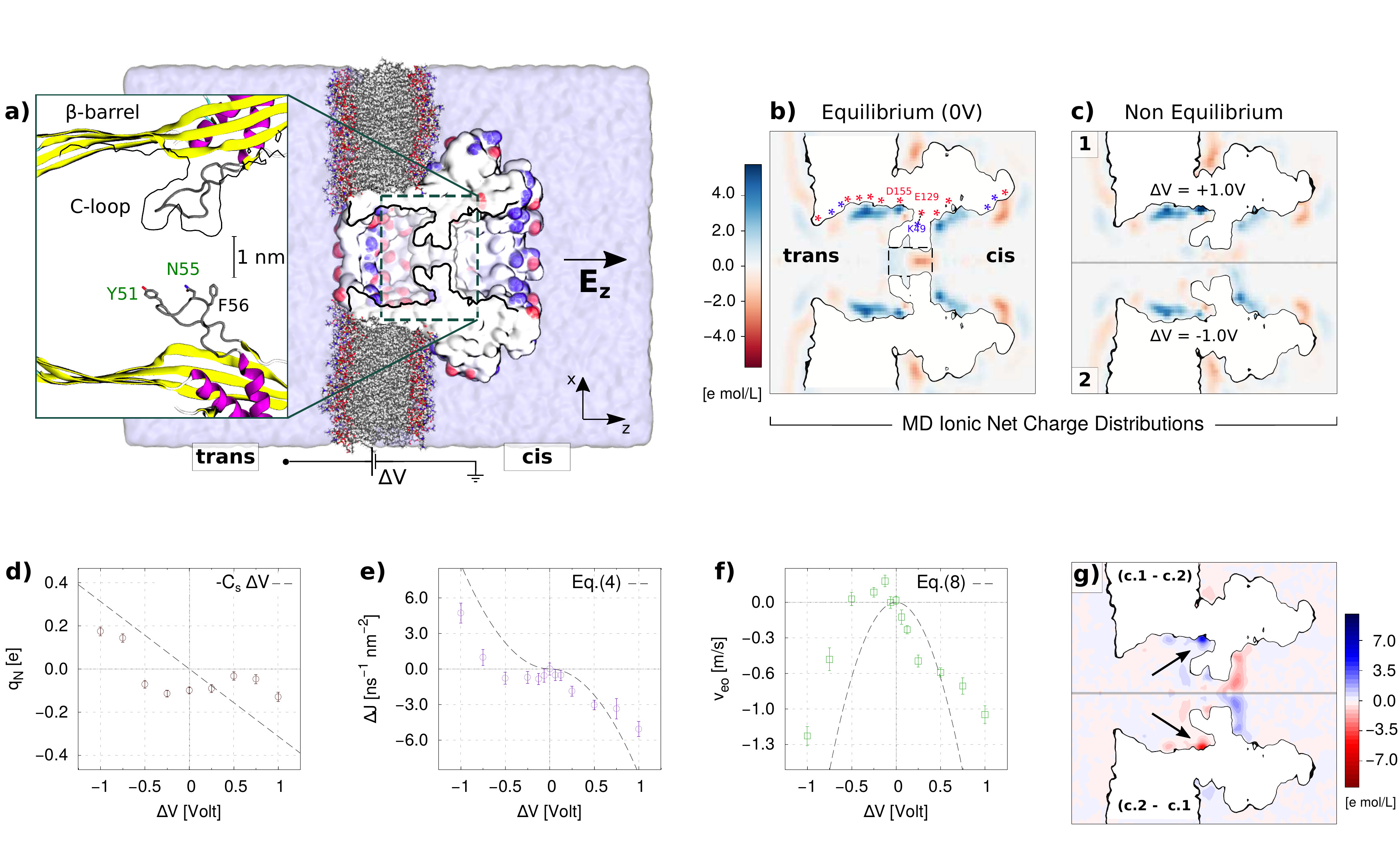}
        \caption{
		\label{fig:csgg}
		{\bf The CsgG biological nanopore in 2M KCl water solution.}
		{\bf a)}~MD set-up.
		A volume rendering representation of the pore cross-section (white) 
		embedded in a lipid membrane, with exposed charged residues colored 
		(blue positive, red negative).
		Water and ions are omitted for clarity. 
		The inset shows a zoom of the 
		pore constriction with the cartoon representation of 
		secondary structure on the top side, 
		and licorice representation of
		the residues forming the constriction surface Y51, N55 (hydrophilic, green labels) 
		and F56 (hydrophobic, black label) on the bottom. 
		{\bf b)}~Equilibrium ($\Delta V=0~\mathrm{V}$) and 
		{\bf c)}~non-equilibrium 
		($\Delta V=\pm 1~\mathrm{V}$) 
		MD ionic net charge density distributions.
		The asterisks in {\bf b)} indicate the charged residues 
		exposed toward the nanopore lumen. 
		{\bf d)}~Charge in the constriction, 
		{\bf e)}~selectivity and 
		{\bf f)}~electroosmotic velocity as functions of the applied voltage $\Delta V$. 
		Dashed lines represent the theoretical prediction 
		($L=18$~\AA, $R=6$~\AA, $s=9$~\AA, $h=5$~\AA~and $\varepsilon_S=6$);
		The other parameters for the solvent are the same used in Fig.~\ref{fig:asymmetric}.
		{\bf g)}~Difference of the panels {\bf c.1} and {\bf c.2},
		pointing out the opposite charge accumulation inside the lateral cavity
		at opposite voltages $\Delta V=\pm 1~\mathrm{V}$.
		In panels {\bf a-c} and {\bf g} 
		the black line delimiting the pore and the membrane 
		is the water density contour level 
		$\rho=0.5~\rho_{bulk}$, with $\rho_{bulk}$ the
		bulk water density. 
		Fluxes and maps 
		are obtained from $280~\mathrm{ns}$ MD production runs.
		All the trajectories are sampled every $20~\mathrm{ps}$ 
		and analyzed discarding the first $10~\mathrm{ns}$.
		Errors are calculated using a block average
		protocol with a block length of 10 ns.
		}
\end{figure*}

{\bf A biological example: the CsgG nanopore}. 
We then verified if the geometrically induced selectivity
switch and the unidirectional EOF also occur
in more complex scenarios such as
biological nanopores where articulate geometries
and surface charge patterns are usually present.
We selected as a possible candidate the curli specific 
gene G (CsgG) protein from {\sl E. coli}. 
This pore is currently used in commercial devices 
for nanopore DNA sequencing~\cite{brown2016nanopore,van2020dual}.
CsgG is a nonameric membrane protein, part of a transport
machinery comprising at least seven proteins encoded 
by two operons ~\cite{Chapman2002RoleofOperons} that excretes functional 
amyloids ~\cite{VANGERVEN2015StructureInCurli}, the curli proteins~\cite{cao2014structure,goyal2014structural}.
The CsgG pore is constituted by two large vestibules on 
the cis and trans side
connected by a constriction of diameter $\simeq 1.2~\mathrm{nm}$, 
formed by the so-called C-loop, Fig.~\ref{fig:csgg}a.

CsgG pore lumen is irregular,
yet the shape of its constriction 
region resembles the cylindrical pore 
surrounded by a coaxial cavity, albeit being more complex. 
For example, the constriction region is not straight but has 
a cleft at about one-third of its length. 
The lateral cavity is formed between the 
transmembrane $\beta$-barrel and the C-loop 
(residues 47-58, see the inset of Fig.~\ref{fig:csgg}a) 
that is held in place by the cis mixed $\alpha\beta$ domains. 
The geometry of the lateral cavity is wedged and inclined, 
with a moderately polar surface composition. 
D155 is the only exposed charged sidechain 
while K49 and E129 form a stable
salt bridge and are only partially solvent accessible, 
Fig.~\ref{fig:csgg}b.
Several surface charges are present
in the lumen and are marked in Fig.~\ref{fig:csgg}b 
with blue and red asterisks.
The $\beta$-barrel is overall negatively charged 
with four acidic residues 
and two basic ones
for each of the nine protomers.
The cis-vestibule has two acidic residues 
near the constriction.
Other charged residues are located at the 
entrances of the cis and trans vestibules. 
Globally, the total pore charge is zero
and the constriction has no charged
residues exposed.

We performed a set of MD simulations 
at different applied voltages,
in 2M KCl water solution.
At equilibrium ($\Delta V=0$) 
the pore exhibits a net negative charge
$q_N$ in the constriction, 
Fig.~\ref{fig:csgg}b,d.
For $\Delta V >0$, 
$q_N$ remains quite constant
and the anion selectivity ($\Delta J < 0$) 
shows a linear scaling with $\Delta V$, Fig.~\ref{fig:csgg}e.
EOF is negative since the water flow follows the motion
of the anions, Fig.~\ref{fig:csgg}f.
For small negative $\Delta V$, the 
pore is still anion selective
($q_N < 0$ and $\Delta J< 0$)
and $v_{eo}$ becomes positive
since, again, the water flow follows the motion
of the anions.
This is the usual behavior of
an electroosmotic flow where the 
charge accumulation in the pore is due to
a wall potential independent of the $\Delta V$.
An inversion of both the accumulated
charge $q_N$ and selectivity 
is observed for large negative voltages, $\Delta V<-0.5~\mathrm{V}$,
consistently with the geometrically
induced selectivity switch mechanism.
Gray dashed lines
in Fig.~\ref{fig:csgg}d-f  report the
predictions of the theoretical model.
For completeness, 
the current-voltage curve is reported in Supplementary Fig.~S14.
Although the pore geometry is quite far
from the ideal model system of Fig.~\ref{fig:nanopore} and
asymmetries are present in the curves,
the simplified model is still able to capture the 
order of magnitude of the EOF.
As in the solid-state nanopore 
with asymmetric electrolyte discussed in 
Fig.~\ref{fig:asymmetric}, the data suggest that
the presence of an equilibrium (intrinsic) net charge 
in the pore results in a sort of shift of the EOF curve
with respect to the theoretical parabolic prediction.
In the solid-state case of Fig.~\ref{fig:asymmetric}, 
the pore is intrinsically cation-selective (at low $\Delta V$)
and the selectivity inversion occurs at 
a positive $\Delta V$. Accordingly,
the maximum of EOF is shifted towards positive $\Delta V$.
Conversely, in CsgG, the pore is intrinsically 
anion-selective (at low $\Delta V$) 
so the selectivity inversion occurs at 
a negative $\Delta V$ and the EOF curve 
is shifted towards the left.

Further details on the charge distributions 
for $\Delta V = 0$ are reported in Fig.~\ref{fig:csgg}b.
The map shows several charge accumulation spots
due to the solvent-exposed charged residues in the two
vestibules.
Another relevant difference with respect 
to the ideal solid-state case is the charge 
distribution in the constriction at equilibrium ($\Delta V = 0$),
that shows a relative accumulation of positive
(negative) ions on the trans (cis) side of the
constriction.
This peculiar 
distribution and the
consequent intrinsic anion selectivity
may reflect the complex shape of
the constriction and the 
different hydropathy of the surface,
composed of hydrophilic (Y51 and N55)
and hydrophobic (F56) parts, 
see the inset in 
Fig.~\ref{fig:csgg}a.
Nevertheless, in agreement with our induced charge model, 
when an external $\Delta V$ is applied,
ions accumulate in the lateral cavity of CsgG
(altering also the charge distribution in the constriction),
as shown in Fig.~\ref{fig:csgg}c.
This voltage dependent behavior is better highlighted by 
Fig.~\ref{fig:csgg}g,
representing the difference of the maps 
at $\Delta V= 1 V$ and $\Delta V = -1 V$.
	An alternative representation of the differential maps
	with respect to the equilibrium (0V, ~Fig.~\ref{fig:csgg}b)
	is reported in Supplementary Fig.~S14.
	For comparison, we also ran simulations for a neutralized
	pore. Charge accumulation spots in the pore vestibules
	are much less evident, nevertheless the 
	charge distribution in the constriction is
	quite similar to the unmodified CgsG and
	consequently, ion currents, selectivity and
	EOF are, in essence, unchanged, see Supplementary Fig.~S15.
	In addition, in Supplementary Fig.~S16 
	we also reported an analysis that attempts to compare the
	induced charge EOF predicted by our geometrical model
	(that scales as $\Delta V^2$)
	and the expected linear EOF 
	due to intrinsic anion selectivity 
	at different voltages.
	This analysis indicates 
	that for $\vert \Delta V \vert \lesssim 0.3$~V the 
	dominant contribution is the intrinsic selectivity,
	while for $\vert  \Delta V \vert \gtrsim 0.3$~V, the induced charge
	mechanism dominates the EOF. MD data for negative $\Delta V$,
	where selectivity inversion is observed, 
	approximatively supports this theoretical threshold. 
Although $0.3$~V is larger than the typical $\Delta V$ 
employed in biopore experiments,
we mention that
polymeric membranes~\cite{morton2015tailored}
allowed biological nanopore experiments
at $\Delta V \sim 0.3$-$0.4$~V.
In addition, peculiar decoration of
solid-state supports for membrane anchoring
permitted to reach the same voltages 
for both lipid~\cite{kang2019one}
and diblock copolymer~\cite{yu2021stable} membranes.

\section{Conclusion}
We presented a mechanism of 
geometrically induced selectivity 
that switches with the applied voltage polarity
in uncharged cylindrical nanopores,
giving rise to unidirectional electroosmotic flow.
We derived an analytical model  
and we tested our predictions against 
Molecular Dynamics simulations.
The phenomenon is robust under variation
of the system geometry (e.g., cavity size, pore length)
and is shown to be applicable in real-word settings, {\sl i.e.}, with asymmetric electrolytes and weakly charged pores.
Our model provides a quantitatively accurate 
estimation of the electroosmotic velocity that can be 
used for nanopore system design.
Unidirectional electroosmotic flow 
also occurs for a biological pore, the CsgG protein, whose 
shape resembles the cavity-nanopore ideal system
but where, as usual for biopores, a complex surface 
charge pattern is present.   
A similar pore structure is also found in 
other secretion-related proteins of known structure, such as 
InvG~\cite{worrall2016near}
and PilQ~\cite{weaver2020cryoem} secretins, 
extending the possibility to use biomolecular scaffolds
to achieve geometrically induced selectivity.
Moreover, the surface patterning needed to
elicit this effect
is achievable by modern nanofabrication technology, 
such as electron-beam decoration of graphene~\cite{jin2013metallized},
focused ion beam~\cite{semple2021patterning},
or  electron beam lithography, reactive ion etching 
of TEM-drilled silicon nitride membranes~\cite{chou2020lifetime}.
The mechanism we unraveled allows to induce a tunable ion selectivity even without altering 
the pore shape, surface charge or chemistry and, consequently, 
it can open the way to a
more flexible design of selective membranes.
The magnitude of the  EOF associated 
to geometrically induced selectivity is comparable to other 
more common sources of EOF such
as fixed surface charges~\cite{huang2017electro,asandei2016electroosmotic,boukhet2016probing,willems2020accurate,huang2020electro} 
and, by appropriate choice of settings, can even dominate them.
Consequently, 
we expect that such a mechanism may find application in
all the technologies where EOF is already used. 
One example is alternate current electroosmotic 
pumps~\cite{hsu2018theory,wu2016alternating,ajdari2000pumping},
where different mechanisms have been exploited 
to induce a net EOF from a zero average oscillating potential 
in micro~\cite{ajdari2000pumping}
 and nanofluidic~\cite{hsu2018theory,wu2016alternating} systems.
In this respect, the average EOF intensity for 
a membrane constituted by conical nanopores~\cite{wu2016alternating} 
is of the same order as the one we observed.
Similarly, our mechanism may be employed 
in nanopore-based single molecule sensing devices,
where calibrating the competition/cooperation 
between electroosmosis and 
electrophoresis~\cite{boukhet2016probing,chinappi2020analytical}
is crucial to control particle capture,
especially for neutral or weakly charged molecules such 
as proteins and peptides~\cite{huang2017electro,asandei2016electroosmotic}.
Since the EOF is induced without modification of the pore interior, 
in principle the geometric mechanism we propose to 
generate selectivity and electroosmotic flow may allow to  
separately and independently engineer 
the pore lumen to improve the sensing performance 
and the external cavity to control EOF.

\section{Methods}

{\bf General Molecular Dynamics Simulations Methods.}
All MD runs were carried out using NAMD~\cite{phillips2005scalable},
using a time-step of $\Delta t=2.0~\mathrm{fs}$
and Particle Mesh Ewald~\cite{essmann1995smooth}
method with $1.0~\mathrm{\AA}$ spaced grid 
for long range electrostatic interactions.
A cutoff of $12~\mathrm{\AA}$ with switching distance of $14~\mathrm{\AA}$
was set for the short range non-bonded interactions.
Periodic boundary conditions with hexagonal prism cell are used 
unless otherwise stated.
Langevin thermostat was used for all the simulations.
Nos\'e-Hoover Langevin piston pressure control was used for 
constant pressure simulations~\cite{martyna1994constant}.

{\bf Solid-state pore set-up.}
Our model system, 
represented in 
Fig.~\ref{fig:nanopore}d,
is composed of a hexagonal solid membrane of thickness $L$
with a cylindrical nanopore of radius $R$,
surrounded by a coaxial cavity of width $w$ and height $d=L-h$, 
at a distance $s$ from the nanopore wall.
The hexagon apothem $a_h$,
(see the top-view inset of the membrane in
Fig.~\ref{fig:nanopore}d, green line)
is $a_h=2.1(R+s+w)$.
The membrane is composed of hexagonally packed uncharged atoms,
see Supplementary Fig.~S4.
	For Figs.~(\ref{fig:nanopore}-\ref{fig:MDresults}),
	the membrane is immersed 
	into a $2\mathrm{M}$ electrolyte solution, 
	composed of a symmetrical polar fluid (see below)
	in which oppositely charged ions are dissolved.
	For Fig.~\ref{fig:asymmetric},
	the membrane is immersed in a 2M KCl water solution,
	using standard CHARMM parameters for TIP3P water molecules and 
	potassium ($\mathrm{K}^+$) and chloride ($\mathrm{Cl}^-$) ions.
	The $z$-dimension of each simulation cell is about
	$H_z=2a_h + L$, with $L$ height of the membrane,
	to ensure that the liquid height surrounding the pore entrance
	is greater than two times the pore diameter.
	The system is equilibrated with a constant pressure 
	(flexible cell NPT) run at $P=1~\mathrm{atm}$ and $T=250~\mathrm{K}$,
	keeping the x,y plane area fixed.
	The production runs are conducted at 
	constant volume, temperature and particle number (NVT ensemble),
	with a constant and homogeneous electric field 
	$\mathbf{E}=(0,0,E_z)$ applied to charged atoms.

{\bf Model Dipolar Fluid.}
The model fluid is composed of diatomic molecules,
each formed by two atoms of mass $m=10~\mathrm{Da}$, 
of opposite charge 
$q^+=0.5\mathrm{e}$ and $q^-=-0.5\mathrm{e}$,
covalently bound through a harmonic potential
$U = k_b (r - r_0)^2$ where $r$ is the distance between the two atoms,
$r_0=1~\mathrm{\AA}$ the equilibrium distance and 
$k_b = 450~\mathrm{ kcal/(mol \AA^2) }$
the spring constant, see Supplementary Fig.~S4.
Intramolecular interactions are modeled via
a standard Coulomb potential plus a
Lennard-Jones (LJ) potential, with
$\epsilon_{LL}=0.1~\mathrm{kcal/mol}$ and
$\sigma_{LL}=2.68~\mathrm{\AA}$.
The above parameters were chosen to have volume, 
dipole moment and mass similar to those 
of TIP3P water~\cite{jorgensen1983comparison}.
The fluid exhibits a stable liquid phase 
in the temperature range
$200\le T \le 400~\mathrm{K}$, 
under a pressure of $P=1~\mathrm{atm}$,
see the phase diagram in Supplementary Fig.~S5. 
At $T=250~\mathrm{K}$,
the liquid density is $\rho=55.5~\mathrm{mol/L}$
while the 
relative electric permittivity is 
$\varepsilon_L = 83.2 \pm 4.6$
and dynamic viscosity $\eta=0.35 \pm 0.02~\mathrm{mPa\, s}$.
Relative permittivity $\varepsilon_L$
was assessed by computing the dipole moment fluctuations
in equilibrium NVT MD simulations~\cite{raabe2011molecular};
non-equilibrium estimations lead to similar results, Supplementary Fig.~S6.
Viscosity $\eta$ 
was estimated by applying a shear stress on the top of a liquid volume
and measuring the slope of the resulting velocity profile (Couette flow),
Supplementary Fig.~S7.

Non-bonded interactions between fluid and solid molecules
were modeled using a LJ potential,
with $\epsilon_{SL}=0.8\epsilon_{LL}$ and
$\sigma_{SL}=\sigma_{LL}$
(SL, solid-liquid),
resulting in a hydrophilic pore.
The wettability of the solid was assessed 
by evaluating the contact angle $\theta$ of 
a cylindrical drop of fluid onto 
the surface as a function of temperature and 
liquid-solid interaction potential 
$\epsilon_{SL}$~\cite{weijs2011origin},
see Supplementary Fig.~S8.
For the selected $\epsilon_{SL}/\epsilon_{LL}$ ratio,
the contact angle is $\theta \simeq 60^\circ$. 

The dissolved ions are composed of 
monovalent charged particles with
charges $q^\pm = \pm 1\mathrm{e}$
and mass $40~\mathrm{Da}$.
Non-bonded interactions of each ion with other atoms 
are described in Supplementary Fig.~S4.
Ion diffusion coefficient for a $2\mathrm{M}$ solution
at $P=1~\mathrm{atm}$ and $T=250~\mathrm{K}$, 
is $D=94.4 \pm 0.7~\mathrm{\AA^2/ns}$, 
corresponding to an ion mobility 
$\mu=4.4\times10^3~\mathrm{\AA^2/(V~ns)}$;
the diffusion coefficient $D$ is estimated  
from the mean squared displacement (MSD),
see Supplementary Fig.~S9.

{\bf CsgG pore set-up.}
The membrane-CsgG system was assembled using a protocol similar to 
the one reported in~\cite{aksimentiev2005imaging,bonome2017electroosmotic}.
The system was built starting from the CsgG x-ray crystal
structure taken from the Protein Data Bank, PDB\_ID: 4UV3~\cite{goyal2014structural}
downloaded from the OPM database~\cite{lomize2006opm}.
The beta-barrel missing fragments (F144, F193 to L199)
are modeled by using SWISS-MODEL server~\cite{schwede2003swiss}.
Other missing fragments (V258 to S262), 
located in the periphery of the cis side of the pore,
were deemed to be not important for the ion and EOF 
transport and were not taken into account.
The POPC lipid membrane, the water molecules, and the ions to
neutralize the system were added using VMD~\cite{humphrey1996vmd}.
Salt concentration was set to $2\mathrm{M}$ KCl.
The CHARMM36 force field~\cite{brooks2009charmm}
was employed to model lipid, protein, and
TIP3P water molecules~\cite{jorgensen1983comparison}.
Non-bonded fix corrections were applied for ions~\cite{yoo2011improved}.
All covalent bonds with hydrogen were kept rigid,
using SETTLE~\cite{miyamoto1992settle} for water molecules
and SHAKE/RATTLE~\cite{andersen1983rattle} for the rest of the system.

The energy of the system was first minimized for $10\,000$~steps 
using the conjugate gradient method.
Then a pre-equilibration of $1~\mathrm{ns}$ is performed
to let the lipid tails melt and the electrolyte relax:
the temperature was increased from $0$ to $300$~K in $100$~ps,
and then the Langevin thermostat with a damping coefficient of $1~\mathrm{ps^{-1}}$
was applied to all non-hydrogen atoms;
external forces were applied to the water molecules
to avoid their penetration into the membrane, 
while the backbone of the protein and the lipid heads were constrained
to their initial positions by means of harmonic springs,
$k_b = 1~\mathrm{ kcal/(mol \AA^2)}$;
Nose-Hoover Langevin method,
with a period of $100~\mathrm{fs}$ and decay of $50~\mathrm{fs}$, 
was used to keep a pressure of $1~\mathrm{atm}$,
allowing the unit cell volume to fluctuate,
by keeping the ratio between the x,y axis constant.
A second equilibration run of $1.3~\mathrm{ns}$ 
was performed to compact the membrane,
letting the lipid heads unconstrained, 
and reducing the spring constant on the protein backbone 
to $k_b = 0.5~\mathrm{kcal/(mol \AA^2)}$,
until the three unit cell vectors reach a stationary value.
The last equilibration step consisted of a $3~\mathrm{ns}$ NPT run 
(as in the previous step, keeping the ratio between the x,y axis constant)
where all the atoms were unconstrained
and no external forces were applied to the water molecules.
At the end of the equilibration procedure, 
the hexagonal periodic box has the following basis vectors: 
$v_x = (179,0,0)~\mathrm{\AA}$, 
$v_y = (89,155,0)~\mathrm{\AA}$, and 
$v_z = (243,0,0)~\mathrm{\AA}$, 
for a total of $680\,827$ atoms.

{\bf Current measurements.}
The production runs were performed at 
constant volume, temperature and particle number (NVT ensemble).
The length of each simulation is indicated 
in the caption of the figures.
For each case, a uniform and constant
external electric field ${\bf E} = (0,0,E_z)$ was applied
perpendicularly to the membrane.
This protocol was shown to be equivalent to the application of a constant voltage
$\Delta V = E_z L_z$~\cite{gumbart2012constant}, 
($E_z >0$ for $\Delta V > 0$, 
as indicated in Fig.~\ref{fig:nanopore}b).
In the solid-state nanopores, 
the solid atoms are constrained to
initial lattice positions with a harmonic spring,
$k_b = 100~\mathrm{kcal/ (mol \AA^2)}$,
the solid membrane is thermostated and 
coordinates are saved every $\Delta t =50~\mathrm{ps}$.
In the CsgG case, lipid head phosphorus are harmonically constrained 
to the position of the last configuration of the equilibration phase,
with $k_b = 10~\mathrm{kcal/ (mol \AA^2)}$,
and a thermostat is applied to the lipid and protein atoms
(not hydrogens).
Snapshots are saved every $\Delta t = 40$~ps.
The average current in the interval $[t, t + \Delta t]$
is estimated as~\cite{crozier2001model,aksimentiev2005imaging,bonome2017electroosmotic}
\begin{equation}
	I(t) = 
	\frac{1}{\Delta t \, L_z} 
	\sum_{i = 1}^{N} q_i \left [ z_i (t + \Delta t) - z_i(t) \right ]
	\label{eq:current}
\end{equation}
where $q_i$ and $z_i$ are the charge and the z-coordinate of the $i$-th atom,
respectively.
Ionic currents (either K$^+$ and Cl$^-$ or model ions) 
were computed by restricting the sum over the atoms
of corresponding type~\cite{aksimentiev2005imaging}.
The mean current is obtained via a block average of $I(t)$
(each block corresponding to $10~\mathrm{ns}$)
after discarding a transient of $30$~ns.
The EOF is measured similarly, 
computing the summation over the fluid atoms
and using the mass instead of the charge in Eq.~\eqref{eq:current}.
The results are then converted from mass flow rate to 
volumetric flow rate using the bulk liquid density.

{\bf Charge density, velocity fields and potential maps.}
Using the VMD {\it Volmap} plug-in~\cite{humphrey1996vmd}, 
we divided the system in cubic cells of size 
$\Delta x = \Delta y = \Delta z= 1$~\AA, 
and we calculated the average 
charge in each cell using the frames
of the stationary state of the production run.
A similar protocol is applied for the velocity  
profiles. In a given frame $f$, the 
velocity of the i-th  atom is computed as 
${\bf v_i}(f) = ({\bf x_i}(f+1) - {\bf x_i}(f-1))/(2\Delta t)$,
with ${\bf x_i}(f)$ its position and 
$\Delta t$ the sampling interval.
The average velocity in each cell is then calculated by
averaging over the particles belonging to the cell and over time.
	The electric potential maps are computed by 
	using the {\it pmepot} plug-in of VMD~\cite{aksimentiev2005imaging}
	based on the particle-mesh Ewald method (PME).
We then transformed the charge density and the velocity fields
from the $(x,y,z)$ Cartesian coordinate system to 
a cylindrical coordinate system $(r,z,\alpha)$
and performed a further averaging on $\alpha$
to get density and velocity fields in the $(r,z)$ plane
as the ones showed in Fig.~\ref{fig:nanopore}e-f and Fig.~\ref{fig:csgg}b-c.
Confidence intervals in Fig.~\ref{fig:nanopore}e were
obtained using a block average with
each block corresponding to $10~\mathrm{ns}$.

{\bf Surface Charge Models.}
Functional models for the pH dependence
of the surface charge $\sigma_w$
for solid-state SiN nanopores, 
used in Fig.~\ref{fig:soglie},
were taken from the experimental works of
Lin {\sl et al.}~\cite{lin2021surface} and
Bandara {\sl et al.}~\cite{bandara2019chemically}.
These models are used to fit experimental conductance data
measured at different pH for different nanopore set-ups.
In particular, for the black curve of Fig.~\ref{fig:soglie}b
we used the expression reported in Eq.~(8) of~\cite{lin2021surface}
together with the fitted values reported in the Fig.~3a of the same paper.
For the red curve of our Fig.~\ref{fig:soglie}b and all the curves of Fig.~\ref{fig:soglie}c,
we used the expression Eq.~(3) of~\cite{bandara2019chemically},
using for each system the respective fitted parameters reported in 
the Supplementary Information of the same work.

\vspace{0.5 cm}

\section*{Associated content}

{\bf Supplementary Information.}
Details for the calculation of Induced Debye layer 
capacitance for the cavity-nanopore system.
PNP-NS model for EOF and comments 
on the model assumptions.
Characterization of our atomistic model for symmetric electrolyte solution
in terms of phase diagram, relative electrical permittivity, wetting, ion mobility and viscosity.
Ion currents as a function of the voltage for our model system and for the CsgG nanopore.
Electric potential for different cavity sizes.
EOF prediction for a neutral Silicon Nitride nanopore of radius $20\,$~nm.
MD simulations of weakly charged nanopores.
Alternative maps of Fig.~\ref{fig:csgg}g.
Fluxes and charge density maps for Neutral Model of CsgG nanopore.
Comparison between parabolic induced charge and linear fixed charge EOF.
Table reporting surface charges for solid-state nanopores.\\

{\bf Acknowledgments}
The authors acknowledge
supercomputer time provided 
through HP10BGBB69 Iscra B Grant by CINECA and 
s958, s1103 Production Grants by CSCS.\\

{\bf Competing interest}
The authors declare no competing interests.\\

\bibliography{nanopore2021}

\providecommand{\latin}[1]{#1}
\providecommand*\mcitethebibliography{\thebibliography}
\csname @ifundefined\endcsname{endmcitethebibliography}
  {\let\endmcitethebibliography\endthebibliography}{}
\begin{mcitethebibliography}{84}
\providecommand*\natexlab[1]{#1}
\providecommand*\mciteSetBstSublistMode[1]{}
\providecommand*\mciteSetBstMaxWidthForm[2]{}
\providecommand*\mciteBstWouldAddEndPuncttrue
  {\def\EndOfBibitem{\unskip.}}
\providecommand*\mciteBstWouldAddEndPunctfalse
  {\let\EndOfBibitem\relax}
\providecommand*\mciteSetBstMidEndSepPunct[3]{}
\providecommand*\mciteSetBstSublistLabelBeginEnd[3]{}
\providecommand*\EndOfBibitem{}
\mciteSetBstSublistMode{f}
\mciteSetBstMaxWidthForm{subitem}{(\alph{mcitesubitemcount})}
\mciteSetBstSublistLabelBeginEnd
  {\mcitemaxwidthsubitemform\space}
  {\relax}
  {\relax}

\bibitem[Bocquet(2020)]{bocquet2020nanofluidics}
Bocquet,~L. Nanofluidics coming of age. \emph{Nature Materials} \textbf{2020},
  \emph{19}, 254--256\relax
\mciteBstWouldAddEndPuncttrue
\mciteSetBstMidEndSepPunct{\mcitedefaultmidpunct}
{\mcitedefaultendpunct}{\mcitedefaultseppunct}\relax
\EndOfBibitem
\bibitem[Hong \latin{et~al.}(2017)Hong, Constans, Surmani~Martins, Seow,
  Guevara~Carrio, and Garaj]{hong2017scalable}
Hong,~S.; Constans,~C.; Surmani~Martins,~M.~V.; Seow,~Y.~C.;
  Guevara~Carrio,~J.~A.; Garaj,~S. Scalable graphene-based membranes for ionic
  sieving with ultrahigh charge selectivity. \emph{Nano letters} \textbf{2017},
  \emph{17}, 728--732\relax
\mciteBstWouldAddEndPuncttrue
\mciteSetBstMidEndSepPunct{\mcitedefaultmidpunct}
{\mcitedefaultendpunct}{\mcitedefaultseppunct}\relax
\EndOfBibitem
\bibitem[Siwy(2006)]{siwy2006ion}
Siwy,~Z.~S. Ion-current rectification in nanopores and nanotubes with broken
  symmetry. \emph{Advanced Functional Materials} \textbf{2006}, \emph{16},
  735--746\relax
\mciteBstWouldAddEndPuncttrue
\mciteSetBstMidEndSepPunct{\mcitedefaultmidpunct}
{\mcitedefaultendpunct}{\mcitedefaultseppunct}\relax
\EndOfBibitem
\bibitem[Karnik \latin{et~al.}(2007)Karnik, Duan, Castelino, Daiguji, and
  Majumdar]{karnik2007rectification}
Karnik,~R.; Duan,~C.; Castelino,~K.; Daiguji,~H.; Majumdar,~A. Rectification of
  ionic current in a nanofluidic diode. \emph{Nano letters} \textbf{2007},
  \emph{7}, 547--551\relax
\mciteBstWouldAddEndPuncttrue
\mciteSetBstMidEndSepPunct{\mcitedefaultmidpunct}
{\mcitedefaultendpunct}{\mcitedefaultseppunct}\relax
\EndOfBibitem
\bibitem[Beckstein \latin{et~al.}(2001)Beckstein, Biggin, and
  Sansom]{beckstein2001hydrophobic}
Beckstein,~O.; Biggin,~P.~C.; Sansom,~M.~S. A hydrophobic gating mechanism for
  nanopores. \emph{The Journal of Physical Chemistry B} \textbf{2001},
  \emph{105}, 12902--12905\relax
\mciteBstWouldAddEndPuncttrue
\mciteSetBstMidEndSepPunct{\mcitedefaultmidpunct}
{\mcitedefaultendpunct}{\mcitedefaultseppunct}\relax
\EndOfBibitem
\bibitem[Powell \latin{et~al.}(2011)Powell, Cleary, Davenport, Shea, and
  Siwy]{powell2011electric}
Powell,~M.~R.; Cleary,~L.; Davenport,~M.; Shea,~K.~J.; Siwy,~Z.~S.
  Electric-field-induced wetting and dewetting in single hydrophobic nanopores.
  \emph{Nature nanotechnology} \textbf{2011}, \emph{6}, 798--802\relax
\mciteBstWouldAddEndPuncttrue
\mciteSetBstMidEndSepPunct{\mcitedefaultmidpunct}
{\mcitedefaultendpunct}{\mcitedefaultseppunct}\relax
\EndOfBibitem
\bibitem[Wilson and Aksimentiev(2018)Wilson, and Aksimentiev]{wilson2018water}
Wilson,~J.; Aksimentiev,~A. Water-compression gating of nanopore transport.
  \emph{Physical review letters} \textbf{2018}, \emph{120}, 268101\relax
\mciteBstWouldAddEndPuncttrue
\mciteSetBstMidEndSepPunct{\mcitedefaultmidpunct}
{\mcitedefaultendpunct}{\mcitedefaultseppunct}\relax
\EndOfBibitem
\bibitem[Camisasca \latin{et~al.}(2020)Camisasca, Tinti, and
  Giacomello]{camisasca2020gas}
Camisasca,~G.; Tinti,~A.; Giacomello,~A. Gas-Induced Drying of Nanopores.
  \emph{The Journal of Physical Chemistry Letters} \textbf{2020}, \emph{11},
  9171--9177\relax
\mciteBstWouldAddEndPuncttrue
\mciteSetBstMidEndSepPunct{\mcitedefaultmidpunct}
{\mcitedefaultendpunct}{\mcitedefaultseppunct}\relax
\EndOfBibitem
\bibitem[Agre(2004)]{agre2004aquaporin}
Agre,~P. Aquaporin water channels (Nobel lecture). \emph{Angewandte Chemie
  International Edition} \textbf{2004}, \emph{43}, 4278--4290\relax
\mciteBstWouldAddEndPuncttrue
\mciteSetBstMidEndSepPunct{\mcitedefaultmidpunct}
{\mcitedefaultendpunct}{\mcitedefaultseppunct}\relax
\EndOfBibitem
\bibitem[Gravelle \latin{et~al.}(2014)Gravelle, Joly, Ybert, and
  Bocquet]{gravelle2014large}
Gravelle,~S.; Joly,~L.; Ybert,~C.; Bocquet,~L. Large permeabilities of
  hourglass nanopores: From hydrodynamics to single file transport. \emph{The
  Journal of chemical physics} \textbf{2014}, \emph{141}, 18C526\relax
\mciteBstWouldAddEndPuncttrue
\mciteSetBstMidEndSepPunct{\mcitedefaultmidpunct}
{\mcitedefaultendpunct}{\mcitedefaultseppunct}\relax
\EndOfBibitem
\bibitem[Secchi \latin{et~al.}(2016)Secchi, Marbach, Nigu{\`e}s, Stein, Siria,
  and Bocquet]{secchi2016massive}
Secchi,~E.; Marbach,~S.; Nigu{\`e}s,~A.; Stein,~D.; Siria,~A.; Bocquet,~L.
  Massive radius-dependent flow slippage in carbon nanotubes. \emph{Nature}
  \textbf{2016}, \emph{537}, 210--213\relax
\mciteBstWouldAddEndPuncttrue
\mciteSetBstMidEndSepPunct{\mcitedefaultmidpunct}
{\mcitedefaultendpunct}{\mcitedefaultseppunct}\relax
\EndOfBibitem
\bibitem[Holt \latin{et~al.}(2006)Holt, Park, Wang, Stadermann, Artyukhin,
  Grigoropoulos, Noy, and Bakajin]{holt2006fast}
Holt,~J.~K.; Park,~H.~G.; Wang,~Y.; Stadermann,~M.; Artyukhin,~A.~B.;
  Grigoropoulos,~C.~P.; Noy,~A.; Bakajin,~O. Fast mass transport through
  sub-2-nanometer carbon nanotubes. \emph{Science} \textbf{2006}, \emph{312},
  1034--1037\relax
\mciteBstWouldAddEndPuncttrue
\mciteSetBstMidEndSepPunct{\mcitedefaultmidpunct}
{\mcitedefaultendpunct}{\mcitedefaultseppunct}\relax
\EndOfBibitem
\bibitem[Kavokine \latin{et~al.}(2020)Kavokine, Netz, and
  Bocquet]{kavokine2020fluids}
Kavokine,~N.; Netz,~R.~R.; Bocquet,~L. Fluids at the Nanoscale: From Continuum
  to Subcontinuum Transport. \emph{Annual Review of Fluid Mechanics}
  \textbf{2020}, \emph{53}, 377--410\relax
\mciteBstWouldAddEndPuncttrue
\mciteSetBstMidEndSepPunct{\mcitedefaultmidpunct}
{\mcitedefaultendpunct}{\mcitedefaultseppunct}\relax
\EndOfBibitem
\bibitem[B{\'e}termier \latin{et~al.}(2020)B{\'e}termier, Cressiot, Di~Muccio,
  Jarroux, Bacri, Della~Rocca, Chinappi, Pelta, and
  Tarascon]{betermier2020single}
B{\'e}termier,~F.; Cressiot,~B.; Di~Muccio,~G.; Jarroux,~N.; Bacri,~L.;
  Della~Rocca,~B.~M.; Chinappi,~M.; Pelta,~J.; Tarascon,~J.-M. Single-sulfur
  atom discrimination of polysulfides with a protein nanopore for improved
  batteries. \emph{Communications Materials} \textbf{2020}, \emph{1},
  1--11\relax
\mciteBstWouldAddEndPuncttrue
\mciteSetBstMidEndSepPunct{\mcitedefaultmidpunct}
{\mcitedefaultendpunct}{\mcitedefaultseppunct}\relax
\EndOfBibitem
\bibitem[Gu \latin{et~al.}(1999)Gu, Braha, Conlan, Cheley, and
  Bayley]{gu1999stochastic}
Gu,~L.-Q.; Braha,~O.; Conlan,~S.; Cheley,~S.; Bayley,~H. Stochastic sensing of
  organic analytes by a pore-forming protein containing a molecular adapter.
  \emph{Nature} \textbf{1999}, \emph{398}, 686--690\relax
\mciteBstWouldAddEndPuncttrue
\mciteSetBstMidEndSepPunct{\mcitedefaultmidpunct}
{\mcitedefaultendpunct}{\mcitedefaultseppunct}\relax
\EndOfBibitem
\bibitem[Feng \latin{et~al.}(2016)Feng, Graf, Liu, Ovchinnikov, Dumcenco,
  Heiranian, Nandigana, Aluru, Kis, and Radenovic]{feng2016single}
Feng,~J.; Graf,~M.; Liu,~K.; Ovchinnikov,~D.; Dumcenco,~D.; Heiranian,~M.;
  Nandigana,~V.; Aluru,~N.~R.; Kis,~A.; Radenovic,~A. Single-layer MoS 2
  nanopores as nanopower generators. \emph{Nature} \textbf{2016}, \emph{536},
  197--200\relax
\mciteBstWouldAddEndPuncttrue
\mciteSetBstMidEndSepPunct{\mcitedefaultmidpunct}
{\mcitedefaultendpunct}{\mcitedefaultseppunct}\relax
\EndOfBibitem
\bibitem[Siria \latin{et~al.}(2013)Siria, Poncharal, Biance, Fulcrand, Blase,
  Purcell, and Bocquet]{siria2013giant}
Siria,~A.; Poncharal,~P.; Biance,~A.-L.; Fulcrand,~R.; Blase,~X.;
  Purcell,~S.~T.; Bocquet,~L. Giant osmotic energy conversion measured in a
  single transmembrane boron nitride nanotube. \emph{Nature} \textbf{2013},
  \emph{494}, 455--458\relax
\mciteBstWouldAddEndPuncttrue
\mciteSetBstMidEndSepPunct{\mcitedefaultmidpunct}
{\mcitedefaultendpunct}{\mcitedefaultseppunct}\relax
\EndOfBibitem
\bibitem[Tu \latin{et~al.}(2020)Tu, Song, Ren, Shen, Chowdhury, Rajapaksha,
  Culp, Samineni, Lang, Thokkadam, \latin{et~al.} others]{tu2020rapid}
Tu,~Y.-M.; Song,~W.; Ren,~T.; Shen,~Y.-x.; Chowdhury,~R.; Rajapaksha,~P.;
  Culp,~T.~E.; Samineni,~L.; Lang,~C.; Thokkadam,~A. \latin{et~al.}  Rapid
  fabrication of precise high-throughput filters from membrane protein
  nanosheets. \emph{Nature Materials} \textbf{2020}, \emph{19}, 347--354\relax
\mciteBstWouldAddEndPuncttrue
\mciteSetBstMidEndSepPunct{\mcitedefaultmidpunct}
{\mcitedefaultendpunct}{\mcitedefaultseppunct}\relax
\EndOfBibitem
\bibitem[Schoch \latin{et~al.}(2008)Schoch, Han, and
  Renaud]{schoch2008transport}
Schoch,~R.~B.; Han,~J.; Renaud,~P. Transport phenomena in nanofluidics.
  \emph{Reviews of modern physics} \textbf{2008}, \emph{80}, 839\relax
\mciteBstWouldAddEndPuncttrue
\mciteSetBstMidEndSepPunct{\mcitedefaultmidpunct}
{\mcitedefaultendpunct}{\mcitedefaultseppunct}\relax
\EndOfBibitem
\bibitem[Boukhet \latin{et~al.}(2016)Boukhet, Piguet, Ouldali,
  Pastoriza-Gallego, Pelta, and Oukhaled]{boukhet2016probing}
Boukhet,~M.; Piguet,~F.; Ouldali,~H.; Pastoriza-Gallego,~M.; Pelta,~J.;
  Oukhaled,~A. Probing driving forces in aerolysin and $\alpha$-hemolysin
  biological nanopores: electrophoresis versus electroosmosis. \emph{Nanoscale}
  \textbf{2016}, \emph{8}, 18352--18359\relax
\mciteBstWouldAddEndPuncttrue
\mciteSetBstMidEndSepPunct{\mcitedefaultmidpunct}
{\mcitedefaultendpunct}{\mcitedefaultseppunct}\relax
\EndOfBibitem
\bibitem[Chinappi \latin{et~al.}(2020)Chinappi, Yamaji, Kawano, and
  Cecconi]{chinappi2020analytical}
Chinappi,~M.; Yamaji,~M.; Kawano,~R.; Cecconi,~F. Analytical Model for Particle
  Capture in Nanopores Elucidates Competition among Electrophoresis,
  Electroosmosis, and Dielectrophoresis. \emph{ACS nano} \textbf{2020},
  \emph{14}, 15816--15828\relax
\mciteBstWouldAddEndPuncttrue
\mciteSetBstMidEndSepPunct{\mcitedefaultmidpunct}
{\mcitedefaultendpunct}{\mcitedefaultseppunct}\relax
\EndOfBibitem
\bibitem[Huang \latin{et~al.}(2017)Huang, Willems, Soskine, Wloka, and
  Maglia]{huang2017electro}
Huang,~G.; Willems,~K.; Soskine,~M.; Wloka,~C.; Maglia,~G. Electro-osmotic
  capture and ionic discrimination of peptide and protein biomarkers with FraC
  nanopores. \emph{Nature communications} \textbf{2017}, \emph{8}, 1--11\relax
\mciteBstWouldAddEndPuncttrue
\mciteSetBstMidEndSepPunct{\mcitedefaultmidpunct}
{\mcitedefaultendpunct}{\mcitedefaultseppunct}\relax
\EndOfBibitem
\bibitem[Asandei \latin{et~al.}(2016)Asandei, Schiopu, Chinappi, Seo, Park, and
  Luchian]{asandei2016electroosmotic}
Asandei,~A.; Schiopu,~I.; Chinappi,~M.; Seo,~C.~H.; Park,~Y.; Luchian,~T.
  Electroosmotic trap against the electrophoretic force near a protein nanopore
  reveals peptide dynamics during capture and translocation. \emph{ACS applied
  materials \& interfaces} \textbf{2016}, \emph{8}, 13166--13179\relax
\mciteBstWouldAddEndPuncttrue
\mciteSetBstMidEndSepPunct{\mcitedefaultmidpunct}
{\mcitedefaultendpunct}{\mcitedefaultseppunct}\relax
\EndOfBibitem
\bibitem[Ram{\'\i}rez \latin{et~al.}(2003)Ram{\'\i}rez, Mafe, Alcaraz, and
  Cervera]{ramirez2003modeling}
Ram{\'\i}rez,~P.; Mafe,~S.; Alcaraz,~A.; Cervera,~J. Modeling of pH-switchable
  ion transport and selectivity in nanopore membranes with fixed charges.
  \emph{The Journal of Physical Chemistry B} \textbf{2003}, \emph{107},
  13178--13187\relax
\mciteBstWouldAddEndPuncttrue
\mciteSetBstMidEndSepPunct{\mcitedefaultmidpunct}
{\mcitedefaultendpunct}{\mcitedefaultseppunct}\relax
\EndOfBibitem
\bibitem[Small \latin{et~al.}(2015)Small, Wheeler, and
  Spoerke]{small2015nanoporous}
Small,~L.~J.; Wheeler,~D.~R.; Spoerke,~E.~D. Nanoporous membranes with
  electrochemically switchable, chemically stabilized ionic selectivity.
  \emph{Nanoscale} \textbf{2015}, \emph{7}, 16909--16920\relax
\mciteBstWouldAddEndPuncttrue
\mciteSetBstMidEndSepPunct{\mcitedefaultmidpunct}
{\mcitedefaultendpunct}{\mcitedefaultseppunct}\relax
\EndOfBibitem
\bibitem[Zeng \latin{et~al.}(2015)Zeng, Yeh, Zhang, and Qian]{zeng2015ion}
Zeng,~Z.; Yeh,~L.-H.; Zhang,~M.; Qian,~S. Ion transport and selectivity in
  biomimetic nanopores with pH-tunable zwitterionic polyelectrolyte brushes.
  \emph{Nanoscale} \textbf{2015}, \emph{7}, 17020--17029\relax
\mciteBstWouldAddEndPuncttrue
\mciteSetBstMidEndSepPunct{\mcitedefaultmidpunct}
{\mcitedefaultendpunct}{\mcitedefaultseppunct}\relax
\EndOfBibitem
\bibitem[Nishizawa \latin{et~al.}(1995)Nishizawa, Menon, and
  Martin]{nishizawa1995metal}
Nishizawa,~M.; Menon,~V.~P.; Martin,~C.~R. Metal nanotubule membranes with
  electrochemically switchable ion-transport selectivity. \emph{Science}
  \textbf{1995}, \emph{268}, 700--702\relax
\mciteBstWouldAddEndPuncttrue
\mciteSetBstMidEndSepPunct{\mcitedefaultmidpunct}
{\mcitedefaultendpunct}{\mcitedefaultseppunct}\relax
\EndOfBibitem
\bibitem[Kalman \latin{et~al.}(2009)Kalman, Sudre, Vlassiouk, and
  Siwy]{kalman2009control}
Kalman,~E.~B.; Sudre,~O.; Vlassiouk,~I.; Siwy,~Z.~S. Control of ionic transport
  through gated single conical nanopores. \emph{Analytical and bioanalytical
  chemistry} \textbf{2009}, \emph{394}, 413--419\relax
\mciteBstWouldAddEndPuncttrue
\mciteSetBstMidEndSepPunct{\mcitedefaultmidpunct}
{\mcitedefaultendpunct}{\mcitedefaultseppunct}\relax
\EndOfBibitem
\bibitem[Guan \latin{et~al.}(2014)Guan, Li, and Reed]{guan2014voltage}
Guan,~W.; Li,~S.~X.; Reed,~M.~A. Voltage gated ion and molecule transport in
  engineered nanochannels: theory, fabrication and applications.
  \emph{Nanotechnology} \textbf{2014}, \emph{25}, 122001\relax
\mciteBstWouldAddEndPuncttrue
\mciteSetBstMidEndSepPunct{\mcitedefaultmidpunct}
{\mcitedefaultendpunct}{\mcitedefaultseppunct}\relax
\EndOfBibitem
\bibitem[Cheng \latin{et~al.}(2018)Cheng, Jiang, Simon, Liu, and
  Li]{cheng2018low}
Cheng,~C.; Jiang,~G.; Simon,~G.~P.; Liu,~J.~Z.; Li,~D. Low-voltage
  electrostatic modulation of ion diffusion through layered graphene-based
  nanoporous membranes. \emph{Nature nanotechnology} \textbf{2018}, \emph{13},
  685--690\relax
\mciteBstWouldAddEndPuncttrue
\mciteSetBstMidEndSepPunct{\mcitedefaultmidpunct}
{\mcitedefaultendpunct}{\mcitedefaultseppunct}\relax
\EndOfBibitem
\bibitem[Fuest \latin{et~al.}(2015)Fuest, Boone, Rangharajan, Conlisk, and
  Prakash]{fuest2015three}
Fuest,~M.; Boone,~C.; Rangharajan,~K.~K.; Conlisk,~A.~T.; Prakash,~S. A
  three-state nanofluidic field effect switch. \emph{Nano letters}
  \textbf{2015}, \emph{15}, 2365--2371\relax
\mciteBstWouldAddEndPuncttrue
\mciteSetBstMidEndSepPunct{\mcitedefaultmidpunct}
{\mcitedefaultendpunct}{\mcitedefaultseppunct}\relax
\EndOfBibitem
\bibitem[Ren \latin{et~al.}(2017)Ren, Zhang, Nadappuram, Akpinar, Klenerman,
  Ivanov, Edel, and Korchev]{ren2017nanopore}
Ren,~R.; Zhang,~Y.; Nadappuram,~B.~P.; Akpinar,~B.; Klenerman,~D.;
  Ivanov,~A.~P.; Edel,~J.~B.; Korchev,~Y. Nanopore extended field-effect
  transistor for selective single-molecule biosensing. \emph{Nature
  communications} \textbf{2017}, \emph{8}, 1--9\relax
\mciteBstWouldAddEndPuncttrue
\mciteSetBstMidEndSepPunct{\mcitedefaultmidpunct}
{\mcitedefaultendpunct}{\mcitedefaultseppunct}\relax
\EndOfBibitem
\bibitem[Bazant and Squires(2010)Bazant, and Squires]{bazant2010induced}
Bazant,~M.~Z.; Squires,~T.~M. Induced-charge electrokinetic phenomena.
  \emph{Current Opinion in Colloid \& Interface Science} \textbf{2010},
  \emph{15}, 203--213\relax
\mciteBstWouldAddEndPuncttrue
\mciteSetBstMidEndSepPunct{\mcitedefaultmidpunct}
{\mcitedefaultendpunct}{\mcitedefaultseppunct}\relax
\EndOfBibitem
\bibitem[Yao \latin{et~al.}(2020)Yao, Wen, Pham, and Zhang]{yao2020induced}
Yao,~Y.; Wen,~C.; Pham,~N.~H.; Zhang,~S.-L. On induced surface charge in
  solid-state nanopores. \emph{Langmuir} \textbf{2020}, \emph{36},
  8874--8882\relax
\mciteBstWouldAddEndPuncttrue
\mciteSetBstMidEndSepPunct{\mcitedefaultmidpunct}
{\mcitedefaultendpunct}{\mcitedefaultseppunct}\relax
\EndOfBibitem
\bibitem[Hsu \latin{et~al.}(2018)Hsu, Hwang, and Daiguji]{hsu2018theory}
Hsu,~W.-L.; Hwang,~J.; Daiguji,~H. Theory of Transport-Induced-Charge
  Electroosmotic Pumping toward Alternating Current Resistive Pulse Sensing.
  \emph{ACS sensors} \textbf{2018}, \emph{3}, 2320--2326\relax
\mciteBstWouldAddEndPuncttrue
\mciteSetBstMidEndSepPunct{\mcitedefaultmidpunct}
{\mcitedefaultendpunct}{\mcitedefaultseppunct}\relax
\EndOfBibitem
\bibitem[Cao \latin{et~al.}(2014)Cao, Zhao, Kou, Ni, Zhang, and
  Huang]{cao2014structure}
Cao,~B.; Zhao,~Y.; Kou,~Y.; Ni,~D.; Zhang,~X.~C.; Huang,~Y. Structure of the
  nonameric bacterial amyloid secretion channel. \emph{Proceedings of the
  National Academy of Sciences} \textbf{2014}, \emph{111}, E5439--E5444\relax
\mciteBstWouldAddEndPuncttrue
\mciteSetBstMidEndSepPunct{\mcitedefaultmidpunct}
{\mcitedefaultendpunct}{\mcitedefaultseppunct}\relax
\EndOfBibitem
\bibitem[Goyal \latin{et~al.}(2014)Goyal, Krasteva, Van~Gerven, Gubellini,
  Van~den Broeck, Troupiotis-Tsa{\"\i}laki, Jonckheere, P{\'e}hau-Arnaudet,
  Pinkner, Chapman, \latin{et~al.} others]{goyal2014structural}
Goyal,~P.; Krasteva,~P.~V.; Van~Gerven,~N.; Gubellini,~F.; Van~den Broeck,~I.;
  Troupiotis-Tsa{\"\i}laki,~A.; Jonckheere,~W.; P{\'e}hau-Arnaudet,~G.;
  Pinkner,~J.~S.; Chapman,~M.~R. \latin{et~al.}  Structural and mechanistic
  insights into the bacterial amyloid secretion channel CsgG. \emph{Nature}
  \textbf{2014}, \emph{516}, 250\relax
\mciteBstWouldAddEndPuncttrue
\mciteSetBstMidEndSepPunct{\mcitedefaultmidpunct}
{\mcitedefaultendpunct}{\mcitedefaultseppunct}\relax
\EndOfBibitem
\bibitem[Van~der Verren \latin{et~al.}(2020)Van~der Verren, Van~Gerven,
  Jonckheere, Hambley, Singh, Kilgour, Jordan, Wallace, Jayasinghe, and
  Remaut]{van2020dual}
Van~der Verren,~S.~E.; Van~Gerven,~N.; Jonckheere,~W.; Hambley,~R.; Singh,~P.;
  Kilgour,~J.; Jordan,~M.; Wallace,~E.~J.; Jayasinghe,~L.; Remaut,~H. A
  dual-constriction biological nanopore resolves homonucleotide sequences with
  high fidelity. \emph{Nature biotechnology} \textbf{2020}, \emph{38},
  1415--1420\relax
\mciteBstWouldAddEndPuncttrue
\mciteSetBstMidEndSepPunct{\mcitedefaultmidpunct}
{\mcitedefaultendpunct}{\mcitedefaultseppunct}\relax
\EndOfBibitem
\bibitem[Herr \latin{et~al.}(2000)Herr, Molho, Santiago, Mungal, Kenny, and
  Garguilo]{herr2000electroosmotic}
Herr,~A.; Molho,~J.; Santiago,~J.; Mungal,~M.; Kenny,~T.; Garguilo,~M.
  Electroosmotic capillary flow with nonuniform zeta potential.
  \emph{Analytical chemistry} \textbf{2000}, \emph{72}, 1053--1057\relax
\mciteBstWouldAddEndPuncttrue
\mciteSetBstMidEndSepPunct{\mcitedefaultmidpunct}
{\mcitedefaultendpunct}{\mcitedefaultseppunct}\relax
\EndOfBibitem
\bibitem[Bruus()]{bruustheoretical}
Bruus,~H. Theoretical microfluidics. 2008\relax
\mciteBstWouldAddEndPuncttrue
\mciteSetBstMidEndSepPunct{\mcitedefaultmidpunct}
{\mcitedefaultendpunct}{\mcitedefaultseppunct}\relax
\EndOfBibitem
\bibitem[Zeng \latin{et~al.}(2019)Zeng, Wen, Solomon, Zhang, and
  Zhang]{zeng2019rectification}
Zeng,~S.; Wen,~C.; Solomon,~P.; Zhang,~S.-L.; Zhang,~Z. Rectification of
  protein translocation in truncated pyramidal nanopores. \emph{Nature
  nanotechnology} \textbf{2019}, \emph{14}, 1056--1062\relax
\mciteBstWouldAddEndPuncttrue
\mciteSetBstMidEndSepPunct{\mcitedefaultmidpunct}
{\mcitedefaultendpunct}{\mcitedefaultseppunct}\relax
\EndOfBibitem
\bibitem[Houghtaling \latin{et~al.}(2019)Houghtaling, Ying, Eggenberger,
  Fennouri, Nandivada, Acharjee, Li, Hall, and
  Mayer]{houghtaling2019estimation}
Houghtaling,~J.; Ying,~C.; Eggenberger,~O.~M.; Fennouri,~A.; Nandivada,~S.;
  Acharjee,~M.; Li,~J.; Hall,~A.~R.; Mayer,~M. Estimation of shape, volume, and
  dipole moment of individual proteins freely transiting a synthetic nanopore.
  \emph{ACS nano} \textbf{2019}, \emph{13}, 5231--5242\relax
\mciteBstWouldAddEndPuncttrue
\mciteSetBstMidEndSepPunct{\mcitedefaultmidpunct}
{\mcitedefaultendpunct}{\mcitedefaultseppunct}\relax
\EndOfBibitem
\bibitem[Chou \latin{et~al.}(2020)Chou, Masih~Das, Monos, and
  Drndi\'c]{chou2020lifetime}
Chou,~Y.-C.; Masih~Das,~P.; Monos,~D.~S.; Drndi\'c,~M. Lifetime and stability
  of silicon nitride nanopores and nanopore arrays for ionic measurements.
  \emph{ACS nano} \textbf{2020}, \emph{14}, 6715--6728\relax
\mciteBstWouldAddEndPuncttrue
\mciteSetBstMidEndSepPunct{\mcitedefaultmidpunct}
{\mcitedefaultendpunct}{\mcitedefaultseppunct}\relax
\EndOfBibitem
\bibitem[Lin \latin{et~al.}(2021)Lin, Li, Tao, Li, Yang, Ma, Li, Sha, and
  Chen]{lin2021surface}
Lin,~K.; Li,~Z.; Tao,~Y.; Li,~K.; Yang,~H.; Ma,~J.; Li,~T.; Sha,~J.; Chen,~Y.
  Surface Charge Density Inside a Silicon Nitride Nanopore. \emph{Langmuir}
  \textbf{2021}, \emph{37}, 10521--10528\relax
\mciteBstWouldAddEndPuncttrue
\mciteSetBstMidEndSepPunct{\mcitedefaultmidpunct}
{\mcitedefaultendpunct}{\mcitedefaultseppunct}\relax
\EndOfBibitem
\bibitem[Bandara \latin{et~al.}(2019)Bandara, Karawdeniya, Hagan, Chevalier,
  and Dwyer]{bandara2019chemically}
Bandara,~Y. N.~D.; Karawdeniya,~B.~I.; Hagan,~J.~T.; Chevalier,~R.~B.;
  Dwyer,~J.~R. Chemically functionalizing controlled dielectric breakdown
  silicon nitride nanopores by direct photohydrosilylation. \emph{ACS applied
  materials \& interfaces} \textbf{2019}, \emph{11}, 30411--30420\relax
\mciteBstWouldAddEndPuncttrue
\mciteSetBstMidEndSepPunct{\mcitedefaultmidpunct}
{\mcitedefaultendpunct}{\mcitedefaultseppunct}\relax
\EndOfBibitem
\bibitem[Yeh and Hummer(2004)Yeh, and Hummer]{yeh2004system}
Yeh,~I.-C.; Hummer,~G. System-size dependence of diffusion coefficients and
  viscosities from molecular dynamics simulations with periodic boundary
  conditions. \emph{The Journal of Physical Chemistry B} \textbf{2004},
  \emph{108}, 15873--15879\relax
\mciteBstWouldAddEndPuncttrue
\mciteSetBstMidEndSepPunct{\mcitedefaultmidpunct}
{\mcitedefaultendpunct}{\mcitedefaultseppunct}\relax
\EndOfBibitem
\bibitem[Hoogerheide \latin{et~al.}(2009)Hoogerheide, Garaj, and
  Golovchenko]{hoogerheide2009probing}
Hoogerheide,~D.~P.; Garaj,~S.; Golovchenko,~J.~A. Probing surface charge
  fluctuations with solid-state nanopores. \emph{Physical review letters}
  \textbf{2009}, \emph{102}, 256804\relax
\mciteBstWouldAddEndPuncttrue
\mciteSetBstMidEndSepPunct{\mcitedefaultmidpunct}
{\mcitedefaultendpunct}{\mcitedefaultseppunct}\relax
\EndOfBibitem
\bibitem[Larkin \latin{et~al.}(2014)Larkin, Henley, Muthukumar, Rosenstein, and
  Wanunu]{larkin2014high}
Larkin,~J.; Henley,~R.~Y.; Muthukumar,~M.; Rosenstein,~J.~K.; Wanunu,~M.
  High-Bandwidth Protein Analysis Using Solid-State Nanopores. \emph{Biophys.
  J.} \textbf{2014}, \emph{106}, 696--704\relax
\mciteBstWouldAddEndPuncttrue
\mciteSetBstMidEndSepPunct{\mcitedefaultmidpunct}
{\mcitedefaultendpunct}{\mcitedefaultseppunct}\relax
\EndOfBibitem
\bibitem[Kosmulski(1997)]{kosmulski1997attempt}
Kosmulski,~M. Attempt to Determine Pristine Points of Zero Charge of Nb2O5,
  Ta2O5, and HfO2. \emph{Langmuir} \textbf{1997}, \emph{13}, 6315--6320\relax
\mciteBstWouldAddEndPuncttrue
\mciteSetBstMidEndSepPunct{\mcitedefaultmidpunct}
{\mcitedefaultendpunct}{\mcitedefaultseppunct}\relax
\EndOfBibitem
\bibitem[Dukhin \latin{et~al.}(2005)Dukhin, Dukhin, and
  Goetz]{dukhin2005electrokinetics}
Dukhin,~A.; Dukhin,~S.; Goetz,~P. Electrokinetics at high ionic strength and
  hypothesis of the double layer with zero surface charge. \emph{Langmuir}
  \textbf{2005}, \emph{21}, 9990--9997\relax
\mciteBstWouldAddEndPuncttrue
\mciteSetBstMidEndSepPunct{\mcitedefaultmidpunct}
{\mcitedefaultendpunct}{\mcitedefaultseppunct}\relax
\EndOfBibitem
\bibitem[Kim and Darve(2009)Kim, and Darve]{kim2009high}
Kim,~D.; Darve,~E. High-ionic-strength electroosmotic flows in uncharged
  hydrophobic nanochannels. \emph{Journal of colloid and interface science}
  \textbf{2009}, \emph{330}, 194--200\relax
\mciteBstWouldAddEndPuncttrue
\mciteSetBstMidEndSepPunct{\mcitedefaultmidpunct}
{\mcitedefaultendpunct}{\mcitedefaultseppunct}\relax
\EndOfBibitem
\bibitem[Mucha \latin{et~al.}(2005)Mucha, Frigato, Levering, Allen, Tobias,
  Dang, and Jungwirth]{mucha2005unified}
Mucha,~M.; Frigato,~T.; Levering,~L.~M.; Allen,~H.~C.; Tobias,~D.~J.;
  Dang,~L.~X.; Jungwirth,~P. Unified Molecular Picture of the Surfaces of
  Aqueous Acid, Base, and Salt Solutions. \emph{J. Phys. Chem. B}
  \textbf{2005}, \emph{109}, 7617--7623\relax
\mciteBstWouldAddEndPuncttrue
\mciteSetBstMidEndSepPunct{\mcitedefaultmidpunct}
{\mcitedefaultendpunct}{\mcitedefaultseppunct}\relax
\EndOfBibitem
\bibitem[Brown and Clarke(2016)Brown, and Clarke]{brown2016nanopore}
Brown,~C.~G.; Clarke,~J. Nanopore development at Oxford nanopore. \emph{Nature
  biotechnology} \textbf{2016}, \emph{34}, 810--811\relax
\mciteBstWouldAddEndPuncttrue
\mciteSetBstMidEndSepPunct{\mcitedefaultmidpunct}
{\mcitedefaultendpunct}{\mcitedefaultseppunct}\relax
\EndOfBibitem
\bibitem[Chapman \latin{et~al.}(2002)Chapman, Robinson, Pinkner, Roth, Heuser,
  Hammar, Normark, and Hultgren]{Chapman2002RoleofOperons}
Chapman,~M.~R.; Robinson,~L.~S.; Pinkner,~J.~S.; Roth,~R.; Heuser,~J.;
  Hammar,~M.; Normark,~S.; Hultgren,~S.~J. Role of Escherichia coli Curli
  Operons in Directing Amyloid Fiber Formation. \emph{Science} \textbf{2002},
  \emph{295}, 851--855\relax
\mciteBstWouldAddEndPuncttrue
\mciteSetBstMidEndSepPunct{\mcitedefaultmidpunct}
{\mcitedefaultendpunct}{\mcitedefaultseppunct}\relax
\EndOfBibitem
\bibitem[{Van Gerven} \latin{et~al.}(2015){Van Gerven}, Klein, Hultgren, and
  Remaut]{VANGERVEN2015StructureInCurli}
{Van Gerven},~N.; Klein,~R.~D.; Hultgren,~S.~J.; Remaut,~H. Bacterial Amyloid
  Formation: Structural Insights into Curli Biogensis. \emph{Trends in
  Microbiology} \textbf{2015}, \emph{23}, 693--706\relax
\mciteBstWouldAddEndPuncttrue
\mciteSetBstMidEndSepPunct{\mcitedefaultmidpunct}
{\mcitedefaultendpunct}{\mcitedefaultseppunct}\relax
\EndOfBibitem
\bibitem[Morton \latin{et~al.}(2015)Morton, Mortezaei, Yemenicioglu, Isaacman,
  Nova, Gundlach, and Theogarajan]{morton2015tailored}
Morton,~D.; Mortezaei,~S.; Yemenicioglu,~S.; Isaacman,~M.~J.; Nova,~I.~C.;
  Gundlach,~J.~H.; Theogarajan,~L. Tailored polymeric membranes for
  Mycobacterium smegmatis porin A (MspA) based biosensors. \emph{Journal of
  Materials Chemistry B} \textbf{2015}, \emph{3}, 5080--5086\relax
\mciteBstWouldAddEndPuncttrue
\mciteSetBstMidEndSepPunct{\mcitedefaultmidpunct}
{\mcitedefaultendpunct}{\mcitedefaultseppunct}\relax
\EndOfBibitem
\bibitem[Kang \latin{et~al.}(2019)Kang, Alibakhshi, and Wanunu]{kang2019one}
Kang,~X.; Alibakhshi,~M.~A.; Wanunu,~M. One-pot species release and nanopore
  detection in a voltage-stable lipid bilayer platform. \emph{Nano Letters}
  \textbf{2019}, \emph{19}, 9145--9153\relax
\mciteBstWouldAddEndPuncttrue
\mciteSetBstMidEndSepPunct{\mcitedefaultmidpunct}
{\mcitedefaultendpunct}{\mcitedefaultseppunct}\relax
\EndOfBibitem
\bibitem[Yu \latin{et~al.}(2021)Yu, Kang, Alibakhshi, Pavlenok, Niederweis, and
  Wanunu]{yu2021stable}
Yu,~L.; Kang,~X.; Alibakhshi,~M.~A.; Pavlenok,~M.; Niederweis,~M.; Wanunu,~M.
  Stable polymer bilayers for protein channel recordings at high guanidinium
  chloride concentrations. \emph{Biophysical Journal} \textbf{2021},
  \emph{120}, 1537--1541\relax
\mciteBstWouldAddEndPuncttrue
\mciteSetBstMidEndSepPunct{\mcitedefaultmidpunct}
{\mcitedefaultendpunct}{\mcitedefaultseppunct}\relax
\EndOfBibitem
\bibitem[Worrall \latin{et~al.}(2016)Worrall, Hong, Vuckovic, Deng, Bergeron,
  Majewski, Huang, Spreter, Finlay, Yu, \latin{et~al.} others]{worrall2016near}
Worrall,~L.; Hong,~C.; Vuckovic,~M.; Deng,~W.; Bergeron,~J.; Majewski,~D.;
  Huang,~R.; Spreter,~T.; Finlay,~B.; Yu,~Z. \latin{et~al.}
  Near-atomic-resolution cryo-EM analysis of the Salmonella T3S injectisome
  basal body. \emph{Nature} \textbf{2016}, \emph{540}, 597--601\relax
\mciteBstWouldAddEndPuncttrue
\mciteSetBstMidEndSepPunct{\mcitedefaultmidpunct}
{\mcitedefaultendpunct}{\mcitedefaultseppunct}\relax
\EndOfBibitem
\bibitem[Weaver \latin{et~al.}(2020)Weaver, Ortega, Sazinsky, Dalia, Dalia, and
  Jensen]{weaver2020cryoem}
Weaver,~S.~J.; Ortega,~D.~R.; Sazinsky,~M.~H.; Dalia,~T.~N.; Dalia,~A.~B.;
  Jensen,~G.~J. CryoEM structure of the type IVa pilus secretin required for
  natural competence in Vibrio cholerae. \emph{Nature communications}
  \textbf{2020}, \emph{11}, 1--13\relax
\mciteBstWouldAddEndPuncttrue
\mciteSetBstMidEndSepPunct{\mcitedefaultmidpunct}
{\mcitedefaultendpunct}{\mcitedefaultseppunct}\relax
\EndOfBibitem
\bibitem[Jin \latin{et~al.}(2013)Jin, Sun, Ke, Shih, Paulus, Wang, Mu, Yin, and
  Strano]{jin2013metallized}
Jin,~Z.; Sun,~W.; Ke,~Y.; Shih,~C.-J.; Paulus,~G.~L.; Wang,~Q.~H.; Mu,~B.;
  Yin,~P.; Strano,~M.~S. Metallized DNA nanolithography for encoding and
  transferring spatial information for graphene patterning. \emph{Nature
  communications} \textbf{2013}, \emph{4}, 1--9\relax
\mciteBstWouldAddEndPuncttrue
\mciteSetBstMidEndSepPunct{\mcitedefaultmidpunct}
{\mcitedefaultendpunct}{\mcitedefaultseppunct}\relax
\EndOfBibitem
\bibitem[Semple \latin{et~al.}(2021)Semple, Hryciw, Li, Flaim, and
  Iyer]{semple2021patterning}
Semple,~M.; Hryciw,~A.~C.; Li,~P.; Flaim,~E.; Iyer,~A.~K. Patterning of
  Complex, Nanometer-Scale Features in Wide-Area Gold Nanoplasmonic Structures
  Using Helium Focused Ion Beam Milling. \emph{ACS Applied Materials \&
  Interfaces} \textbf{2021}, \emph{13}, 43209–--43220\relax
\mciteBstWouldAddEndPuncttrue
\mciteSetBstMidEndSepPunct{\mcitedefaultmidpunct}
{\mcitedefaultendpunct}{\mcitedefaultseppunct}\relax
\EndOfBibitem
\bibitem[Willems \latin{et~al.}(2020)Willems, Rui{\'c}, Lucas, Barman,
  Verellen, Hofkens, Maglia, and Van~Dorpe]{willems2020accurate}
Willems,~K.; Rui{\'c},~D.; Lucas,~F.~L.; Barman,~U.; Verellen,~N.; Hofkens,~J.;
  Maglia,~G.; Van~Dorpe,~P. Accurate modeling of a biological nanopore with an
  extended continuum framework. \emph{Nanoscale} \textbf{2020}, \emph{12},
  16775--16795\relax
\mciteBstWouldAddEndPuncttrue
\mciteSetBstMidEndSepPunct{\mcitedefaultmidpunct}
{\mcitedefaultendpunct}{\mcitedefaultseppunct}\relax
\EndOfBibitem
\bibitem[Huang \latin{et~al.}(2020)Huang, Willems, Bartelds, van Dorpe,
  Soskine, and Maglia]{huang2020electro}
Huang,~G.; Willems,~K.; Bartelds,~M.; van Dorpe,~P.; Soskine,~M.; Maglia,~G.
  Electro-osmotic vortices promote the capture of folded proteins by PlyAB
  nanopores. \emph{Nano letters} \textbf{2020}, \emph{20}, 3819--3827\relax
\mciteBstWouldAddEndPuncttrue
\mciteSetBstMidEndSepPunct{\mcitedefaultmidpunct}
{\mcitedefaultendpunct}{\mcitedefaultseppunct}\relax
\EndOfBibitem
\bibitem[Wu \latin{et~al.}(2016)Wu, Ramiah~Rajasekaran, and
  Martin]{wu2016alternating}
Wu,~X.; Ramiah~Rajasekaran,~P.; Martin,~C.~R. An alternating current
  electroosmotic pump based on conical nanopore membranes. \emph{Acs Nano}
  \textbf{2016}, \emph{10}, 4637--4643\relax
\mciteBstWouldAddEndPuncttrue
\mciteSetBstMidEndSepPunct{\mcitedefaultmidpunct}
{\mcitedefaultendpunct}{\mcitedefaultseppunct}\relax
\EndOfBibitem
\bibitem[Ajdari(2000)]{ajdari2000pumping}
Ajdari,~A. Pumping liquids using asymmetric electrode arrays. \emph{Physical
  Review E} \textbf{2000}, \emph{61}, R45\relax
\mciteBstWouldAddEndPuncttrue
\mciteSetBstMidEndSepPunct{\mcitedefaultmidpunct}
{\mcitedefaultendpunct}{\mcitedefaultseppunct}\relax
\EndOfBibitem
\bibitem[Phillips \latin{et~al.}(2005)Phillips, Braun, Wang, Gumbart,
  Tajkhorshid, Villa, Chipot, Skeel, Kale, and Schulten]{phillips2005scalable}
Phillips,~J.~C.; Braun,~R.; Wang,~W.; Gumbart,~J.; Tajkhorshid,~E.; Villa,~E.;
  Chipot,~C.; Skeel,~R.~D.; Kale,~L.; Schulten,~K. Scalable molecular dynamics
  with NAMD. \emph{Journal of computational chemistry} \textbf{2005},
  \emph{26}, 1781--1802\relax
\mciteBstWouldAddEndPuncttrue
\mciteSetBstMidEndSepPunct{\mcitedefaultmidpunct}
{\mcitedefaultendpunct}{\mcitedefaultseppunct}\relax
\EndOfBibitem
\bibitem[Essmann \latin{et~al.}(1995)Essmann, Perera, Berkowitz, Darden, Lee,
  and Pedersen]{essmann1995smooth}
Essmann,~U.; Perera,~L.; Berkowitz,~M.~L.; Darden,~T.; Lee,~H.; Pedersen,~L.~G.
  A smooth particle mesh Ewald method. \emph{The Journal of chemical physics}
  \textbf{1995}, \emph{103}, 8577--8593\relax
\mciteBstWouldAddEndPuncttrue
\mciteSetBstMidEndSepPunct{\mcitedefaultmidpunct}
{\mcitedefaultendpunct}{\mcitedefaultseppunct}\relax
\EndOfBibitem
\bibitem[Martyna \latin{et~al.}(1994)Martyna, Tobias, and
  Klein]{martyna1994constant}
Martyna,~G.~J.; Tobias,~D.~J.; Klein,~M.~L. Constant pressure molecular
  dynamics algorithms. \emph{The Journal of chemical physics} \textbf{1994},
  \emph{101}, 4177--4189\relax
\mciteBstWouldAddEndPuncttrue
\mciteSetBstMidEndSepPunct{\mcitedefaultmidpunct}
{\mcitedefaultendpunct}{\mcitedefaultseppunct}\relax
\EndOfBibitem
\bibitem[Jorgensen \latin{et~al.}(1983)Jorgensen, Chandrasekhar, Madura, Impey,
  and Klein]{jorgensen1983comparison}
Jorgensen,~W.~L.; Chandrasekhar,~J.; Madura,~J.~D.; Impey,~R.~W.; Klein,~M.~L.
  Comparison of simple potential functions for simulating liquid water.
  \emph{The Journal of chemical physics} \textbf{1983}, \emph{79},
  926--935\relax
\mciteBstWouldAddEndPuncttrue
\mciteSetBstMidEndSepPunct{\mcitedefaultmidpunct}
{\mcitedefaultendpunct}{\mcitedefaultseppunct}\relax
\EndOfBibitem
\bibitem[Raabe and Sadus(2011)Raabe, and Sadus]{raabe2011molecular}
Raabe,~G.; Sadus,~R.~J. Molecular dynamics simulation of the dielectric
  constant of water: The effect of bond flexibility. \emph{The Journal of
  chemical physics} \textbf{2011}, \emph{134}, 234501\relax
\mciteBstWouldAddEndPuncttrue
\mciteSetBstMidEndSepPunct{\mcitedefaultmidpunct}
{\mcitedefaultendpunct}{\mcitedefaultseppunct}\relax
\EndOfBibitem
\bibitem[Weijs \latin{et~al.}(2011)Weijs, Marchand, Andreotti, Lohse, and
  Snoeijer]{weijs2011origin}
Weijs,~J.~H.; Marchand,~A.; Andreotti,~B.; Lohse,~D.; Snoeijer,~J.~H. Origin of
  line tension for a Lennard-Jones nanodroplet. \emph{Physics of fluids}
  \textbf{2011}, \emph{23}, 022001\relax
\mciteBstWouldAddEndPuncttrue
\mciteSetBstMidEndSepPunct{\mcitedefaultmidpunct}
{\mcitedefaultendpunct}{\mcitedefaultseppunct}\relax
\EndOfBibitem
\bibitem[Aksimentiev and Schulten(2005)Aksimentiev, and
  Schulten]{aksimentiev2005imaging}
Aksimentiev,~A.; Schulten,~K. Imaging $\alpha$-hemolysin with molecular
  dynamics: ionic conductance, osmotic permeability, and the electrostatic
  potential map. \emph{Biophysical journal} \textbf{2005}, \emph{88},
  3745--3761\relax
\mciteBstWouldAddEndPuncttrue
\mciteSetBstMidEndSepPunct{\mcitedefaultmidpunct}
{\mcitedefaultendpunct}{\mcitedefaultseppunct}\relax
\EndOfBibitem
\bibitem[Bonome \latin{et~al.}(2017)Bonome, Cecconi, and
  Chinappi]{bonome2017electroosmotic}
Bonome,~E.~L.; Cecconi,~F.; Chinappi,~M. Electroosmotic flow through an
  $\alpha$-hemolysin nanopore. \emph{Microfluidics and Nanofluidics}
  \textbf{2017}, \emph{21}, 96\relax
\mciteBstWouldAddEndPuncttrue
\mciteSetBstMidEndSepPunct{\mcitedefaultmidpunct}
{\mcitedefaultendpunct}{\mcitedefaultseppunct}\relax
\EndOfBibitem
\bibitem[Lomize \latin{et~al.}(2006)Lomize, Lomize, Pogozheva, and
  Mosberg]{lomize2006opm}
Lomize,~M.~A.; Lomize,~A.~L.; Pogozheva,~I.~D.; Mosberg,~H.~I. OPM:
  orientations of proteins in membranes database. \emph{Bioinformatics}
  \textbf{2006}, \emph{22}, 623--625\relax
\mciteBstWouldAddEndPuncttrue
\mciteSetBstMidEndSepPunct{\mcitedefaultmidpunct}
{\mcitedefaultendpunct}{\mcitedefaultseppunct}\relax
\EndOfBibitem
\bibitem[Schwede \latin{et~al.}(2003)Schwede, Kopp, Guex, and
  Peitsch]{schwede2003swiss}
Schwede,~T.; Kopp,~J.; Guex,~N.; Peitsch,~M.~C. SWISS-MODEL: an automated
  protein homology-modeling server. \emph{Nucleic acids research}
  \textbf{2003}, \emph{31}, 3381--3385\relax
\mciteBstWouldAddEndPuncttrue
\mciteSetBstMidEndSepPunct{\mcitedefaultmidpunct}
{\mcitedefaultendpunct}{\mcitedefaultseppunct}\relax
\EndOfBibitem
\bibitem[Humphrey \latin{et~al.}(1996)Humphrey, Dalke, and
  Schulten]{humphrey1996vmd}
Humphrey,~W.; Dalke,~A.; Schulten,~K. VMD: visual molecular dynamics.
  \emph{Journal of molecular graphics} \textbf{1996}, \emph{14}, 33--38\relax
\mciteBstWouldAddEndPuncttrue
\mciteSetBstMidEndSepPunct{\mcitedefaultmidpunct}
{\mcitedefaultendpunct}{\mcitedefaultseppunct}\relax
\EndOfBibitem
\bibitem[Brooks \latin{et~al.}(2009)Brooks, Brooks~III, Mackerell~Jr, Nilsson,
  Petrella, Roux, Won, Archontis, Bartels, Boresch, \latin{et~al.}
  others]{brooks2009charmm}
Brooks,~B.~R.; Brooks~III,~C.~L.; Mackerell~Jr,~A.~D.; Nilsson,~L.;
  Petrella,~R.~J.; Roux,~B.; Won,~Y.; Archontis,~G.; Bartels,~C.; Boresch,~S.
  \latin{et~al.}  CHARMM: the biomolecular simulation program. \emph{Journal of
  computational chemistry} \textbf{2009}, \emph{30}, 1545--1614\relax
\mciteBstWouldAddEndPuncttrue
\mciteSetBstMidEndSepPunct{\mcitedefaultmidpunct}
{\mcitedefaultendpunct}{\mcitedefaultseppunct}\relax
\EndOfBibitem
\bibitem[Yoo and Aksimentiev(2011)Yoo, and Aksimentiev]{yoo2011improved}
Yoo,~J.; Aksimentiev,~A. Improved parametrization of Li+, Na+, K+, and Mg2+
  ions for all-atom molecular dynamics simulations of nucleic acid systems.
  \emph{The journal of physical chemistry letters} \textbf{2011}, \emph{3},
  45--50\relax
\mciteBstWouldAddEndPuncttrue
\mciteSetBstMidEndSepPunct{\mcitedefaultmidpunct}
{\mcitedefaultendpunct}{\mcitedefaultseppunct}\relax
\EndOfBibitem
\bibitem[Miyamoto and Kollman(1992)Miyamoto, and Kollman]{miyamoto1992settle}
Miyamoto,~S.; Kollman,~P.~A. Settle: An analytical version of the SHAKE and
  RATTLE algorithm for rigid water models. \emph{Journal of computational
  chemistry} \textbf{1992}, \emph{13}, 952--962\relax
\mciteBstWouldAddEndPuncttrue
\mciteSetBstMidEndSepPunct{\mcitedefaultmidpunct}
{\mcitedefaultendpunct}{\mcitedefaultseppunct}\relax
\EndOfBibitem
\bibitem[Andersen(1983)]{andersen1983rattle}
Andersen,~H.~C. Rattle: A “velocity” version of the shake algorithm for
  molecular dynamics calculations. \emph{Journal of Computational Physics}
  \textbf{1983}, \emph{52}, 24--34\relax
\mciteBstWouldAddEndPuncttrue
\mciteSetBstMidEndSepPunct{\mcitedefaultmidpunct}
{\mcitedefaultendpunct}{\mcitedefaultseppunct}\relax
\EndOfBibitem
\bibitem[Gumbart \latin{et~al.}(2012)Gumbart, Khalili-Araghi, Sotomayor, and
  Roux]{gumbart2012constant}
Gumbart,~J.; Khalili-Araghi,~F.; Sotomayor,~M.; Roux,~B. Constant electric
  field simulations of the membrane potential illustrated with simple systems.
  \emph{Biochimica et Biophysica Acta (BBA)-Biomembranes} \textbf{2012},
  \emph{1818}, 294--302\relax
\mciteBstWouldAddEndPuncttrue
\mciteSetBstMidEndSepPunct{\mcitedefaultmidpunct}
{\mcitedefaultendpunct}{\mcitedefaultseppunct}\relax
\EndOfBibitem
\bibitem[Crozier \latin{et~al.}(2001)Crozier, Henderson, Rowley, and
  Busath]{crozier2001model}
Crozier,~P.~S.; Henderson,~D.; Rowley,~R.~L.; Busath,~D.~D. Model channel ion
  currents in NaCl-extended simple point charge water solution with
  applied-field molecular dynamics. \emph{Biophysical journal} \textbf{2001},
  \emph{81}, 3077--3089\relax
\mciteBstWouldAddEndPuncttrue
\mciteSetBstMidEndSepPunct{\mcitedefaultmidpunct}
{\mcitedefaultendpunct}{\mcitedefaultseppunct}\relax
\EndOfBibitem
\end{mcitethebibliography}


\begin{thebibliography}{10}

\bibitem{lauger1967electrical}
P~L{\"a}uger, W~Lesslauer, E~Marti, and J~Richter.
\newblock Electrical properties of bimolecular phospholipid membranes.
\newblock {\em Biochimica et Biophysica Acta (BBA)-Biomembranes},
  135(1):20--32, 1967.

\bibitem{schoch2008transport}
Reto~B Schoch, Jongyoon Han, and Philippe Renaud.
\newblock Transport phenomena in nanofluidics.
\newblock {\em Reviews of modern physics}, 80(3):839, 2008.

\bibitem{bazant2010induced}
Martin~Z Bazant and Todd~M Squires.
\newblock Induced-charge electrokinetic phenomena.
\newblock {\em Current Opinion in Colloid \& Interface Science},
  15(3):203--213, 2010.

\bibitem{yao2020induced}
Yao Yao, Chenyu Wen, Ngan~H Pham, and Shi-Li Zhang.
\newblock On induced surface charge in solid-state nanopores.
\newblock {\em Langmuir}, 36(30):8874--8882, 2020.

\bibitem{bruus2008theoretical}
Henrik Bruus.
\newblock {\em Theoretical microfluidics}.
\newblock Oxford university press Oxford, 2008.

\bibitem{kavokine2020fluids}
Nikita Kavokine, Roland~R Netz, and Lyd{\'e}ric Bocquet.
\newblock Fluids at the nanoscale: From continuum to subcontinuum transport.
\newblock {\em Annual Review of Fluid Mechanics}, 53:377--410, 2020.

\bibitem{chinappi2018protein}
Mauro Chinappi and Fabio Cecconi.
\newblock Protein sequencing via nanopore based devices: a nanofluidics
  perspective.
\newblock {\em Journal of Physics: Condensed Matter}, 30(20):204002, 2018.

\bibitem{bonome2017electroosmotic}
Emma~Letizia Bonome, Fabio Cecconi, and Mauro Chinappi.
\newblock Electroosmotic flow through an $\alpha$-hemolysin nanopore.
\newblock {\em Microfluidics and Nanofluidics}, 21(5):96, 2017.

\bibitem{willems2020accurate}
Kherim Willems, Dino Rui{\'c}, Florian~LR Lucas, Ujjal Barman, Niels Verellen,
  Johan Hofkens, Giovanni Maglia, and Pol Van~Dorpe.
\newblock Accurate modeling of a biological nanopore with an extended continuum
  framework.
\newblock {\em Nanoscale}, 12(32):16775--16795, 2020.

\bibitem{wilson2019rapid}
James Wilson, Kumar Sarthak, Wei Si, Luyu Gao, and Aleksei Aksimentiev.
\newblock Rapid and accurate determination of nanopore ionic current using a
  steric exclusion model.
\newblock {\em ACS sensors}, 4(3):634--644, 2019.

\bibitem{humphrey1996vmd}
William Humphrey, Andrew Dalke, and Klaus Schulten.
\newblock Vmd: visual molecular dynamics.
\newblock {\em Journal of molecular graphics}, 14(1):33--38, 1996.

\bibitem{2020SciPy}
Pauli {Virtanen}, Ralf {Gommers}, Travis~E. {Oliphant}, Matt {Haberland}, Tyler
  {Reddy}, David {Cournapeau}, Evgeni {Burovski}, Pearu {Peterson}, Warren
  {Weckesser}, Jonathan {Bright}, St{\'e}fan~J. {van der Walt}, Matthew
  {Brett}, Joshua {Wilson}, K.~{Jarrod Millman}, Nikolay {Mayorov}, Andrew
  R.~J. {Nelson}, Eric {Jones}, Robert {Kern}, Eric {Larson}, CJ~{Carey},
  {\.I}lhan {Polat}, Yu~{Feng}, Eric~W. {Moore}, Jake {Vand erPlas}, Denis
  {Laxalde}, Josef {Perktold}, Robert {Cimrman}, Ian {Henriksen}, E.~A.
  {Quintero}, Charles~R {Harris}, Anne~M. {Archibald}, Ant{\^o}nio~H.
  {Ribeiro}, Fabian {Pedregosa}, Paul {van Mulbregt}, and SciPy 1.~0
  {Contributors}.
\newblock {SciPy 1.0: Fundamental Algorithms for Scientific Computing in
  Python}.
\newblock {\em Nature Methods}, 17:261--272, 2020.

\bibitem{Hunter2007}
J.~D. Hunter.
\newblock Matplotlib: A 2d graphics environment.
\newblock {\em Computing in Science \& Engineering}, 9(3):90--95, 2007.

\bibitem{bonthuis2011dielectric}
Douwe~Jan Bonthuis, Stephan Gekle, and Roland~R Netz.
\newblock Dielectric profile of interfacial water and its effect on
  double-layer capacitance.
\newblock {\em Physical review letters}, 107(16):166102, 2011.

\bibitem{raabe2011molecular}
Gabriele Raabe and Richard~J Sadus.
\newblock Molecular dynamics simulation of the dielectric constant of water:
  The effect of bond flexibility.
\newblock {\em The Journal of chemical physics}, 134(23):234501, 2011.

\bibitem{weijs2011origin}
Joost~H Weijs, Antonin Marchand, Bruno Andreotti, Detlef Lohse, and Jacco~H
  Snoeijer.
\newblock Origin of line tension for a lennard-jones nanodroplet.
\newblock {\em Physics of fluids}, 23(2):022001, 2011.

\bibitem{frenkel2001understanding}
Daan Frenkel and Berend Smit.
\newblock {\em Understanding molecular simulation: from algorithms to
  applications}, volume~1.
\newblock Elsevier, 2001.

\bibitem{atkins2011physical}
Peter Atkins and Julio De~Paula.
\newblock {\em Physical chemistry for the life sciences}.
\newblock Oxford University Press, USA, 2011.

\bibitem{lin2021surface}
Kabin Lin, Zhongwu Li, Yi~Tao, Kun Li, Haojie Yang, Jian Ma, Tie Li, Jingjie
  Sha, and Yunfei Chen.
\newblock Surface charge density inside a silicon nitride nanopore.
\newblock {\em Langmuir}, 37(35):10521--10528, 2021.

\bibitem{hoogerheide2009probing}
David~P Hoogerheide, Slaven Garaj, and Jene~A Golovchenko.
\newblock Probing surface charge fluctuations with solid-state nanopores.
\newblock {\em Physical review letters}, 102(25):256804, 2009.

\bibitem{bandara2019chemically}
YM~Nuwan~DY Bandara, Buddini~I Karawdeniya, James~T Hagan, Robert~B Chevalier,
  and Jason~R Dwyer.
\newblock Chemically functionalizing controlled dielectric breakdown silicon
  nitride nanopores by direct photohydrosilylation.
\newblock {\em ACS applied materials \& interfaces}, 11(33):30411--30420, 2019.

\bibitem{kosmulski1997attempt}
M~Kosmulski.
\newblock Attempt to determine pristine points of zero charge of nb2o5, ta2o5,
  and hfo2.
\newblock {\em Langmuir}, 13(23):6315--6320, 1997.

\bibitem{larkin2014high}
J~Larkin, R~Y Henley, M~Muthukumar, J~K Rosenstein, and M~Wanunu.
\newblock High-bandwidth protein analysis using solid-state nanopores.
\newblock {\em Biophys. J.}, 106(3):696--704, 2014.

\bibitem{xie2009surface}
Yanbo Xie, Jianming Xue, Lin Wang, Xinwei Wang, Ke~Jin, Long Chen, and Yugang
  Wang.
\newblock Surface modification of single track-etched nanopores with surfactant
  ctab.
\newblock {\em Langmuir}, 25(16):8870--8874, 2009.

\bibitem{kosmulski2002ph}
Marek Kosmulski.
\newblock The ph-dependent surface charging and the points of zero charge.
\newblock {\em Journal of Colloid and Interface Science}, 253(1):77--87, 2002.

\end{thebibliography}

\end{document}



\title{
\large{
	Supplementary Information: 
	Geometrically Induced Selectivity and 
	Unidirectional Electroosmosis in Uncharged Nanopores}
}

\author{\normalsize{
	Giovanni Di Muccio$^1$,
	Blasco Morozzo della Rocca$^2$, 
	Mauro Chinappi$^1$
	}}
 
 \email{mauro.chinappi@uniroma2.it}

 \affiliation{$^1$ Dipartimento di Ingegneria Industriale, Universit\`a di Roma Tor Vergata, Via del Politecnico 1,
                 00133, Rome, Italy. \\
 	      $^2$ Dipartimento di Biologia, Universit\`a di Roma Tor Vergata, Via della Ricerca Scientifica 1, 
                00133, Rome, Italy.\\  }

\date{\today}

\begin{abstract}
\bigskip
\bigskip

\noindent {\bf Contents:} \linebreak

\begin{tabular}{p{13cm} p{4 cm}}
Supplementary Note S1: Induced Debye layer capacitance for the cavity-nanopore.        & p. S-2  \\
{Supplementary Note S2: Comments on the PNP-NS model} 				& p. S-7  \\
\end{tabular}

\vspace{0.3cm}

\begin{tabular}{p{13cm} p{4 cm}}
Supplementary Figure S1: Induced charge model and equivalent capacitance.                                         & p. S-4  \\
{Supplementary Figure S2: Electroosmotic velocity profile and Flow rate.} & p. S-8  \\
{Supplementary Figure S3: Role of prefactor on EOF and charge estimations.}         & p. S-9  \\
Supplementary Figure S4: Solid and dipolar fluid models.                               & p. S-10  \\
Supplementary Figure S5: Phase diagram for model dipolar fluid.                        & p. S-11  \\
Supplementary Figure S6: Dielectric constant for model dipolar fluid.                  & p. S-12  \\ 
Supplementary Figure S7: Viscosity of the model dipolar fluid in the liquid state.     & p. S-13  \\
Supplementary Figure S8: Wettability of solid-state membrane by model dipolar fluid.   & p. S-14 \\
Supplementary Figure S9: Ion diffusion coefficients and mobility for the model solution.& p. S-15 \\
Supplementary Figure S10: Ionic currents for model system and symmetric electrolyte.   & p. S-16 \\
{Supplementary Figure S11: Electric potential and Field lines for different cavity sizes}   	       & p. S-17 \\
Supplementary Figure S12: EOF predictions for a silicon nitride nanopore.              & p. S-18 \\
{Supplementary Figure S13: MD simulation of a weakly charged model pore}                 & p. S-19 \\
{Supplementary Figure S14: Net charge distribution and ionic currents for CsgG.}          & p. S-20  \\
{Supplementary Figure S15: WT vs. Neutral Model of CsgG nanopore.}                         & p. S-21 \\
{Supplementary Figure S16: Comparison of Induced Charge and Intrinsic Selectivity of CsgG nanopore.}                         & p. S-22 \\
{Supplementary Table S1: Examples of surface charges and point of zero charge for some solid-state nanopores.}                                                                   & p. S-23 \\
Additional References:                                                                 & p. S-24 \\
\end{tabular}

\end{abstract}

\maketitle


\section*{Supplementary Note S1: Induced Debye Layer Capacitance}

Here we provide details on the derivation
of the expression of the equivalent
capacitance $C_s$ between the lateral cavity and the pore lumen, 
Eq. (2) of the manuscript.
We first derive 
the capacitance of a planar wall separating two reservoirs 
containing an electrolyte solution, then we extend 
the discussion for a cylinder and, finally, 
we calculate the equivalent capacitance for the cavity-nanopore system.\\

\noindent {\bf Planar membrane.}
The capacitance of a planar membrane is a classical 
problem, see, {\sl e.g.} L{\"a}uger {\sl et al.}~\cite{lauger1967electrical},
 here revised for reader convenience.
Let us consider the system, represented in Supplementary Fig.~S1a,
composed of an infinite neutral solid membrane
of relative permittivity $\varepsilon_S$ and thickness $h$,
separating two reservoirs of a perfectly symmetric 1:1 electrolyte
solution
with relative permittivity $\varepsilon_L$
and bulk concentration $c_0$.
A voltage $\Delta V$ is applied between the two reservoirs.
Without loss of generality, we assumed that the left side (G) is 
grounded.
The membrane is parallel to the $Oxy$ plane, 
and $z=0$ ($z=-h$) corresponds 
to the interface between the right (left) reservoir
and the solid membrane.
The problem is hence one dimensional and 
all the variables depend only on the $z$ coordinate.
For small surface potential 
$\zeta_w \ll k_B T / e$,
the variation of the electric potential $\phi(z)$ 
into the electrolyte solution (A and G domains) 
is ruled by the Debye-H{\"u}ckel approximation of 
the Poisson-Boltzmann equation~\cite{schoch2008transport}
\begin{equation}
	\frac{d^2 \phi}{d z^2} =
	  \frac{1}{\lambda_D^2} 
	  \phi
	\; , \quad
	\text{(electrolyte solution G)}
	\; ,
	\label{eq:debyeG}
\end{equation}
\begin{equation}
	\frac{d^2 \phi}{d z^2} =
	  \frac{1}{\lambda_D^2} 
	  \left(\phi - \Delta V \right) 
	\; , \quad
	\text{(electrolyte solution A)}
	\; ,
	\label{eq:debyeA}
\end{equation}
where
\begin{equation*}
	\lambda_D = \sqrt{
		\frac{\varepsilon_0\varepsilon_L k_B T}{2(\nu e)^2 c_0}
		}
\end{equation*}
is the Debye length, $\varepsilon_0$ the vacuum electrical permittivity,
$\nu$ the valence of the ions (1 in our case),
$e$ the elementary charge, 
$k_B$ the Boltzmann constant and $T$ the temperature.
Inside the membrane, instead, 
the Poisson equation reduces to
\begin{equation}
	\frac{d^2 \phi}{d z^2} 
	= 
	0
	\; , \quad
	\text{(membrane S)}
	\; .
	\label{eq:laplace}
\end{equation}
Equations~\eqref{eq:debyeG}-\eqref{eq:laplace} 
are solved with the following boundary conditions,
\begin{equation}
	\phi(z) = 0 \quad \text{for} \quad z \to \infty  
	\; ,
	\label{eq:boundaryBCG}
\end{equation}
\begin{equation}
	\phi(z) = \Delta V \quad \text{for} \quad z \to -\infty  
	\; ,
	\label{eq:boundaryBCA}
\end{equation}
\begin{equation}
	\varepsilon_S    
	\frac{d\phi_S}{d z} \Big |_{z=0} 
	= 
	\varepsilon_L 
	\frac{d\phi_G}{d z} \Big |_{z=0} 
	\; ,
	\label{eq:boundaryBCSG}
\end{equation}
\begin{equation}
	\varepsilon_S    
	\frac{d \phi_S}{d z} \Big |_{z=-h} 
	= 
	\varepsilon_L 
	\frac{d \phi_A}{d z} \Big |_{z=-h} 
	\; ,
	\label{eq:boundaryBCSA}
\end{equation}
where \eqref{eq:boundaryBCSG} and \eqref{eq:boundaryBCSA} impose the 
continuity of the normal component of the electrical displacement
vector ${ \bf D} = \varepsilon_0 \varepsilon_{L/S} {\bf E}$ at the interfaces
between the domains G and S, Eq.~\eqref{eq:boundaryBCSG}, 
and A and S, Eq.~\eqref{eq:boundaryBCSA}, respectively.
The solution of the system~\eqref{eq:debyeG}-\eqref{eq:laplace} 
with the boundary conditions~\eqref{eq:boundaryBCG}-\eqref{eq:boundaryBCSA}
is composed of two exponential branches in the liquid 
reservoirs and a linear branch in the solid membrane. 
In particular, we get the following solution
\begin{equation}
	\phi(z) = \zeta_w \exp\left[ -\frac{z}{\lambda_D} \right]
	\; ,  \; \;	
	\text{for} \,
	z \ge 0 
	\quad  
	\text{(electrolyte solution G)}
	,
	\label{eq:debye-solG}
\end{equation}
\begin{equation}
	\phi(z) = \Delta V 
        -  \zeta_w \exp\left[ \frac{(z+h)}{\lambda_D} \right]
	\; ,  \; \;	
	\text{for} \,
	z \le -h 
	\quad  
	\text{(electrolyte solution A)}
	,
	\label{eq:debye-solA}
\end{equation}
\begin{equation}
	\phi(z) = \zeta_w - \left( \frac{\Delta V - 2\zeta_w}{h}\right) z
	\; ,  \; \;	
	\text{for} \,
	-h < z < 0 
	\quad  
	\text{(membrane S)}
	\; .
	\label{eq:laplace-sol}
\end{equation}
where 
\begin{equation}
	\zeta_w = 
	\left( 
	\frac{\varepsilon_S}{\varepsilon_L}
	\right)
	\left( 
	\frac{\lambda_D}{h}
	\right)
	\left( 
	\frac{\Delta V}
	{1 + \frac{2\lambda_D}{h}
	     \frac{\varepsilon_S}{\varepsilon_L}}
	\right)
	\label{eq:zetaw}
\end{equation}       
is the magnitude of the difference between the bulk potential
and the surface potential. 
In essence, $\zeta_w$ is the magnitude of the 
induced surface potential.

From now on, since the system is symmetric with 
respect to the membrane ({\sl e.g.} the two electrolyte solutions in the reservoirs 
G and A are identical), we will focus only on the ground reservoir G.
Once $\phi(z)$ is known, the charge density
in the liquid, $\rho_{el}(z)$, is obtained as
\begin{equation}
	\rho_{el}(z)=
		-\frac{\varepsilon_0\varepsilon_L \zeta_w}{\lambda_D^2}
		\exp\left[ -\frac{z}{\lambda_D} \right]
	\; ,  \; \;	
	\text{for} \,
	z \ge 0 
	\; ,
	\label{eq:rhoel}
\end{equation}
where, again, we are using the Debye-H{\"u}ckel approximation.
The region in the electrolyte solution domain 
where the charge accumulates under the action
of the applied voltage $\Delta V$ is known as induced 
Debye layer (IDL, colored area in Supplementary Fig.~S1 and 
manuscript Fig.~1c). 
Integrating Eq.~\eqref{eq:rhoel} in the liquid reservoir,
the total charge $q$ for unit of surface in the IDL is
\begin{equation}
	q_{pla} 
	= 
	\int_0^\infty \rho_{el}(z) \, dz 
	= 
	-\frac{\varepsilon_0\varepsilon_L}{\lambda_D}\zeta_w 
	=
	-\left( 
	\frac{\varepsilon_0\varepsilon_S}{h}
	\right)
	\left( 
	\frac{\Delta V}
	{1 + \frac{2\lambda_D}{h}
	     \frac{\varepsilon_S}{\varepsilon_L}}
	\right)
	\; .
	\label{eq:total-charge}
\end{equation}
so that the capacitance per unit area of the membrane
is 
\begin{equation}
	C_{pla} 
	=
	\left \vert
	\frac{q_{pla} }{\Delta V}
	\right \vert
	=
	\frac{\varepsilon_0\varepsilon_S}{h}
	\left ( 1 + \frac{2\lambda_D}{h}
	     \frac{\varepsilon_S}{\varepsilon_L}\right )^{-1}
	\, .
	\label{eq:cappla}
\end{equation}
The quantity $q_{pla}$ is also indicated as induced charge 
since its presence depends on 
the application of an external electric field. 
For a review on the various induce charge
electrokinetic phenomena (often indicated as ICEK 
in the literature), we refer the reader to~\cite{bazant2010induced}
and reference therein.
For $\lambda_D \ll h$,
Eq.~\eqref{eq:cappla}
reduces to the formula 
for the geometric capacitance per unit area 
of a planar capacitor of height $h$,
\begin{equation}
	C_{pla} 
	= 
	\frac
	{\varepsilon_0\varepsilon_S}
	{h}
	\; .
	\label{eq:cappla1}
\end{equation}


\begin{figure*}
	\centering
	\includegraphics[width=0.76\textwidth]
	{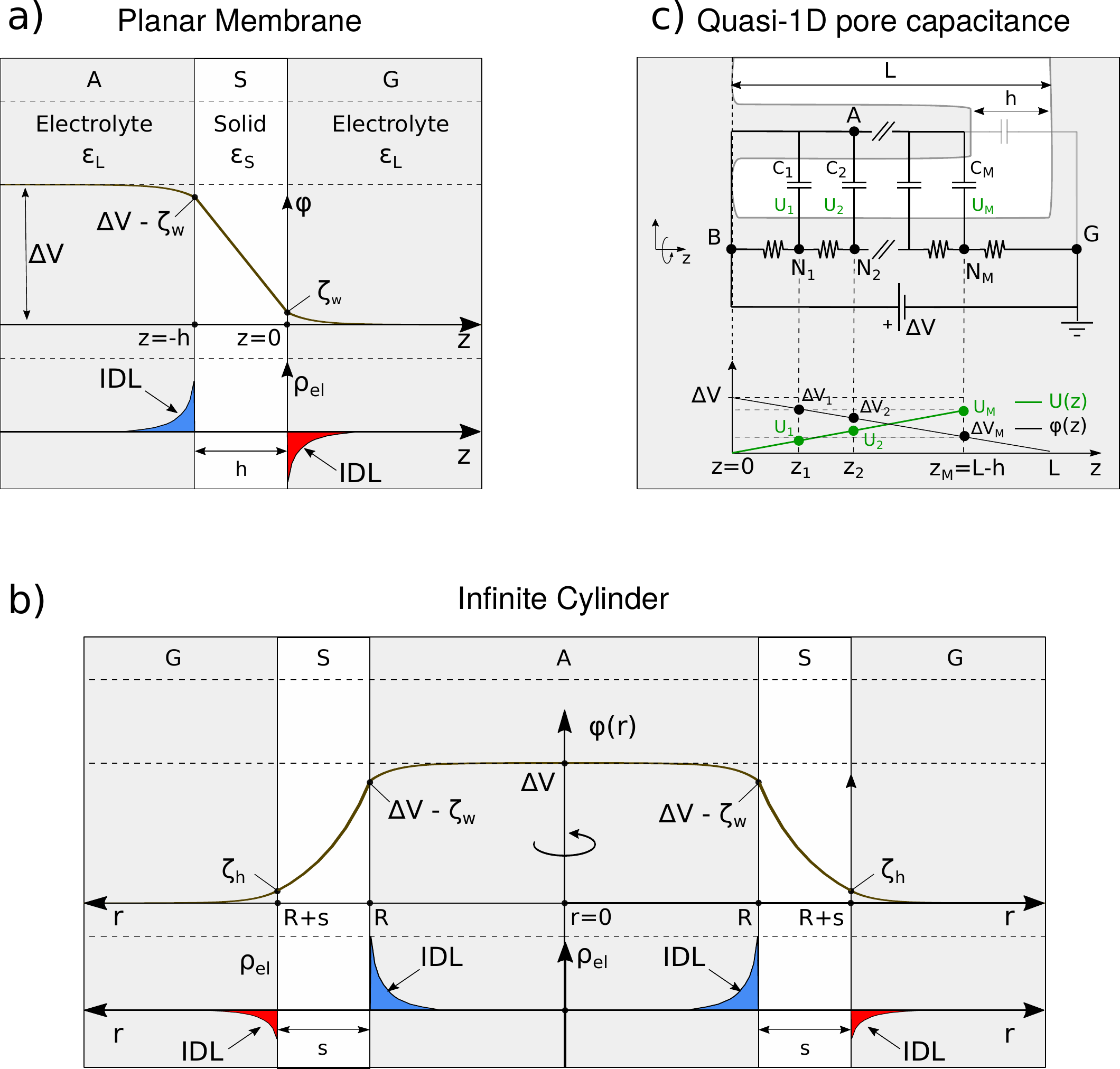}
\end{figure*}

\vspace{-0.6 cm}

\small
\noindent
	{\bf Supplementary Figure S1.}
	{\bf Induced charge model and equivalent cavity-nanopore capacitance.}
	{\bf a-b)} 
	Sketches of the electrical potential $\phi$ and
	net charge distributions $\rho_{el}$
	induced by a voltage drop $\Delta V$ 
	across {\bf a)} a planar membrane
	and {\bf b)} an infinite cylinder, separating two electrolyte solutions 
	(A and G).
	The continuity of the electrical displacement field 
	at the solid-liquid interfaces drives the formation of 
	Induced Debye Layer (IDL) of opposite charge,
	resulting in a capacitance.
	{\bf c)} 
	Circuit model 
	employed for calculating the equivalent capacitance $C_s$ 
	between the pore lumen and the lateral cavity. 
	The lower panel report the 
	electric potential $\phi(z)$
	(black line) 
	that linearly decreases from
	$\Delta V$ (point~B, $z=0$) 
	to $0$ (point~G, $z=L$)
	and the potential difference
	$U(z)$ (green line) between the 
	lateral cavity and the pore lumen
	that linearly increases from 
	$0$ (point~B, $z=0$) to $\Delta V (L-h)/L$ 
	(node $N_m$, $z=L-h$).

\bigskip
\bigskip

\vspace{0.2 cm}

\noindent This is not surprising, indeed,
for $\lambda_D \ll h$, the charge
in the Debye layer accumulates in a very thin region close
to the liquid-solid interface
and $\zeta_w \ll \Delta V$, see Eq.~\eqref{eq:zetaw}.
In our system, $\lambda_D$ and $h$ are 
comparable,
nevertheless, considering a 1M water solution $\varepsilon_L=78$ 
of a 1:1 electrolyte at $T=300~K$,
($\lambda_D=0.304$~nm)
and a $h=1$~nm thick membrane
with $\varepsilon_S=1$,
the second factor of Eq.~\eqref{eq:cappla} is $\simeq0.99$.
So, again, Eq.~\eqref{eq:cappla1} is a very good
approximation for the capacitance per unit surface.
For this reason, Eq.~\eqref{eq:cappla1} is commonly used
as starting point in the theoretical analysis of
some induced charge electrokinetics phenomena, 
see, e.g.~\cite{yao2020induced}, where it is employed
to provide an estimation of the induced charge in a conical nanopore.

As a final comment, we note that, for the 
above mentioned systems (1M 1:1 electrolyte), 
for an applied voltage $\Delta V=1$~V,
Eq.~\eqref{eq:zetaw} predicts a value $\zeta_w \approx 3.4~\mathrm{mV}$
that is lower than the thermal voltage $k_B T/e = 25~\mathrm{mV}$,
implying that the Debye-H{\"u}ckel approximation is valid
even for relatively large voltages (compared to those
often employed in nanopore experiments).\\

\vspace{0.1 cm}
\noindent {\bf Infinite cylinder.}
Applying a similar approach, we derived an 
analogous result for a cylindrical electrolytic capacitance.
Let us consider an infinite insulating cylindrical membrane
of inner radius $R$ and outer radius $R+s$
separating two electrolyte solutions (A and G), see Supplementary Fig.~1b.
Outer domain is grounded (G) 
while the potential on the cylinder axis is $\Delta V$.
As in the planar case, in principle it is possible 
to solve the Debye-H{\"u}ckel approximation of 
the Poisson-Boltzmann equation
in the two domains A and G and the Poisson equation in the 
membrane domain S, Supplementary Fig.~1b.
However, for $\lambda_D \ll R$ and $\lambda_D \ll h$,
the IDL is thin compared to the cylindrical membrane size.
So
the capacitance for unit of length of the
system can be approximated to 
\begin{equation}
	C_{cyl} 
	= 
	\frac{2\pi\varepsilon_0\varepsilon_S}
	{\ln\left( 1 + \frac{s}{R} \right)}
	\label{eq:cylindrical}
	\; ,
\end{equation}
that, similarly to the planar case, 
 is the geometrical capacitance for unit of length 
 for a cylindrical capacitor 
with dielectric constant $\varepsilon_S$,
inner radius $R$ and outer radius $R+s$.

\vspace{0.3 cm}

\noindent{\bf Equivalent cavity-nanopore cylindrical capacitance.}
For the estimation of the capacitance between
the nanopore lumen and the cavity surrounding 
the entrance, we 
developed the quasi-1D model sketched 
in Supplementary Fig.~S1c.
The white domain in Supplementary Fig.~S1c 
corresponds to a half-section of
the solid membrane separating the two electrolyte reservoirs.
The nanopore connects the two 
reservoirs and its axis lies on the $z$ axis
of the coordinate system.
In our circuit model,
the branch BG corresponds to the 
pore lumen and the point~A to the lateral cavity.
The point G  is connected to the ground and 
a potential $\Delta V$ is applied at the point B
(left pore entrance, $z = 0$).
We assume that the 
potential inside the lateral cavity 
is uniform and equal to $\Delta V$,
{\sl i.e.}, equal to the potential at the  
left pore entrance.
Instead, the potential 
$\phi(z)$ along the nanopore 
linearly decays from $\Delta V$ to zero,
when moving from left entrance ($z=0$)
to right entrance ($z=L$)
\begin{equation}
	\phi(z)
	=
	\Delta V
	\left (1 - \frac{z}{L} \right )
	\, .
\end{equation}
The difference in the potential between 
the pore lumen and the cavity (point~A) 
leads to an accumulation of charges inside
the pore lumen that can be estimated as follows.
Let us divide the branch BG into $M$ segments
so that $M-1$ intermediate nodes $N_k$ exist,
each one at potential $V_k=\phi(z_k)$,
with $z_k=k \Delta z$ and $\Delta z=(L-h)/M$, with
$L$ the pore length and $h$ the membrane thickness
in the cavity, see Supplementary Fig.~S1c.
For each node $N_k$,
we assume 
a cylindrical capacitor $C_{k}$ of length $\Delta z$,
subjected to a potential
difference $U_k = \Delta V - V_k$.
The charge $q_N$ is
then
the sum of the charges accumulated in 
each $C_k$ capacity,
\begin{equation}
	q_N
	=
	\sum_{k=1}^{M} C_{cyl} \, U_k \, \Delta z
	\; ,
\end{equation}
with $C_{cyl}=C_k/\Delta z$ the cylindrical capacitance per unit length.

In the limit $M \rightarrow \infty $,
we have $z_k \to z$, $U_k = U(z)$ and
the summation becomes the integral 
\begin{equation}
	q_N =
	\int_0^{L-h} 
	C_{cyl}(z)
	U(z)
	\, 
	dz 
	\, .
	\label{eq:capacitance0}
\end{equation}
In general $C_{cyl}(z)$ depends on the 
radius of the pore, see Eq.~\eqref{eq:cylindrical},
and hence,
for a generic axial-symmetric pore shape,
it is a function of $z$.
However, in our case the pore is cylindrical
so $C_{cyl}$ is constant along the pore
and Eq.~\eqref{eq:capacitance0} 
can be analytically worked out as
\begin{equation}
	q_N =
	C_{cyl}
	\int_0^{L-h} dz \; U(z)
	=
	\frac{1}{2}
	\frac{(L-h)^2}{L}
	\,
	C_{cyl}
	\Delta V
	\label{eq:capacitance01}
\end{equation}
that, using Eq.~\eqref{eq:cylindrical},
reduces to
\begin{equation}
	q_N 
	=
	\frac{\pi \varepsilon_0\varepsilon_S}{L}
	\frac
		{(L-h)^2}
		{\ln\left(1+\frac{s}{R}\right)}
	\Delta V
	\; ,
	\label{eq:capacitance2}
\end{equation}
and consequently the equivalent capacitance is
\begin{equation}
	C_s 
	=
	\frac{\pi \varepsilon_0\varepsilon_S}{L}
	\frac
		{(L-h)^2}
		{\ln\left(1+\frac{s}{R}\right)}
	\; .
	\label{eq:capacitance21}
\end{equation}

In the Eq.~(2) of the manuscript we added a prefactor 
$(L-4\lambda_D)/L$ to take into account 
the effect of pore entrances.
Indeed, as shown in Fig.~1e of the manuscript,
the deformation of the charge distribution at the pore
entrance extends for around $2\lambda_D$ 
inside the pore. 
In our systems,
for 2M ion concentration 
$\lambda_D \simeq 2$~\AA, 
for long pores 
({\sl e.g.} the $L=30$~\AA~ pore discussed in Fig.~1) 
this additional factor
corresponds to a correction of 
$\simeq 25\%$ on pore lumen charge $q_N$ and
electroosmotic flow rate $Q_w$.

\clearpage

\newpage

{
\section*{Supplementary Note S2: Comments on the PNP-NS model}
}
{To achieve practical analytical 
solutions such as Eq.~(4) and Eq.~(8), we needed 
to introduce several hypotheses in the electrohydrodynamical model.
Here, for clarity, we revise the overall
continuum theoretical framework on which our work 
is based. 
This framework is quite standard, we report here the 
main concepts, focusing on some hypotheses 
we find important to discuss, 
referring the interested readers 
to other sources such as~\cite{bruus2008theoretical,schoch2008transport}.
}

{As a first step, we present the continuum model.
In electrohydrodynamics of electrolyte solutions, 
continuum models are derived by the momentum, mass, energy and species
balance equations
plus the Poisson equation for electrostatics.
We indicate as $c_{\alpha}$ the ion concentration, with 
$\alpha \in [1, N_s]$ where $N_s$ is the number of ion species.
The first hypothesis we introduce is the 
dilution, {\sl i.e.}, we assume that the concentration of
ions is so low that 
the properties of the solutions ({\sl e.g.} viscosity $\eta$, 
relative permittivity $\varepsilon_L$, density $\rho$) 
are not affected by $c_{\alpha}$.
Moreover, we assume that the solution is 
incompressible, hence the pressure $p$ does not affect 
$\rho$.
Then, we also assume that the variation 
of temperature $T$, 
due for instance to viscous dissipation 
or to heat flux at system boundaries, 
is so small that it does not affect the solution properties 
($\rho$, $\eta$, $\varepsilon_L$, etc.).
This last hypothesis allows to decouple the energy
equation from the others, {\sl i.e.},
mass, species and momentum balance 
can be solved without considering the 
temperature distribution.
}
Dilution has two other relevant implications: 
i) the terminal velocity of a single ion of the solution 
due to an electric field ${\bf E}$
is strictly proportional to the 
electric field with the proportionality constant
not affected by ion concentration $c_{\alpha}$.
ii) the diffusion contribution to the particle flux
can be modelled using Fick's law with diffusion
coefficient $D$ not depending on $c_{\alpha}$.
(These two claims are linked by fluctuation-dissipation relation).
If we use, as is common for simple fluids,
the Newtonian model for viscous stresses, the final
set of equations reads
{
\begin{align}
\label{eq:continuity}
\frac{\partial c_\alpha}{\partial t}
+
{\bf u} \cdot
\boldsymbol{\nabla}
c_\alpha
=
D
\nabla^2 c_{\alpha}
+
c_\alpha
\mu \nabla^2 \phi
\\
\label{eq:momentum}
\frac{\partial\boldsymbol{u}}{\partial{t}}
+
\boldsymbol{u}\cdot\boldsymbol{\nabla}\boldsymbol{u}
=
   \frac{1}{\rho} \left(
	\eta \nabla^2 {\bf u} 
   -	\nabla p
   -
	\rho_e\boldsymbol{\nabla}\phi
   \right)
	\;,\\
\label{eq:incompressible}
{\nabla}\cdot\boldsymbol{u}=0\;,\\
\label{eq:poisson}
{\nabla^2}\phi=-\frac{\rho_{e}}{\varepsilon_L}\;,
\end{align}
where $\boldsymbol{u}$ and $\phi$ are 
the fluid velocity and the electrostatic potential and $\rho_e$ 
is the electric charge density
which can be written as a function of the ionic 
species concentration $c_\alpha$ by
\begin{equation}
        \rho_e
        =
	\sum\limits_{\alpha=1}^{N_s} c_\alpha \nu_{\alpha} e\;,
\end{equation}
where $\nu_\alpha e$ is the charge of species $\alpha$,
expressed in terms of the elementary charge $e$, so, {\sl e.g.}
 $\nu_{\alpha} = +1$ for $K^+$ and $\nu_{\alpha} = -1$ for $Cl^-$.
Eq.~\eqref{eq:continuity}-\eqref{eq:poisson} 
are commonly indicated as Poisson-Nernst-Planck-Navier-Stokes 
equations (PNP-NS).
At sufficiently small scales, typical of nanopores, the Reynolds number is small and it is safe
to neglect nonlinear and time-dependent terms,
{\sl i.e.}, the left-hand side of Eq.~\eqref{eq:momentum}, leading to the Stokes equation.}
Moreover, PNP-NS is a mean-field model, and, hence, 
correlations between the ions are neglected and well as fluctuation
of the number of ions in the system.
The latter issue are particularly 
relevant in nanopore systems.
In particular, the particle number fluctuations in an open
system of $N$ particles scale as $\sqrt{N}$. 
In a nanopore, the number of ions is,
in general, very low, and, in principle, fluctuations 
cannot be disregarded. Nevertheless,
including fluctuation would lead to theoretical
approaches that are difficult to implement analytically.
For a discussion on the potential impact of 
such an assumption in nanofluidic systems and on 
alternative theoretical approaches,
we refer the reader to the 
two reviews~\cite{schoch2008transport,kavokine2020fluids}
and references therein.
{For generic geometries, the system \eqref{eq:continuity}-\eqref{eq:poisson}
needs to be solved numerically.
However, a specific solution exists
in the Debye-H{\"u}ckel approximation
for the cylindrical pores 
whose surface is at a wall potential $\zeta_w$ 
under the action
of an external electrical field parallel to the pore axis.
}
{This solution can be expressed in terms 
of Bessel functions~\cite{chinappi2018protein,bruus2008theoretical},
but, for $\lambda_D \ll R$, in essence the 
velocity profile is quite similar to a plug flow
whose magnitude in the center of the channel
is given by the Helmholtz-Smoluchowski 
electroosmotic velocity
\begin{equation}
	\vert v_{eo} \vert = 
		\frac{\varepsilon_0 \varepsilon_L \vert \zeta_w \vert }{\eta} 
		\frac{\vert \Delta V \vert}{L} 
	\; ,
	\label{eq:Qeo}
\end{equation}
while the charge profile is zero everywhere, except for 
a thin region close to the wall.
Eq.~\ref{eq:Qeo} corresponds to Eq.~(5) of the manuscript,
the only difference being that in Eq.~(5) we reported
the sign of the velocity in agreement with the 
reference system we selected, Fig.~1a-b.
As $\lambda_D/R$ decreases, the charge accumulated
in the channel converges to 
the product of pore surface $2\pi RL$ 
times the surface charge of a planar Debye layer
$\varepsilon_0 \varepsilon_L \zeta_w / \lambda_D$.
Eq.~(9) of the manuscript is identical to
Eq.~\eqref{eq:Qeo} when 
the expression 
$\sigma_w =\varepsilon_0 \varepsilon_L \zeta_w / \lambda_D$
is used for nanopore surface charge.
}

{
Hence, in the derivation of our model we have 
two different sets of hypotheses.
On the one hand, since we rely on PNP-NS model,
we implicitly assume all the above-cited hypotheses
(continuum, dilution, incompressibility, mean-field).
Although the reliability of these assumptions
at the nanoscale is \textit{a priori} questionable, several 
studies reported an unexpected capability
of PNP-NS model to capture quantitatively 
the current through nanopores.
For instance, in Bonome {\sl et al.}~\cite{bonome2017electroosmotic},
an analytical electroosmotic model based on 
ideal EOF in a cylindrical channel
was shown to be able to capture the order of magnitude of
EOF measured from MD simulation in an $\alpha$HL nanopore.
Other remarkable examples of successful application
of continuum models to nanopores are 
the recent study by Willems {\sl et al.}~\cite{willems2020accurate},
where the EOF through a large biological pore is calculated
using a PNP-NS model,
and the work by
Wilson {\sl et al.}~\cite{wilson2019rapid}
 where a steric exclusion model 
based on Ohm law allowed to calculate current blockades for
proteins confined into a solid-state nanopore, 
open pore currents for biological channels and blockade
currents produced by DNA homopolymers in MspA, 
showing impressive agreement with all-atom MD data.
In both these works, a crucial ingredient to
improve the quantitative matches
of the continuum model 
with experiments or all-atom MD
was the additional calibration of the 
transport coefficient ({\sl e.g.} dependence of
ion mobility on the distance from the 
wall and on the ion concentration).
However, the qualitative trends are expected to
be well captured also without these more detailed
models, see, {\sl e.g.} Supporting Information
of Willems {\sl et al.}~\cite{willems2020accurate}.
}
{The second class of hypotheses, instead, enters after 
PNP-NS and concerns the above-mentioned limit
for $\lambda_D \ll R$. As shown in Fig.~S2, these
hypotheses tend to overestimate both the 
EOF and the charge accumulated in the nanopore. }

\begin{figure*}[!h]
	\centering
	\includegraphics[width=0.48\textwidth]
	{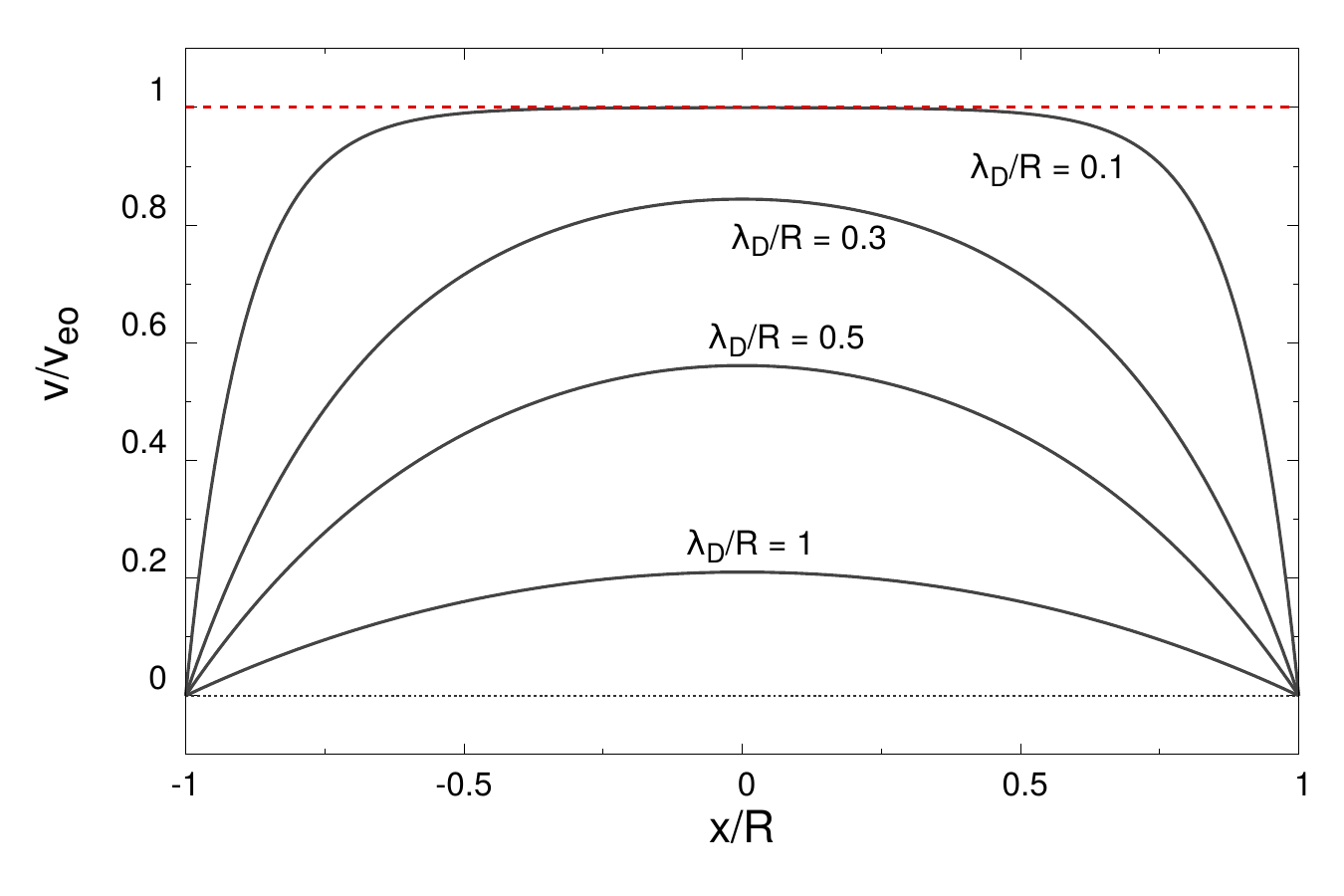}
	\hspace{0.4 cm}
	\includegraphics[width=0.48\textwidth]
	{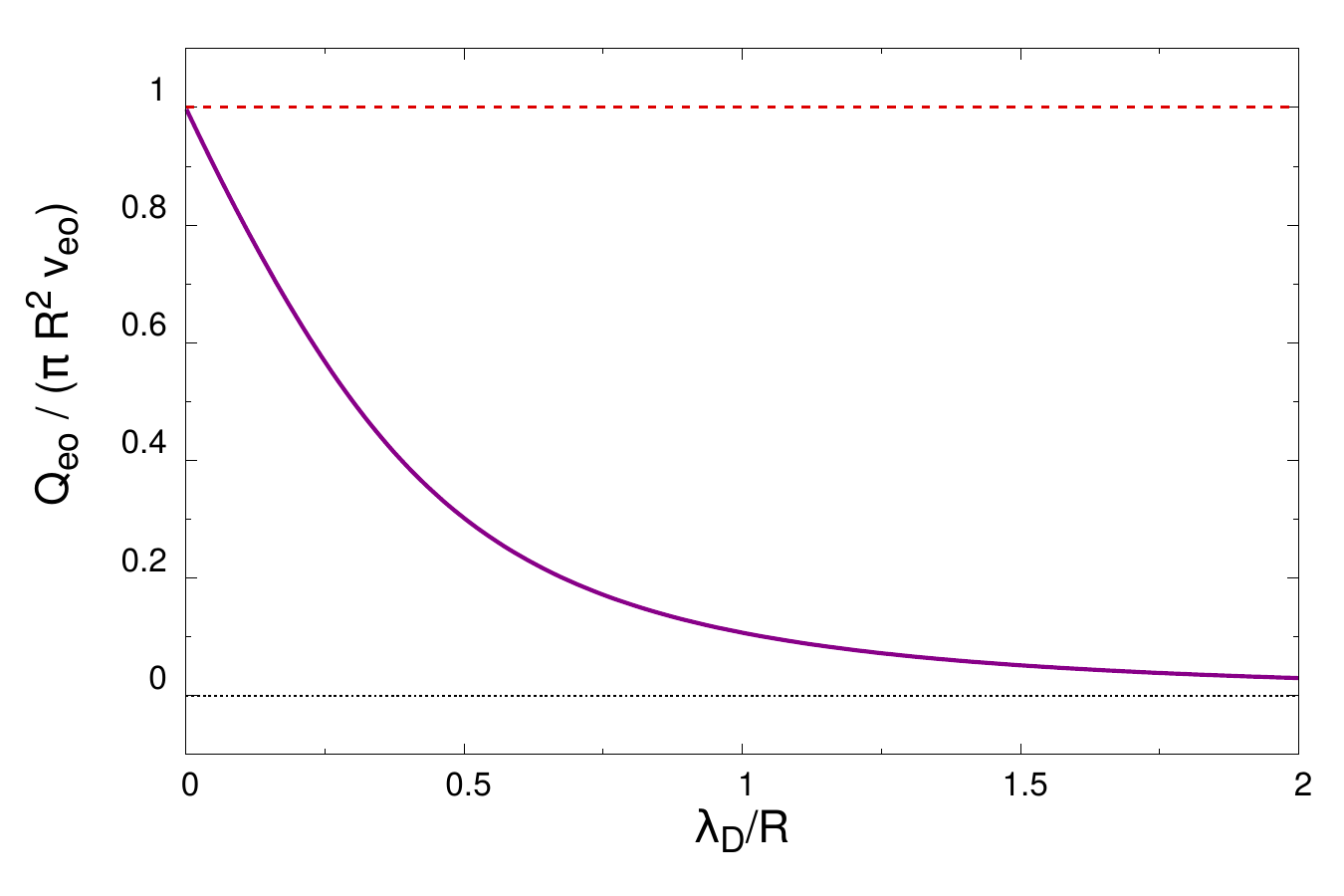}
\end{figure*}

\vspace{-0.55 cm}

\small
\noindent
	{\bf Supplementary Figure S2. Electroosmotic velocity profile and Flow rate.}
	{\bf a)} Electroosmotic velocity profile as a function
	of radial coordinate $r$ for different values of $\lambda_D / R$.
	The curves refer to the classical analytical solution
	in a cylindrical pore of radius $R$ with 
	constant $\zeta_w$ at the wall 
	and external electric field parallel to 
	the pore axis.
	More specifically,
	$v = v_{eo} \left [1 - I_0(r/\lambda_D) / I_0(R/\lambda_D) \right ]$
	where $I_0$ is the modified Bessel function 
	of order $0$~\cite{bruus2008theoretical} and
	$v_{eo}$ is the Helmholtz-Smoluchowski 
	electroosmotic velocity~\cite{bruus2008theoretical}.
	It is evident that for $\lambda_D / R \to 0$, 
	the profile tends to a plug flow of velocity $v_{eo}$.
	{\bf b)} Flow rate through a cylindrical pore 
	as a function of $\lambda_D / R$, obtained integrating
	the velocity profiles $v(r)$ on the pore section.
	For $\lambda_D / R \to 0$, 
	the mass flow rate coincides with the plug flow $Q_{eo} = \pi R^2 v_{eo}$, Eq.~(5) of the manuscript.
\bigskip

\vspace{0.1 cm}

{The above-cited literature  
strongly suggested that our model would be 
able to describe the qualitative trends
or to capture the order of magnitude
of the selectivity and EOF. 
The fact that we also
get a nice quantitative agreement 
with simulation data was not {\sl a priori} expected.
One possibility is that some fortuitous compensation
happens. For instance,
various overestimations of 
the EOF due to the assumption that 
$\lambda_D \ll R$
may be compensated by the prefactor 
$(L-4\lambda_D)/L$ introduced to take into account 
the effect of pore entrances (see Supplementary Note S1).}
{Nevertheless, also without the pore entrance prefactor,
our model provides predictions
that are quite close
to the simulation data
 (the maximum difference is a factor 3 and occurs 
for very short pores where entrance effects are more relevant)
 see Supplementary Fig.~S3, where the same MD data of Fig.~3 of
the manuscript are reported together with model prediction
without pore entrance prefactor.
}
\\

\vspace{1.0cm }

\begin{figure*}[!h]
	\centering
	\includegraphics[width=0.99\textwidth]
	{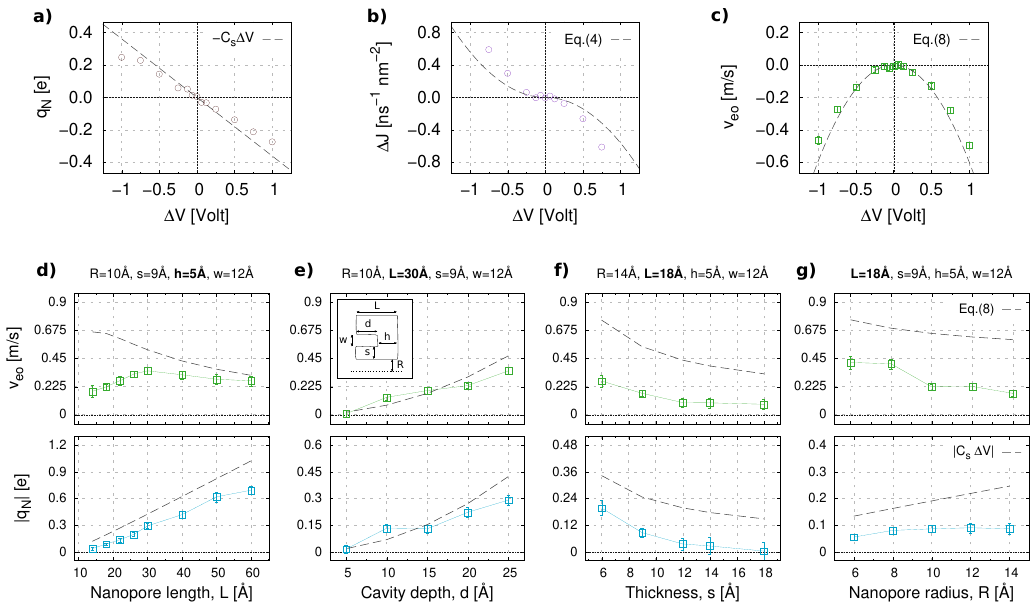}
\end{figure*}
\small
\noindent
	{\bf Supplementary Figure S3.}
	{\bf Role of prefactor on EOF and charge estimations.}
	The MD data reported are the same as in Fig.~3 of the manuscript, while
	the lines refer to the theoretical prediction obtained
	without the prefactor $ (L-4\lambda_D)/L $ introduced 
	to consider the effect of pore entrance.  
	In all cases, the model is able to capture the order
	of magnitude of EOF. The maximum difference is 
	of about a factor 3 and occurs for short pores 
	(see {\sl e.g.} panels {\bf f)} and {\bf g)}, $L=18$~\AA~ and
	panel {\bf a)} for $L < 25$~\AA. 

\clearpage


\begin{figure*}
	\centering
	\includegraphics[width=\textwidth]{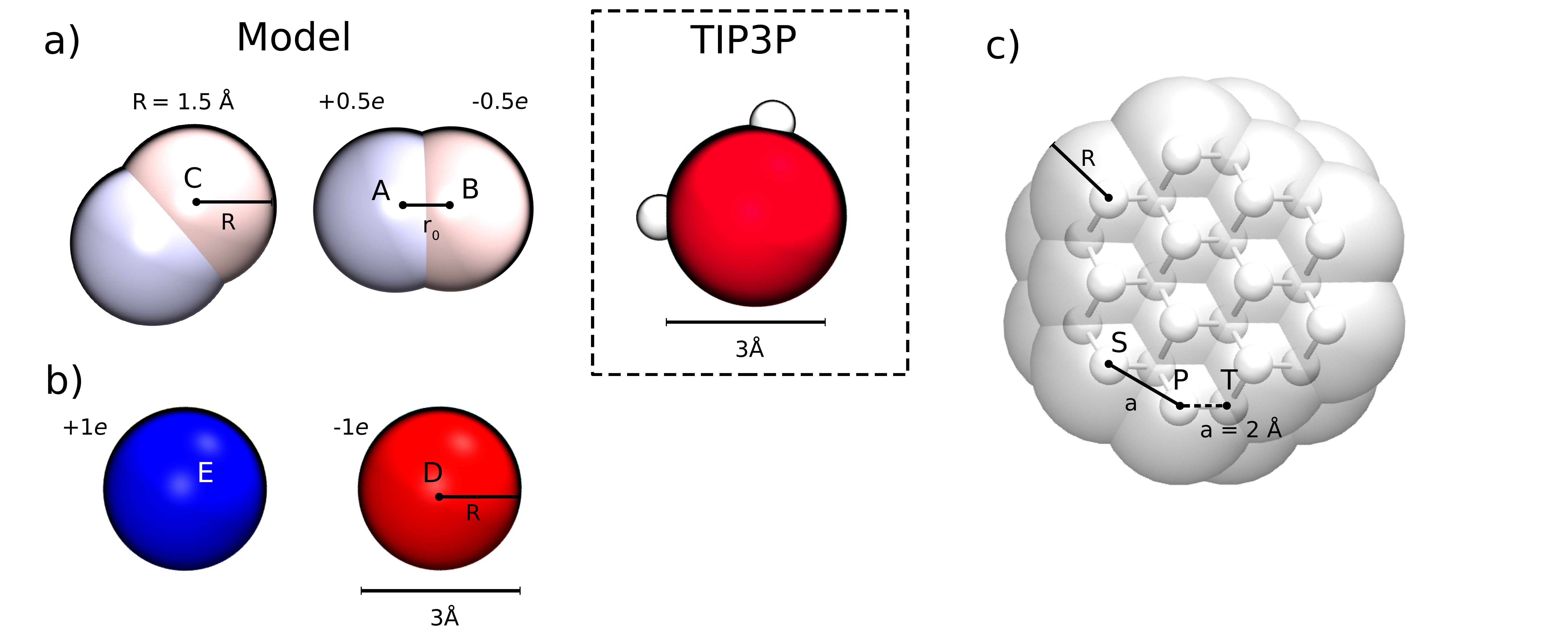}
\end{figure*}

\small
{\noindent
{\bf Supplementary Figure S4.} 
{\bf Solid and dipolar fluid models.}
{\bf a)} Diatomic particle forming our dipolar model fluid (the solvent). 
Each molecule is composed of two atoms 
covalently bound via a harmonic potential
with spring constant~$k_b=450$~kcal/(mol~\AA$^2$)
and equilibrium bond length $r_0=1$~\AA.
The two atoms carry opposite charges 
$q^+=0.5\mathrm{e}$ and $q^-=-0.5\mathrm{e}$, 
giving an overall neutral molecule 
with a dipole intensity 
$\vert p \vert=0.05~\mathrm{e \cdot nm}$.  
The mass of each atom is $m=10~\mathrm{Da}$. 
Non-bonded interactions, {\sl e.g.} between atom A and atom C,
are modeled using 
a standard Coulomb potential for the electrostatic forces and
a Lennard-Jones (LJ) potential  
with parameters $\epsilon_{LL}=0.1$~kcal/mol, $\sigma_{LL}=2.68~\mathrm{\AA}$.
On the right, a TIP3P water is drawn to scale.
{\bf b)}
Ions are formed by atoms
with charge $q^{\pm}=\pm 1\mathrm{e}$ and
mass $m_I=40~\mathrm{Da}$.
Non bonded interactions 
are again modeled as electrostatic
plus LJ potentials.
The LJ parameters 
for the ion-solvent interaction 
are the same as for solvent-solvent,
while for ion-ion interaction 
we used $\sigma_I=3.125~\mathrm{\AA}$.
The slightly larger value $\sigma_I$ compared to 
the solvent-solvent $\sigma_{LL}$ was used 
to avoid the aggregation of ions. 
The resulting electrolyte solution corresponds to a fully dissociated binary symmetric salt in a symmetric liquid.
{\bf c)} Solid crystal forming the membrane. 
Top view of the hexagonal closest packed solid lattice,
only two atomic layers are shown. 
Atoms are uncharged, with mass $m_S=10~\mathrm{Da}$ and
equilibrium distance of $a=2$~\AA. 
To keep the material solid, the  
atoms are harmonically bonded to 
three nearest neighbors belonging to the upper and lower
planes (see {\sl e.g.} atoms P and T) for a total of six bonds for each atom,
with equilibrium distance $r_{0,s}=a$, 
and spring constant $k_{b}=450$~kcal/(mol~\AA$^2$).
LJ model is used 
for non-bonded interactions 
with $\sigma_{SS}=a$ and $\epsilon_{SS}=1$~kcal/mol.
In nanopore simulations the atoms are also constrained 
by a harmonic spring of $k_c=100$~kcal/(mol~\AA$^2$).
LJ parameters for solid-liquid 
interaction are 
$\epsilon_{SL}=0.8\epsilon_{LL}$ and 
$\sigma_{SL}=\sigma$.
The value of $\epsilon_{SL}$ can be modified 
to control the wettability, see Supplementary Fig.~8.
Images are made using VMD~\cite{humphrey1996vmd}. 
}
\clearpage

\begin{figure*}
	\centering
	\includegraphics[width=\textwidth]{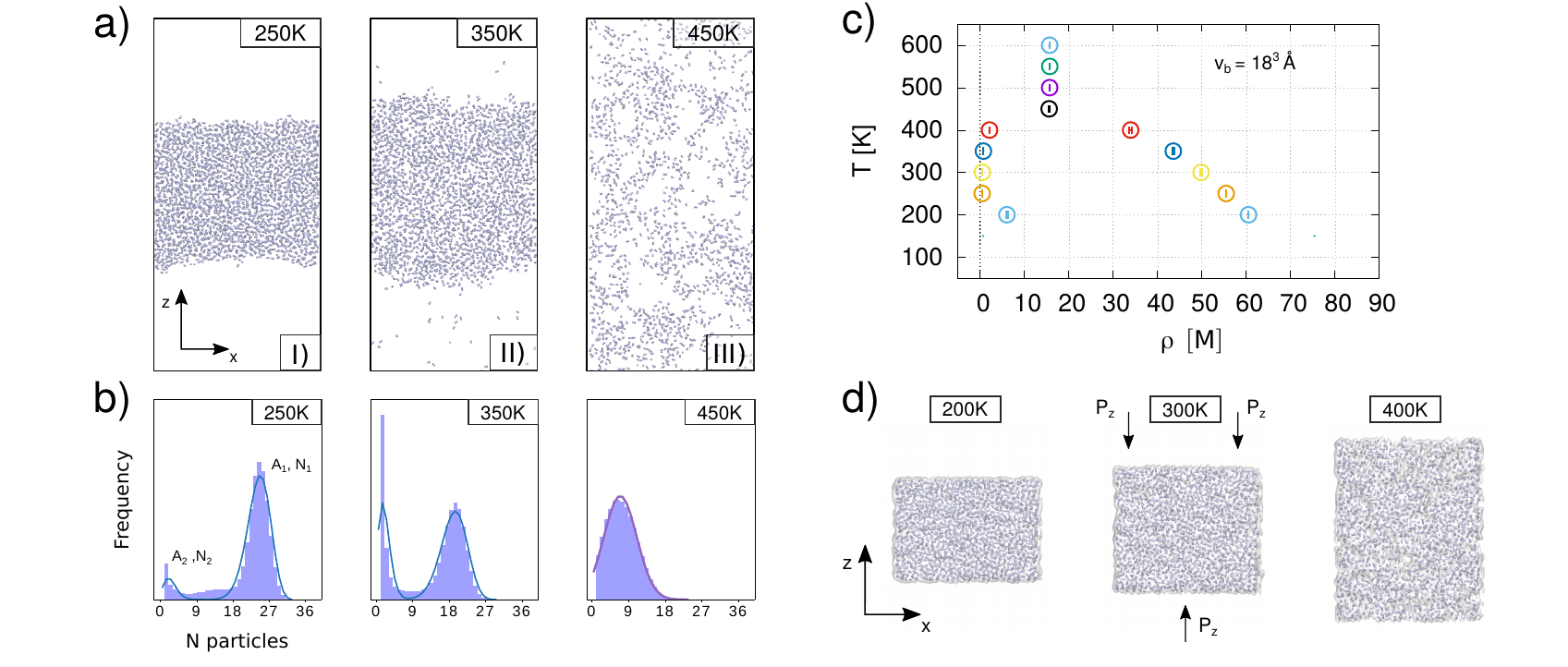}
\end{figure*}

\small
{\noindent
{\bf Supplementary Figure S5.} 
{\bf Phase diagram for model dipolar fluid.}
NVT simulations of $20\,000$ dipolar molecules
in a tripediodic box ($L_x=L_y=90$~\AA, $L_z=270$~\AA)
were run at different temperatures $T$.
A preliminary $200$~ps pre-heating at $2500$~K
is followed by a $500$~ps cooling to the final $T$.
The pre-equilibration stage is used to bring the system away from local
energy minima induced by the initial artificial arrangement of the molecules.
Langevin thermostat with damping coefficient $0.05~\mathrm{ps^{-1}}$
is used in all the runs. 
The equilibrated system is then sampled for $4~\mathrm{ns}$,
saving the coordinates every $20~\mathrm{ps}$.
{\bf a)} Snapshots of the system at three representative $T$. 
{\bf b)} Density distribution computed by
dividing the simulation cell in
cubic blocks of volume $v_b=18^3~\mathrm{\AA^3}$
and counting the molecules
in each block at every sampled frame.
Solid lines refer to a bimodal Poisson fit
(for bimodal distributions) 
or to a Gaussian fit (for unimodal distribution).
{\bf c)} Density-temperature state diagram.
For each temperature, circles correspond to the peaks of the 
fitted distribution with error bars being the corresponding fit errors. 
Unimodal distributions are found for $T > 400~\mathrm{K}$ 
indicating that the critical temperature 
$T_c$ is between $400 < T_c < 450~\mathrm{K}$.
Number density is reported in $\mathrm{mol/L}$,
for $T = 250~\mathrm{K}$ we get $\rho \simeq 55.5~\mathrm{mol/L}$,
a value very similar to water number density.
{\bf d)} Final frame of $12~\mathrm{ns}$
NPT runs ($P=1~\mathrm{bar}$) at different temperatures,
showing that our fluid is in a liquid state 
in the interval 
$200\le T \le 400~\mathrm{K}$. 
Panels {\bf a} and {\bf d} were made using VMD~\cite{humphrey1996vmd}.
Statistical analyses of density distributions were done using
Python modules Scipy~\cite{2020SciPy} and Matplotlib~\cite{Hunter2007}.
}

\clearpage

\begin{figure*}
	\centering
	\includegraphics[width=\textwidth]{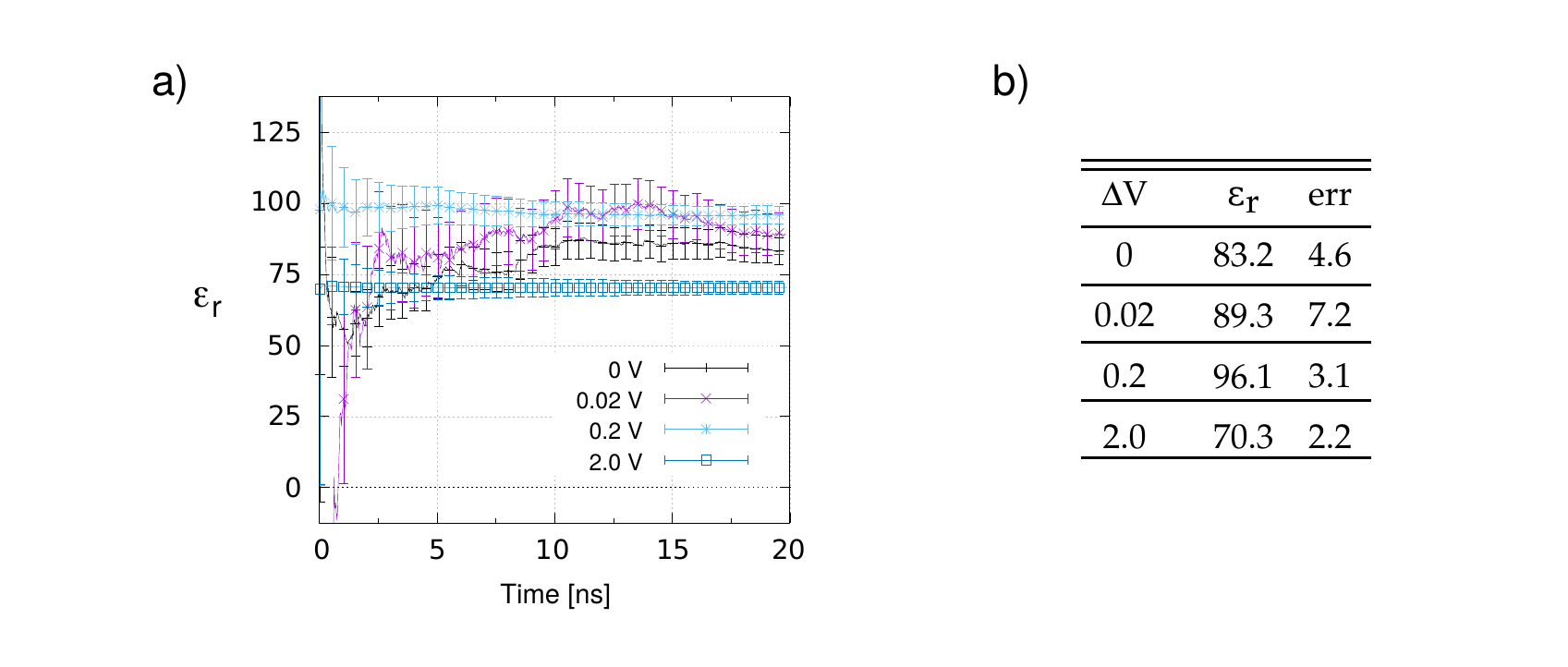}
\end{figure*}

\small
{\noindent
{\bf Supplementary Figure S6.} 
{\bf Dielectric constant for model dipolar fluid.}
{\bf a)} 
The plot represents the MD
relative permittivity estimation, using two different methods, 
as a function of the sampling time, 
for four different applied voltages $\Delta V$.  
The simulation set-up is a rectangular box of 
size $L_x = L_y = 80$~\AA, $L_z=140$~\AA. 
After an NPT equilibration ($T=250$~K, $P=1$~atm),
a $25$~ns NVT production run is started and frames 
are sampled every $20$~ps.
The first $5$~ns were discarded.  
The first permittivity estimation method is a non-equilibrium approach where  
a constant electric field ${\bf E} = (0, 0, E_z)$,
corresponding to a potential $\Delta V= -L_z E_z$
is applied to the system.
Following~\cite{bonthuis2011dielectric,raabe2011molecular},
the relative permittivity $\varepsilon_r$ 
is computed as
\begin{equation*}
	\varepsilon_r =
	1 + 
	\frac{\langle M_z \rangle}{\varepsilon_0 E_z \Gamma} \; ,
	\label{eq:epsirepV}
\end{equation*}
with $\varepsilon_0$ the vacuum permittivity, 
with $M_z$ total dipole moment in the $z$-direction and
$\Gamma=L_x L_y L_z$ the volume of the system.
In the second method (black points)
no electrical field is applied to the system,
and we computed the dielectric constant 
as~\cite{raabe2011molecular}
\begin{equation*}
	\varepsilon_r = 
	1 + 
	\frac{\langle M_z^2 \rangle}{\varepsilon_0 k_B T \Gamma} \; ,
	\label{eq:epsirep0}
\end{equation*}
with $k_B$ the Boltzmann constant and $T$ the temperature.
It is apparent from the figure that the estimation of $\varepsilon_r$ reached a plateau
for all the simulated systems. 
Each data point is calculated as the average
over the previous frames.
The error bars represent the standard errors 
of the respective estimator, computed over the previous frames.
The final estimated values,
after $20~\mathrm{ns}$,
are reported in panel~{\bf b)}.
VMD~\cite{humphrey1996vmd} is used to
calculate $M_z$ and $M_z^2$ at each frame.
}

\clearpage


\begin{figure*}
	\centering
	\includegraphics[width=\textwidth]{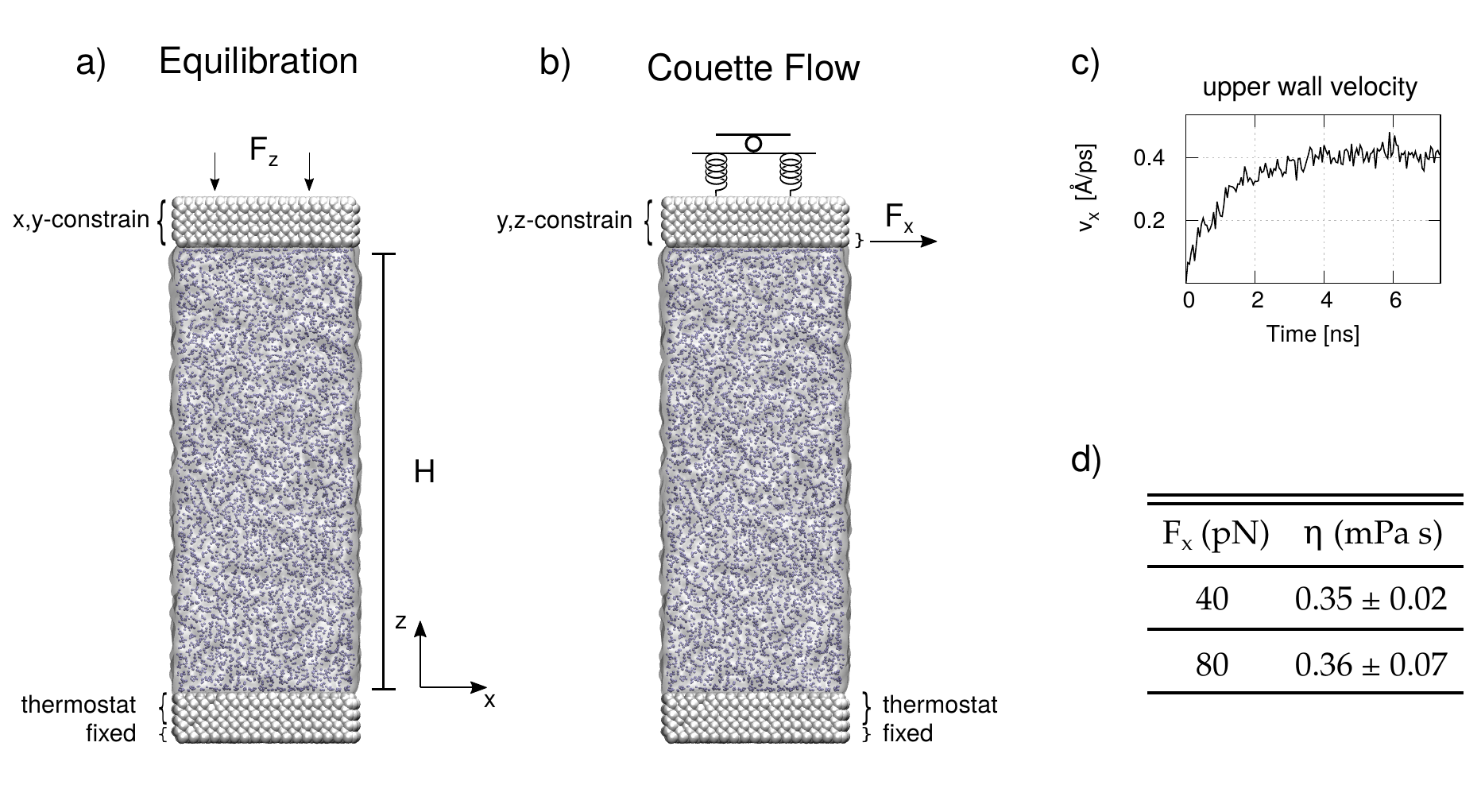}
\end{figure*}

\small
{\noindent
{\bf Supplementary Figure S7.} 
{\bf Viscosity of the model dipolar fluid in the liquid state.} 
The system is composed by $11\,269$ liquid molecules, 
confined between two flat solid slabs, each one composed of $6\,936$ atoms.
The bottom layer of the lower wall is fixed, 
while atoms from the other layers
are thermostated at $T=250~\mathrm{K}$
(Langevin, damping $5~\mathrm{ps^{-1}}$).
During the equilibration, panel {\bf a)},
the atoms of the top layers are constrained
on the x-y plane, while a force $F_z$,
corresponding to a pressure of $1~\mathrm{atm}$,
is applied along the $z$-axis to the topmost layer,
until the liquid reaches a stable height
$H=134.69~\pm~0.02~\mathrm{\AA}$ 
(density $\rho=55.5~\mathrm{mol/L}$). 
The production run is performed by applying a constant force $F_x$ 
to the top solid layer in contact with the liquid,
panel {\bf b)}.
During the production run, 
the atoms of the top layers are harmonically constrained 
on their $y-z$ positions, letting them free to move along the $x$ direction.
The top solid slab reaches a steady-state velocity $v_x$ 
after about $6~\mathrm{ns}$, panel {\bf c)}.
After, a $2~\mathrm{ns}$ sampling is performed to compute the
velocity profile $v_x(z)$ of the liquid, 
saving the coordinates every $50~\mathrm{ps}$.
For each frame, the velocity profile $v_x(z)$ is computed by dividing
the volume height $H$ into slabs of thickness $\Delta z = 1~\mathrm{\AA}$,
using the following procedure:
i) the atoms inside the slab at the frame $f$ are selected;
ii) the x-velocity of each atom $k$ is computed as 
$v_k = (x_k(f+1) - x_k(f-1))/(2\Delta t)$,
with $x_k(f\pm1)$ x-position of the atom $k$ 
at the frame $f\pm1$, and $\Delta t$ sampling interval;
iii) the average velocity in the slab is computed as 
the average of the velocity of the atoms in the slab.
By averaging over all frames after the transient, we get
a linear profile $v_x(z)$ whose slope $\gamma$ is related to the
viscosity of the liquid $\eta=F_x/A\gamma$,
with $A$ surface area where the shear forcing $F_x$ is applied.
{\bf d)} Values of viscosity  $\eta$ estimated at two different $F_x$.
}

\clearpage

\begin{figure*}
	\centering
	\includegraphics[width=\textwidth]{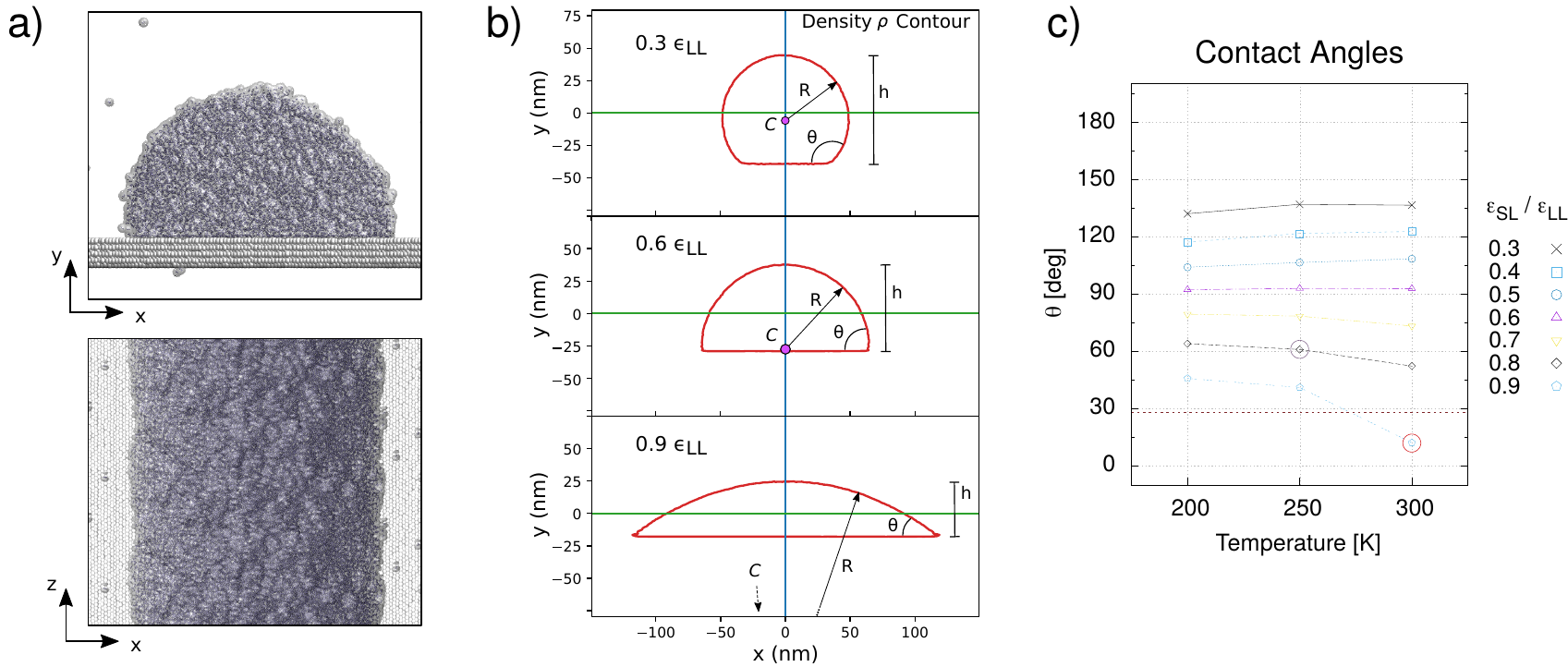}
\end{figure*}

\small
{\noindent
{\bf Supplementary Figure S8.}
{\bf Wettability of solid-state membrane by model dipolar fluid.}
{\bf a)}
System set-up for cylindrical droplet protocol.
A liquid drop formed by $15\,000$ fluid molecules 
is in contact with a solid slab 
($L_x=200~\mathrm{\AA}$,
$L_y=60~\mathrm{\AA}$,
$48\,000$ uncharged atoms, only the central portion of the
system, where the drop is placed, is reported). 
Each solid atom is weakly constrained 
to its lattice position 
(harmonic spring constant~$k_b=100$~kcal/(mol~\AA$^2$).
Wetting can be tuned 
with the liquid-solid LJ parameter $\epsilon_{SL}$.
The solid is eight atomic layers thick 
so that the slab z-height is greater than the LJ
cutoff radius ($r_c~=~12~\mathrm{\AA}$).		
Tri-periodic boundary conditions have been applied. 
To estimate the contact angle $\theta$, NVT simulations are performed 
($2$~ns equilibration, $4$~ns production).
During the production runs, 
a thermostat is applied only to the solid membrane.
{\bf b)}
A procedure similar to one reported in~\cite{weijs2011origin} 
is used to calculate $\theta$.
In brief, we first calculated a 3D density map of the liquid, 
discretizing the system in cells of $1 \times 1 \times 1~\mathrm{\AA^3}$ 
and averaging over time, 
at each step the droplet is centered on its center of mass to
compensate for any collective movements of the fluid parallel to the solid surface. 
After, we computed a 2D density map by averaging over the z-planes.
Droplet boundary is defined as the isodensity contour lines 
corresponding to the half-mode of the 2D density distribution 
(red contour),
and $\theta$ is calculated as $\theta=\arccos(1-h/R)$ with $h$ and $R$ 
the nominal height and the radius of the droplet. 
{\bf c)} Contact angle $\theta$ as a function 
of the temperature $T$ for different 
liquid-solid interaction $\epsilon_{LS}$.
The red circled point under the dashed line 
indicates that the liquid uniformly wets the entire surface.
For our nanopore simulations, we used a $\epsilon_{SL}=0.8 \epsilon_{LL}$,
corresponding to a hydrophilic surface with $\theta\simeq60^{\circ}$,
conditions indicated by the grey circled point.
}

\clearpage

\begin{figure*}
	\centering
	\includegraphics[width=\textwidth]{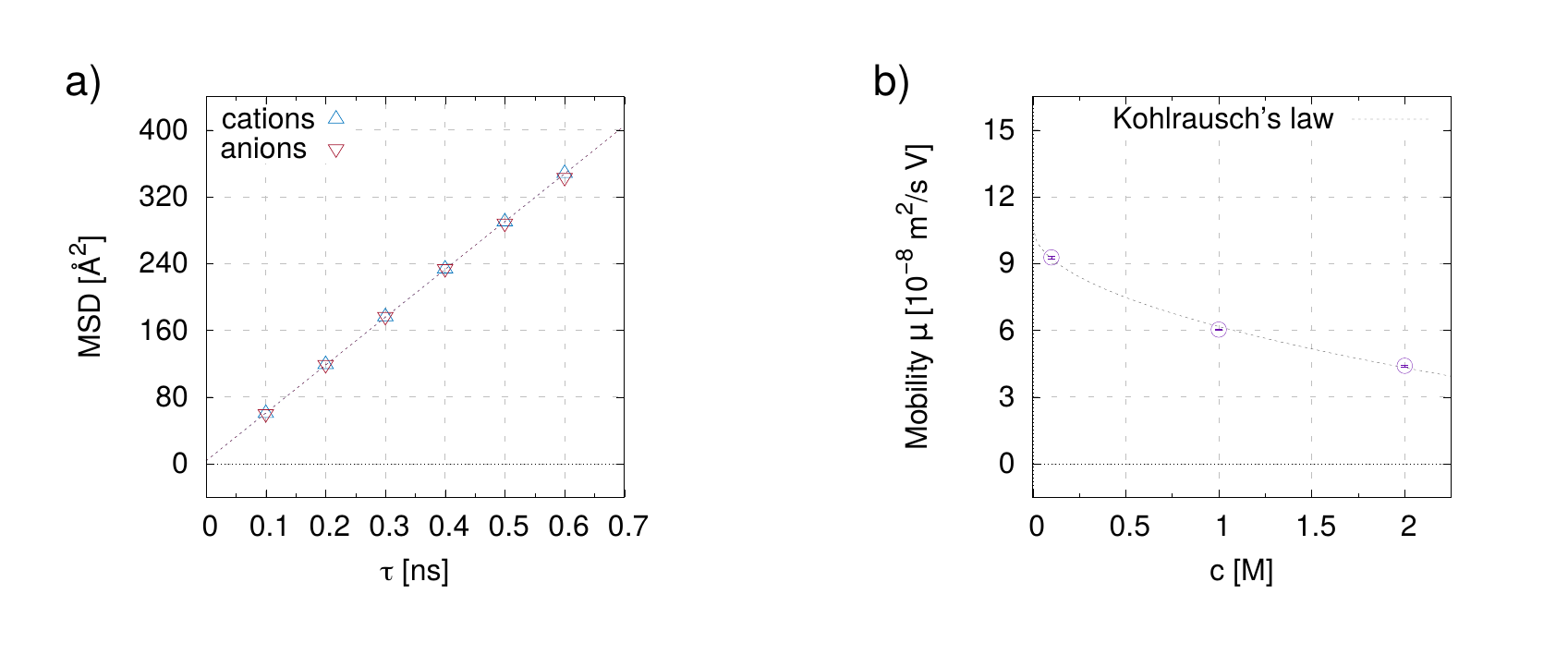}
\end{figure*}

\small
{\noindent
{\bf Supplementary Figure S9. }
{\bf Ion diffusion coefficients and mobility for the model symmetric electrolyte solution.}
{\bf a)} Mean squared displacement (MSD) as a function of simulation time,
for a $2\mathrm{M}$ triperiodic system ($313\,292$ total atoms, $11\,280$ ions).
After an NPT equilibration ($T=250~\mathrm{K}$, $P=1~\mathrm{atm}$),
a $4~\mathrm{ns}$ NVT production run is performed, 
sampling particle positions every $100~\mathrm{ps}$.
The first nanosecond is discarded and then 
the MSD is computed 
over the remaining frames
by using VMD scripts~\cite{humphrey1996vmd}.
The slope of the curve is related to the 
diffusion coefficient $D$
of the ion in the electrolyte solution
by the relation
$
	\mathrm{MSD}(\tau)
	= 6 D \tau 
$
~\cite{frenkel2001understanding},
resulting in the fitted values
$D_{+}=95.6~\pm~0.21~\mathrm{\AA^2/ns}$ and 
$D_{-}=94.4~\pm~0.73~\mathrm{\AA^2/ns}$. 
From the diffusion coefficient, 
the ion mobility is estimated via
the Einstein relation 
$
	\mu_\pm
	= q_\pm D_\pm/(k_B T) 
$.
Results at different concentrations are shown in {\bf b)}.
As predicted by Kohlrausch law~\cite{atkins2011physical},
$
	\mu(c)
	= 
	\mu_0 - K\sqrt{c}
$,
the ionic mobility decreases with increasing ion concentration, 
with $\mu_0$ the mobility at infinite dilution,
and $K$ an empirical constant.
}

\clearpage

\begin{figure*}
	\centering
	\includegraphics[width=0.75\textwidth]{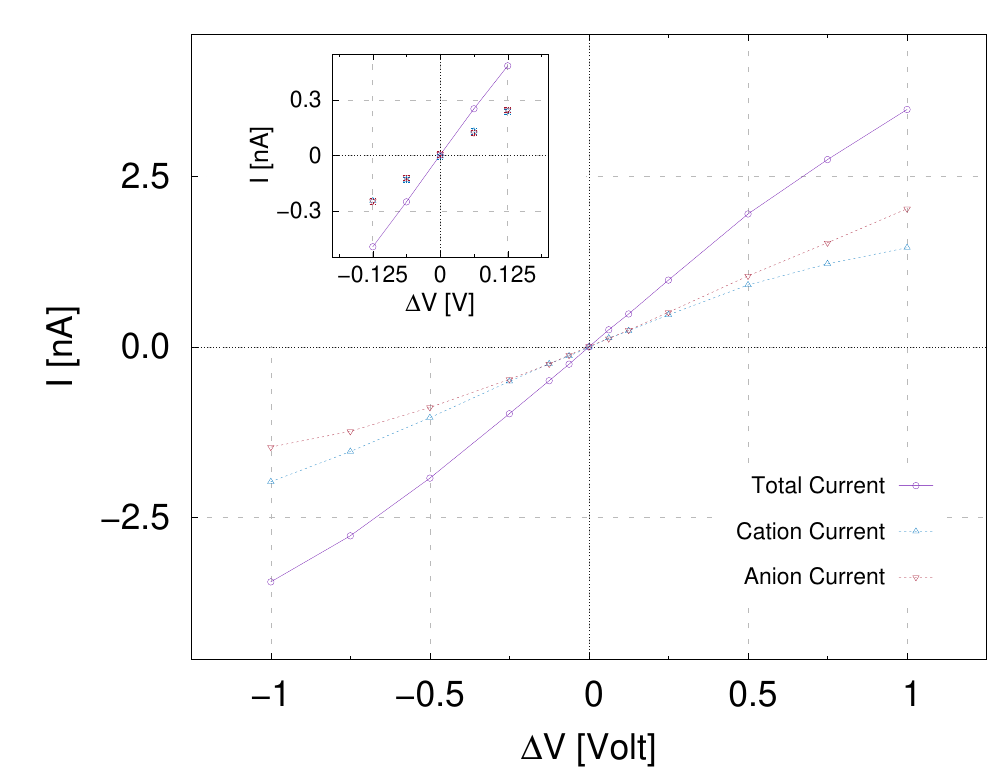}
\end{figure*}

\small
{\noindent
{\bf Supplementary Figure S10.}
{\bf Ionic currents for the model system}
 shown in Fig.~1d-f and Fig.~3a-c of the manuscript.
Currents are computed with the protocol described in the methods,
averaging over $800~\mathrm{ns}$ MD trajectories, 
for a total of $16\,000$ frames.
The first $30~\mathrm{ns}$ are discarded.
Errors are smaller than the point sizes and are calculated
using a block average protocol with a block length of $10$~ns.
It is apparent that for low voltages, $\vert \Delta V \vert \le 0.125 \mathrm{V}$,
the anion and cation currents are equal (see the inset).  
This is expected since, at low $\Delta V$, the external voltage does
not alter the equilibrium ionic distributions.
Since the pore surface is neutral and the salt is symmetric, 
the distributions
of anions and cations are identical at equilibrium.
Moreover, in our model anion and cations have the same mobility, 
consequently the anion and cation currents have the same value. 
}
\clearpage

\begin{figure*}
	\centering
	\includegraphics[width=\textwidth]{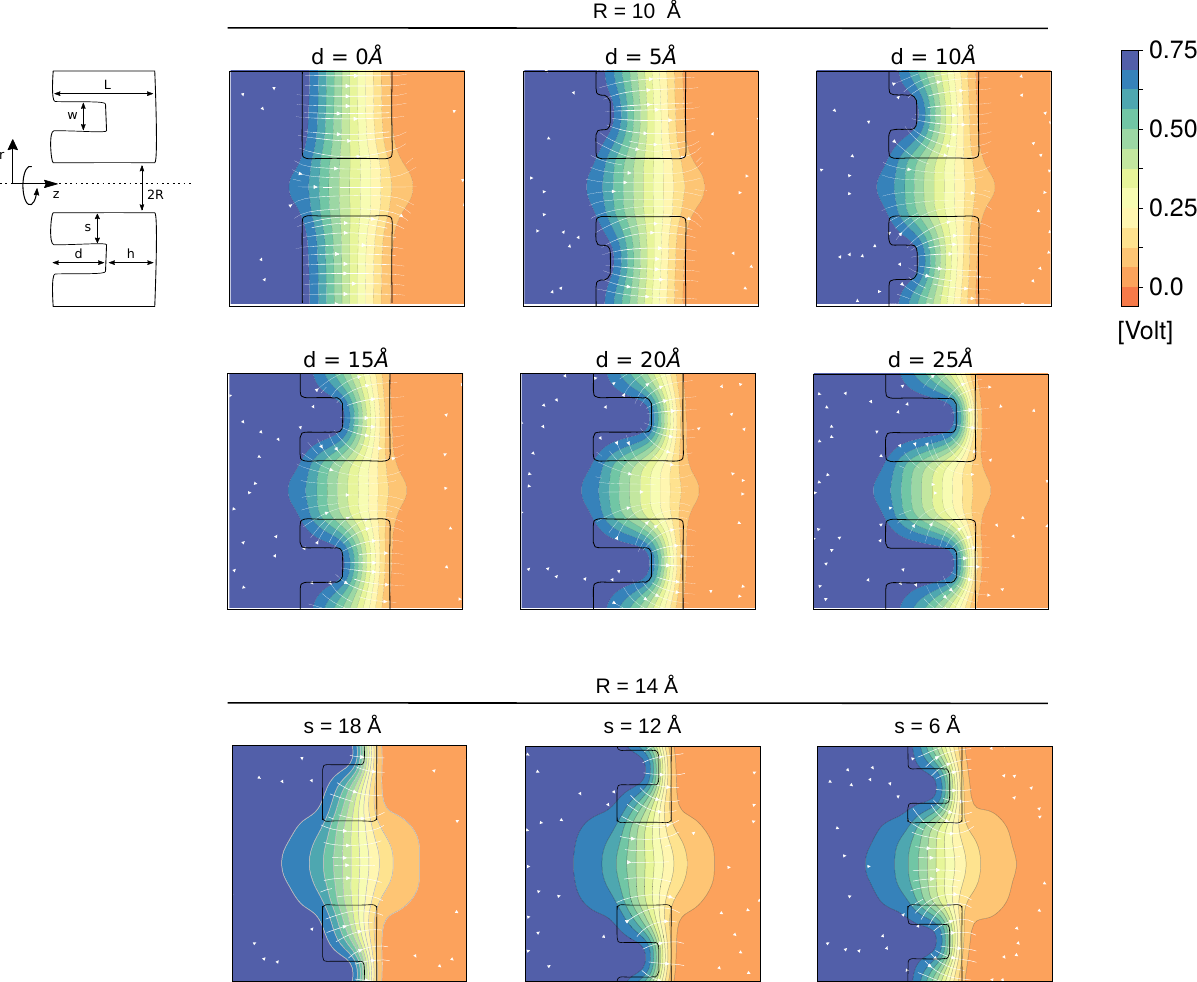}
\end{figure*}
\small
\noindent
{\bf Supplementary Figure S11.}
{\bf Electric Potential and Field lines for different cavity sizes.}
The white arrowed lines represent the electric field
$\mathbf{E}(r,z)=-\nabla V$.
We filtered out the lines where
$|\mathrm{E(r,z)}|<15\%$ of the maximum intensity.
The potential map is averaged
over $800$~ns MD trajectory (16\;000 frames),
at $\Delta V= +0.75\,$V transmembrane applied bias, in the presence of the model symmetric electrolyte.
For the top group,
the other geometric parameters are those
of Fig.~3a-c of the manuscript, namely,
$R=10$\AA, $L=30$\AA, $s=9$\AA, $w=12$\AA.
For the bottom group the geometric parameters are those of
Fig.~3f, namely
$R=14$\AA, $L=18$\AA, $h=5\AA$, $w=12$\AA.
The intensity
of the electric field is larger in the solid
region close to the cavity
(where the membrane thickness is small)
accordingly with the model already reported in Supplementary Note S1.
It can be noted that the flux of the radial component of the electric field
normal to the pore wall $\mathrm{E_r}(R,z)$,
increases as the cavity depth $d$ increases 
and as the coaxial separating membrane thickness $s$ decreases
(both for fixed $L$).
\clearpage

\newpage

\begin{figure*}
	\centering
	\includegraphics[width=\textwidth]{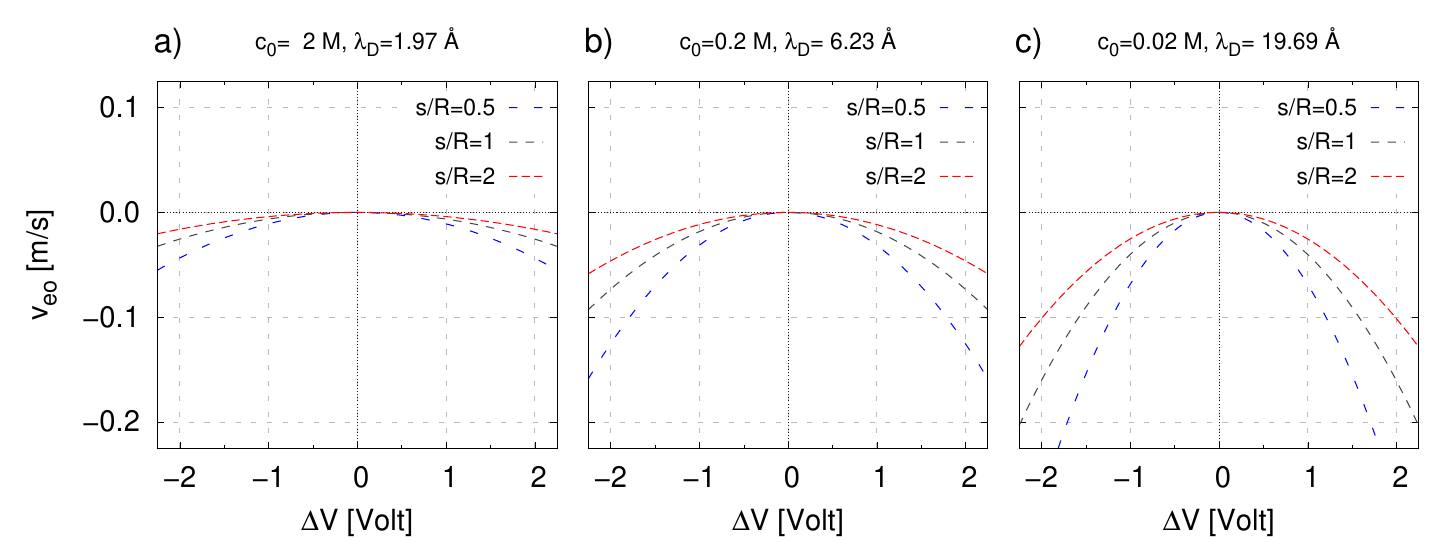}
\end{figure*}

\small
{\noindent
{\bf Supplementary Figure S12.}
{\bf EOF predictions for a neutral silicon nitride nanopore.}
We considered a pore
of radius $R=20~\mathrm{nm}$ and length $L=20~\mathrm{nm}$.
The lateral cavity has a depth $d=10~\mathrm{nm}$ and it
is placed at a distance $s = 40\mathrm{nm}$ (red line, $s/R = 2$),
$s = 20~\mathrm{nm}$ (grey line, $s/R = 1$) and
$s = 10~\mathrm{nm}$ (blue line, $s/R = 0.5$).
The relative electrical permittivity
for solid and liquid are
$\varepsilon_S=7.5$ (silicon nitride) and
$\varepsilon_L=80$ (water).
Panels represent
different electrolyte concentration:
{\bf a)} 2~M, {\bf b)} 0.2~M and {\bf c)} 0.02~M.
The corresponding Debye lengths $\lambda_D$
are also reported on the top of each panel.
The parabolic unidirectional EOF is obtained using Eq.~(8)
of the manuscript.
Even for the lowest concentration (0.02 M), 
a total of $\simeq 600$ ions ($300$ for each species)
are present inside the nanopore, 
{\sl i.e.}, a number high enough to expect that a PNP-NS model
provides a good estimation of the flows.

\clearpage

\begin{figure*}
	\centering
	\includegraphics[width=0.95\textwidth]{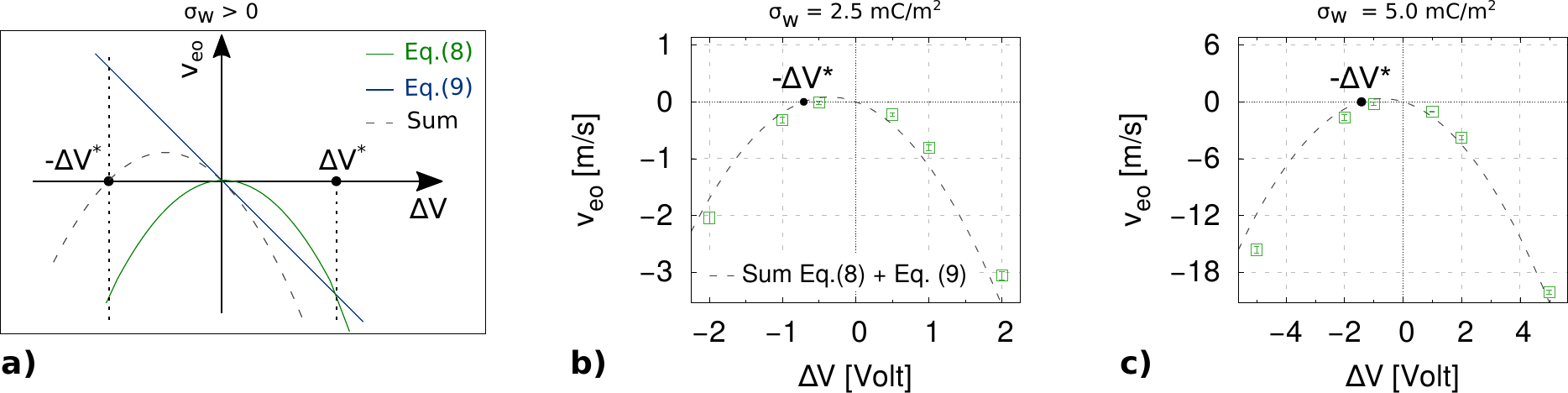}
\end{figure*}
\small
\noindent
{\bf Supplementary Figure S13.}
{\bf 
	MD simulation of a weakly charged model pore and analytical estimates.
}
{\bf a}) Sketch of the superposition of the two contributions of geometrically induced EOF,
 Eq.~(8) of the manuscript (green parabola), 
and of fixed charge EOF, Eq.(9) (blue line), 
as a function of applied voltage $\Delta V$. 
The resulting sum is shown as a dashed curve. Neglecting the trivial case at null voltage,
 there are two voltages in which the effects are equal in magnitude.
For positive $\sigma_w$ at $\Delta V$ = -$\Delta V^*$ the two contributions cancel out, yielding  $v_{eo}=0$.
{\bf b}) and {\bf c}) MD simulation of a nanopore system similar 
to the one shown Fig. 1d of the manuscript,
with geometrical parameters 
$R=10$\AA, $L=30$\AA, $h=5$\AA, $s=9$\AA, $w=12$\AA,
modified by adding a charge density $\sigma_w=2.5$ $\mathrm{mC/m^2}$ 
(panel {\bf b}) or $\sigma_w=5.0$ $\mathrm{mC/m^2}$ (panel {\bf c}).
The pore is immersed in 2M of our symmetric electrolyte and 
non-equilibrium simulations are performed as for the system of Fig.~1d-f of the manuscript,
see Methods. The green squares represent MD data points.
In several cases, error bars are much smaller than the data points.
The dashed curve is an estimate of total EO velocity, calculated as a sum of 
the contribution due to fixed surface charge and the geometrically induced flow,
by using the theoretical estimations reported in Eq.~(8) and Eq.~(9) of the manuscript, 
respectively.
The threshold voltages $-\Delta V^*$, as estimated by Eq.~(10) of the manuscript,
are represented on the curves by a filled circle 
and labeled accordingly.
The good agreement between the MD data and the theoretical predictions 
indicates that the theory, although derived under strong assumptions, provides a 
good description of the competition between fixed and induced charge effect,
 at least in the case of weakly charged pores.

\clearpage

\begin{figure*}
	\centering
	\includegraphics[width=0.65\textwidth]{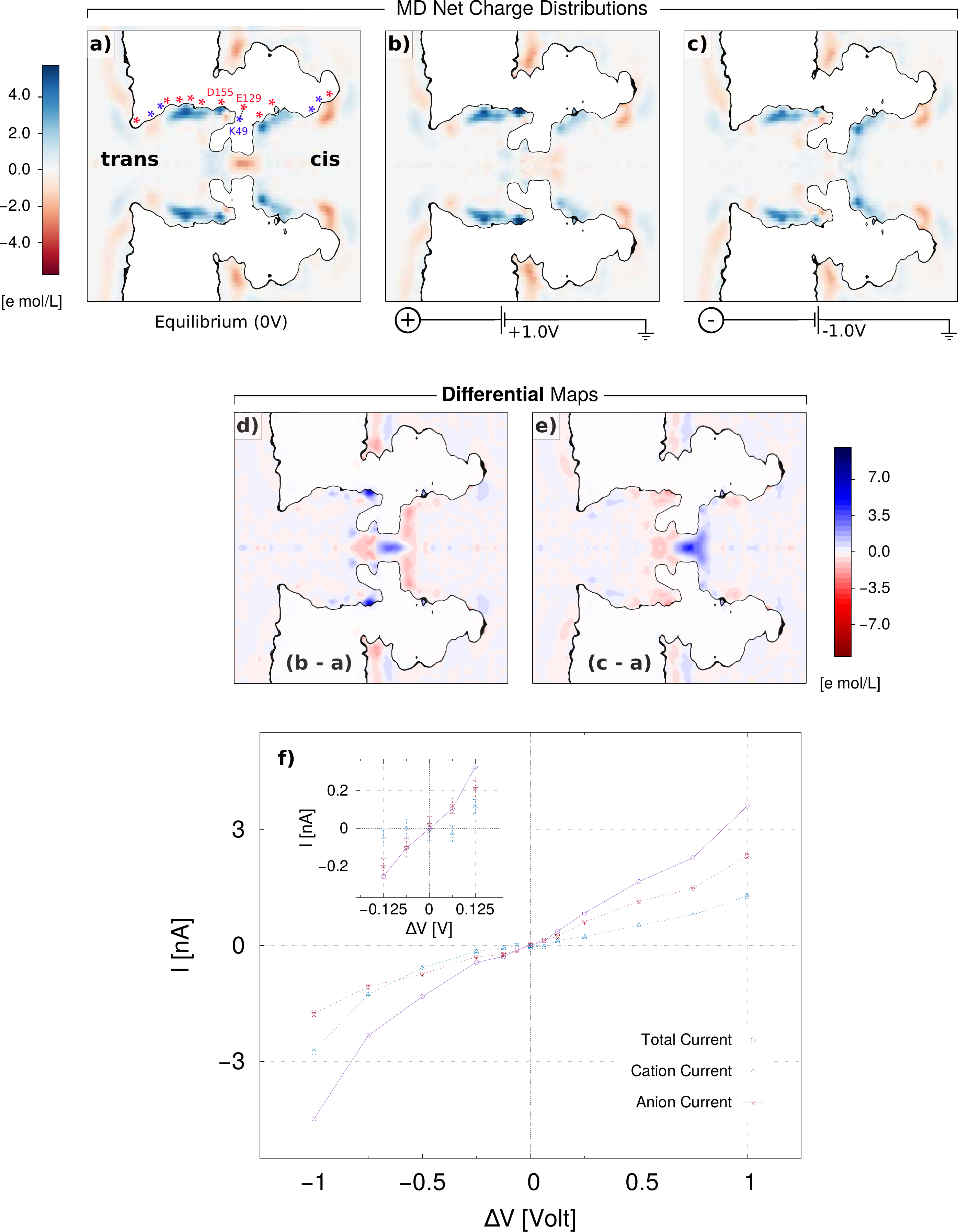}
\end{figure*}
\small
\noindent
{\bf Supplementary Figure S14.}
{\bf 
Net charge distribution and 
ionic currents
for the CsgG nanopore,
}
Fig.~5 of the main text,
computed from MD simulations in 2M KCl water solution.
The first three panels ({\bf a-c}) contain the 
maps already reported in 
Fig.~5{\bf b-c}.
Panels {\bf d-e}, instead, represent the differential maps 
between the non-equilibrium systems
($\Delta V=+1\,$V, panels {\bf b} and {\bf c})
and the equilibrium one ($\Delta V=0\,$V).
The black line delimiting the pore and the membrane is the water density contour
level $\rho = 0.5 \rho_{bulk}$, 
with $\rho_{bulk}$ the bulk water density. 
Maps are obtained from 280 ns MD production runs. 
All the trajectories are sampled each 20 ps, 
and analyzed discarding the first 10 ns.
The applied $\Delta V$ results in a 
strong reduction of the dipole in the pore constriction observed 
at $\Delta V = 0$, for this reason, 
the differential maps 
({\bf c - a})
and 
({\bf b - a})
present a dipole in the constriction.
In the differential maps, it is also apparent 
that the cavity changes 
its charge when reverting the voltage.
In particular, at $\Delta V = 1$V differential maps ({\bf d})
show a positive variation of the cavity charge
while the opposite occurs at $\Delta V = -1$V ({\bf e}).
Panel {\bf f} reports the total electric current and single ionic currents.
Currents and maps are computed with the protocols
described in the methods,
averaging over 
$280~\mathrm{ns}$ MD production trajectories,
for a total of $14\,000$ frames.
All the trajectories are sampled each $20~\mathrm{ps}$,
and the first $10~\mathrm{ns}$ are discarded.
Errors are calculated
using a block average protocol with a block length of $10$~ns.
\clearpage

\begin{figure*}
	\centering
	\includegraphics[width=\textwidth]{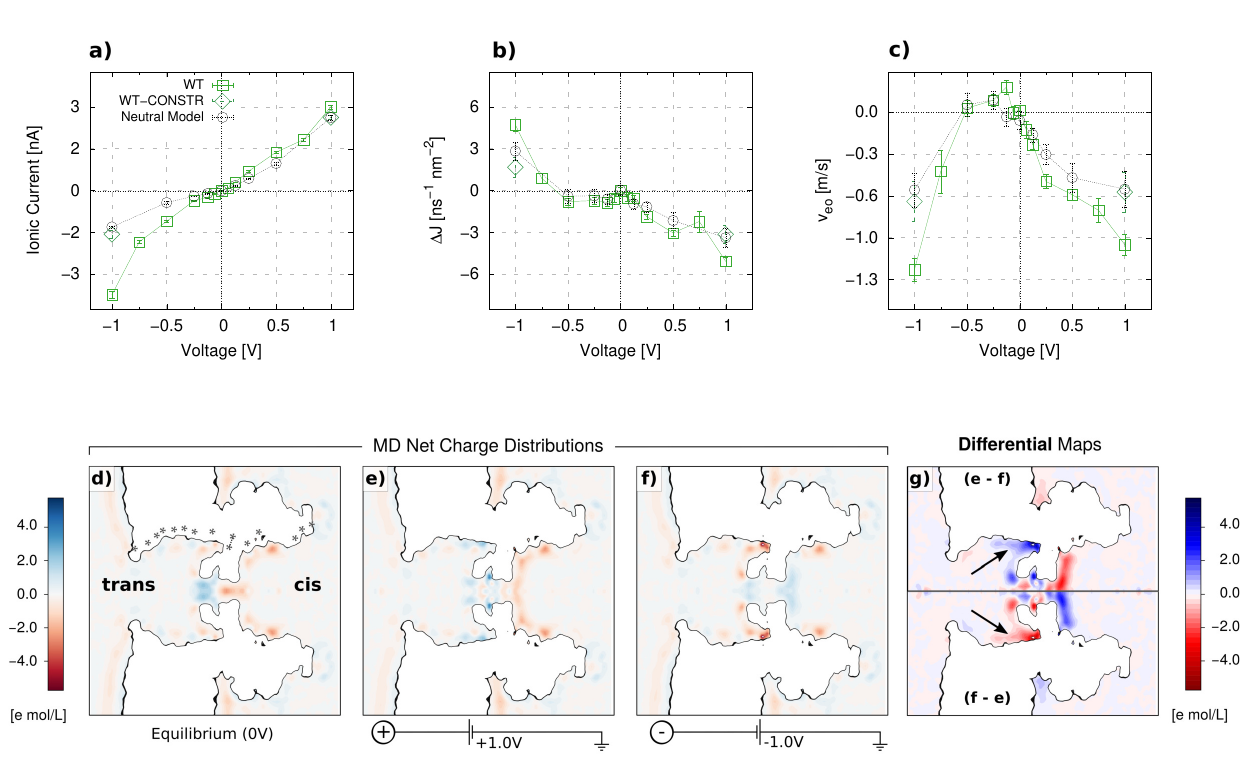}
\end{figure*}
\small
\noindent
{\bf Supplementary Figure S15.}
{\bf Comparison between the CsgG nanopore,
Fig.~5 of the manuscript, 
and a Neutral Model of the same pore.}
The Neutral Model is obtained by neutralizing
the net charges of the acidic and basic residues,
using standard neutral CHARMM patches of the charged residues.
MD simulations are conducted in 2~M~KCl water solution, as for Fig.~5.
The general trends for
{\bf a)} ionic currents, {\bf b)} selectivity and {\bf c)} electroosmotic
velocity are similar to the Wild Type CsgG nanopore.
The quantitative differences 
in the fluxes at high voltages
are due to slight structural changes in the constriction. 
To prove this, we repeated the simulation of the wild type
CsgG by keeping the structure constrained upon the Neutral Model.
The results, green diamond points (WT-CONSTR) at $\Delta V=\pm1.0V$,
overlap with Neutral Models.
Currents are computed with the protocol described in the methods.
Each point is obtained by averaging over $240~\mathrm{ns}$ MD trajectory, 
for a total of $120\,000$ frames. 
The first $10~\mathrm{ns}$ are discarded.
Errors are calculated
using a block average protocol with a block length of $10$~ns.
Panels {\bf d-g} report the charge density map 
for the Neutral Model. 
Charge distribution at the walls of cis and trans vestibule
differs from the Wild Type (Fig.~5.b-c of the manuscript). 
This is expected since the exposed residues  
that were originally charged in the Wild Type are now
neutral. 
Instead, the charge distribution in the constriction 
is only slightly altered with respect to the Wild Type.

\clearpage

\newpage

\begin{figure*}
	\centering
	\includegraphics[width=\textwidth]{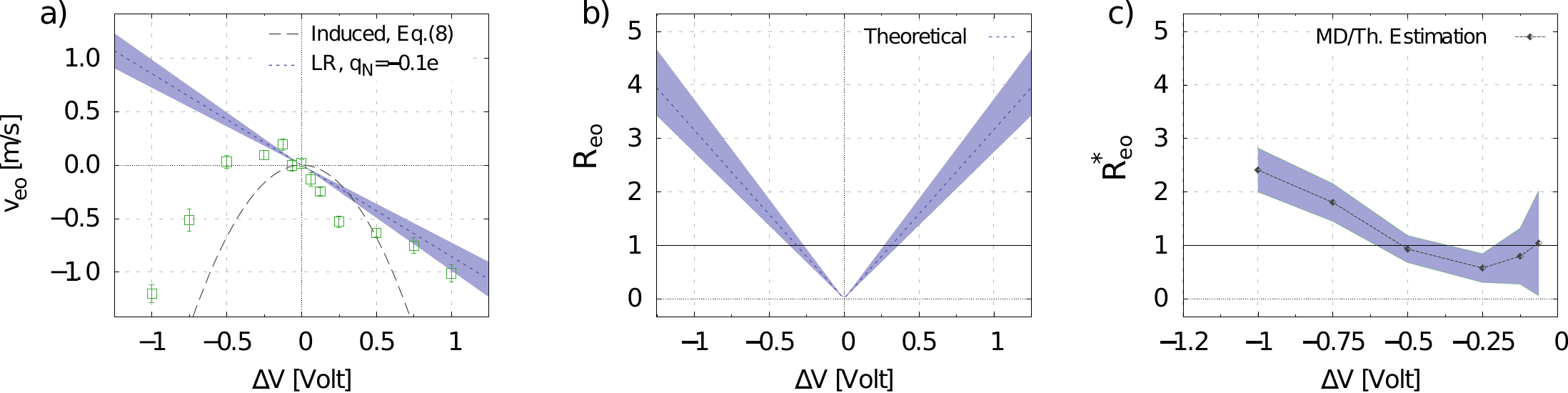}
\end{figure*}
\small
\noindent
{\bf Supplementary Figure S16.}
{\bf Comparison between Induced Charge EOF and Intrinsic Selectivity EOF for the CsgG nanopore.}
	To estimate the relative contribution of our geometrically induced 
	mechanism on observed EOF we applied a simple additive model 
	where the $v_{eo}$ is interpreted as the sum 
	of a linear response (LR) contribution 
	due to intrinsic selectivity and the geometrically induced
	selectivity contribution.
        {\bf a)} Electroosmotic average velocity $v_{eo}$
        as a  function of the applied voltage $\Delta V$.
	Green squared points and dashed grey line are the MD data 
	and theoretical prediction of the geometrically induced charge model 
	as reported in Fig.~5.
        The LR linear curve is obtained from Eqs.~(5-6) of the manuscript,
        considering a fixed $q_N=-0.10 \pm 0.015$e inside the pore lumen.
	{\bf b)} Ratio $R_{eo}$ between the theoretical
        induced charge contribution $v_{eo,ICEO}$ (dashed parabola in panel {\bf a})
        and the intrinsic linear $v_{eo, LR}$ contribution to EOF (dotted line in panel {\bf a}),
	$$
		R_{eo} = \frac{\vert v_{eo,ICEO}\vert }{\vert v_{eo,LR} \vert} \; .
	$$
        {\bf c)} Ratio $R_{eo}^*$ between the
        estimated induced charge contribution $v_{eo,ICEO}^*$ and $v_{eo, LR}$, where
        $v_{eo,ICEO}^*$
        is computed at each $\Delta V$ as the difference between the
        MD measured EOF and the LR value,
        $v_{eo,ICEO}^*=v_{eo, MD} - v_{eo, LR}$. 
	This expression, as well as the overall analysis discussed in this 
	supplementary figure, implicitly assumes 
	a superposition of effects, {\sl i.e.} the total EOF
	can be decomposed as the sum of a linear contribution 
	due to intrinsic selectivity and a quadratic contribution 
	due to induced charge. This approach is similar 
	to the one used in the section 
	``Application to weakly-charged solid-state nanopores'' of the manuscript
	and also discussed in Fig.~S13, although in the case
	of CsggS the selectivity is not due to a net fixed charge present 
	at the pore wall. This hypothesis is quite strong and,
	in general, unjustified. Consequently, these arguments 
	can be used to get preliminary approximate voltage ranges 
	where the intrinsic selectivity or the induced charge mechanism dominate the EOF, 
	while simulations are needed for a reliable non-equilibrium description of the transport.
\clearpage

\newpage

\begin{table}[h]
\begin{center}
\begin{tabular}{ p{3.3cm} | p{2.1cm}  |p{9.3cm}}
\hline
 Paper & Material & Notes on surface charges $\sigma_w$ and point of zero charge\\ 
\hline
\hline
\parbox[t]{3.2cm}{
\raggedright{
 Lin {\sl et al.} \cite{lin2021surface} 
}}
& 
Silicon Nitride 
&
From $\sigma_w = 0.0027~\mathrm{C/m^2}$ ($\mathrm{pH} = 1.2$)
to $\sigma_w = -0.2~\mathrm{C/m^2}$ ($\mathrm{pH} = 11$).
Point of zero charge at $\mathrm{pH}=3.3 - 4.1$ depending on the
salt concentration.
The paper reports an analytical model fitted
on experimental data for the dependence of $\sigma_w$ on 
the pH. This relation is used in Fig.4b of the manuscript. 
\\  
\hline
\parbox[t]{3.2cm}{
\raggedright{
Hoogerheide {\sl et al.}
\cite{hoogerheide2009probing}
 \\
}}
& 
Silicon Nitride
&
From $\sigma_w=-0.04 \mathrm{C/m^2}$ ($\mathrm{pH} = 2$)
to $\sigma_w = 0.02\mathrm{C/m^2}$ ($\mathrm{pH} = 8$).
Point of zero charge at $\mathrm{pH} \simeq 4.1$ measured at 0.11~M KCl. 
 \\  
\hline
%
%
\parbox[t]{3.2cm}{
	\raggedright{
		Bandara {\sl et al.} \cite{bandara2019chemically}
}}
& 
Functiona\-lized Silicon Nitride
&
The manuscript reports expressions for $\sigma_w$ 
as a function of the pH. Using these expressions and the
values reported in the Supporting Information
of that manuscript, it is possible to calculate $\sigma_w$. 
Very small surface charge density is achieved for a wide range 
of pH, for instance, 
$\vert \sigma_w \vert < 0.001~\mathrm{C/m^2}$ for $\mathrm{pH} < 7.5$ 
for $\mathrm{\mbox{-}OH}$ functionalization and
$\vert \sigma_w \vert < 0.003~\mathrm{C/m^2}$ for $\mathrm{pH} > 7.5$ 
 for $\mathrm{\mbox{-}NH_2}$ functionalization.
\\  
\hline
%
%
\parbox[t]{3.2cm}{
\raggedright{
 Kosmulski~\cite{kosmulski1997attempt} 
}}
& 
HfO$_2$ 
&
From $\sigma_w = -0.1 \mathrm{C/m^2}$ to $\sigma_w = 0.1 \mathrm{C/m^2}$ depending on
pH. Point of zero charge at $pH \simeq 7.5$.
This material was used for nanopore experiments in~\cite{larkin2014high}.
 \\  
\hline
%
%
\parbox[t]{3.2cm}{
\raggedright{
 Xie {\sl et al.} \cite{xie2009surface}
}}
& 
PET coated with CTAB
&
From $\sigma_w = -0.01 \mathrm{C/m^2}$ 
to $\sigma_w = 0.01\mathrm{C/m^2}$ with increasing amount of coating with 
(CTAB cetyl trimethyl ammonium bromide), a cationic surfactant.
 \\  
\hline
%
%
\end{tabular}
\end{center}
\vspace{-0.5 cm}
\caption{
{\bf Examples of surface charges 
and point of zero charge 
for some solid-state nanopores materials.}
A first possibility 
to get a neutral charged solid state nanopore 
is to use Silicon Nitride. 
The surface charge $\sigma_w$ varies with the pH. 
The point of zero charge pH is around 4.1 as confirmed by several works
by different groups, see, {\sl e.g.}~\cite{hoogerheide2009probing,lin2021surface}.
Silicon Nitride can also be functionalized 
allowing to get very weak charge ($< 10~\mathrm{mC/m^2}$) at
pH 7~\cite{bandara2019chemically}.
Another material used for nanopores is $\mathrm{Hf O_2}$~\cite{larkin2014high},
the point of zero charge pH of which is $\sim 7.5$~\cite{kosmulski1997attempt}.
Functionalization may also be used on PET nanopores~\cite{xie2009surface}
to tune the surface charge.
A wide list of experimental works
on pH-dependent surface charging and points of zero charge for
several materials is reported in~\cite{kosmulski2002ph}.
Some of these materials are currently used in nanopore experiments.
} 
\end{table}

\clearpage

\newpage

\bibliography{nanopore2021}
\bibliographystyle{unsrt}